\documentclass[twocolumn,notitlepage,showkeys,showpacs]{revtex4-1}
\usepackage[utf8]{inputenc}
\usepackage{stmaryrd}
\usepackage{amsmath}
\usepackage{amssymb}
\usepackage{graphicx}
\usepackage{dcolumn}
\usepackage{bm}
\usepackage{multirow}
\usepackage[colorlinks,linkcolor=blue,citecolor=blue,urlcolor=blue]{hyperref}
\usepackage{color}
\usepackage{ulem}
\usepackage{comment}

\usepackage{booktabs}

\begin{document}
\title{
Distinct magnetic gaps between antiferromagnetic and ferromagnetic orders driven by surface defects in the topological magnet MnBi$_2$Te$_4$
}
\author{Hengxin Tan}
\author{Binghai Yan}
\affiliation{Department of Condensed Matter Physics, Weizmann Institute of Science, Rehovot 7610001, Israel}

\date{\today}

\begin{abstract}
The magnetic topological insulator, MnBi$_2$Te$_4$, shows metallic behavior at zero magnetic fields (antiferromagnetic phase, AFM) in thin film transport, which coincides with gapless surface states observed by angle-resolved photoemission spectroscopy, while it can become a Chern insulator at field larger than 6 T (ferromagnetic phase, FM). Thus, the zero-field surface magnetism was once speculated to be different from the bulk AFM phase. However, recent magnetic force microscopy refutes this assumption by detecting persistent AFM order on the surface. In this work, we propose a mechanism related to surface defects that can rationalize these contradicting observations in different experiments. We find that co-antisites (exchanging Mn and Bi atoms in the surface van der Waals layer) can strongly suppress the magnetic gap down to several meV in the AFM phase without violating the magnetic order but preserve the magnetic gap in the FM phase. The different gap sizes between AFM and FM phases 
are caused by the defect-induced surface charge redistribution among top two van der Waals layers. This theory can be validated by the position- and field-dependent gap in future surface spectroscopy measurements. Our work suggests suppressing related defects in samples to realize the quantum anomalous Hall insulator or axion insulator at zero fields.

\end{abstract}

\maketitle

\textit{Introduction ---} Magnetic topological insulators (TIs) have long been sought for the realization of the quantized anomalous Hall effect (QAHE) \cite{chang2013} and axion insulator (AI) \cite{sciadv.aao1669,PhysRevLett.120.056801}. Recent discovery of an intrinsic magnetic TI, MnBi$_2$Te$_4$ \cite{gong2019experimental,otrokov2019prediction,li2019intrinsic,PhysRevLett.122.206401}, sparks extensive research interest in realizing QAH and AI \cite{deng2020science,nwaa089,liu2020robust,deng2021high,yan2021elusive}, and leads to discovering a general family of magnetic TIs, MnBi$_2$Te$_4$(Bi$_2$Te$_3$)$_n$ \cite{wu2019natural,PhysRevX.11.031032,lei2020magnetized,PhysRevLett.127.236402,WANG2021100098,Zhao2021,li2021large,klimovskikh2020tunable,hu2020sciadv,wu2020toward}.
MnBi$_2$Te$_4$ is a van der Waals (vdW)-type layered material. Its ground state exhibits the A-type antiferromagnetic (AFM) order where each vdW layer shows ferromagnetic (FM) coupling with the easy axis orienting out-of-plane while adjacent layers couple in an AFM way.
MnBi$_2$Te$_4$ was predicted to be a magnetic TI in which the Dirac states open a magnetic gap 
on the vdW surface 
\cite{PhysRevLett.122.107202,PhysRevB.102.035144,PhysRevB.102.245136,zhong2021light,PhysRevX.10.031013}, which is promising for the realization of QAHE and AI.

Angle-resolved photoemission spectroscopy (ARPES) experiments \cite{PhysRevResearch.1.012011,shikin2021sample,Ji2021Detection,PhysRevX.9.041038,PhysRevX.9.041039,PhysRevX.9.041040,PhysRevX.9.041065,PhysRevB.101.161109,XU20202086,PhysRevX.10.031013,hu2020natcom,PhysRevB.101.161113,PhysRevB.102.045130,PhysRevLett.126.176403} usually observed gapless surface states, contradicting theoretical predictions. Thus, it was once speculated that the original surface FM order was broken to rationalize the absence of a magnetic gap \cite{PhysRevX.9.041038,PhysRevX.10.031013,PhysRevB.101.161109}. However, the following magnetic force microscopy \cite{PhysRevLett.125.037201} and ARPES \cite{PhysRevLett.125.117205} experiments revealed a robust A-type AFM order on the surface.

In transport experiments, realizing the bulk-insulating topological states remains a major challenge \cite{zhao2021routes,yan2021elusive,Lee2019}, although the QAHE \cite{deng2020science} and AI \cite{liu2020robust} states were reported in MnBi$_2$Te$_4$ thin films. 
A film with an odd number of MnBi$_2$Te$_4$ layers, which was expected to exhibit QAHE, commonly shows non-quantized Hall resistance at zero fields. 
However, the same film presents a quantized Hall resistance under high magnetic fields $>$ 6 T when the film is polarized to the FM phase \cite{deng2020science,nwaa089,nanolett.0c05117,liu2020robust,liu2021magnetic,ying2021experimental,cai2022electric}.
That is to say, the AFM phase is gapless while the FM phase exhibits a gap in the same thin film, although calculations predicted comparable magnetic gaps for both phases. 

To reconcile the above controversial observations in MnBi$_2$Te$_4$, a potential mechanism should fulfill at least three conditions. (i) It can explain the gapless or nearly gapless surface states to account for ARPES and zero-field transport experiments. (ii) It should respect the A-type AFM order on the surface and bulk region at zero fields. 
(iii) It can provide a sizable energy gap on the surface in the FM phase. 

Because experiments revealed many defects such as antisites in samples \cite{PhysRevB.103.184429,PhysRevMaterials.4.121202,PhysRevX.11.021033,acsnano.0c03149,PhysRevX.11.021033,shikin2021sample}, 
the influence of defects or surface disorders on the magnetic gap has been actively investigated recently \cite{PhysRevB.102.245136,acs.nanolett.0c00031,PhysRevB.102.241406,shikin2020nature,wang220508204,Shikin2022factors,garnica2022native,PhysRevB.103.184429}. Most of these works aim to rationalize the gapless nature of the AFM phase by fulfilling conditions (i)-(ii) but not (iii).
For example, the reference \cite{garnica2022native} showed dramatic magnetic gap reduction on the AFM surface by arguing that the Mn$_{\rm{Bi}}$ antisite (extra Mn replacing Bi) presents a magnetic moment anti-parallel to other Mn moments in the same vdW layer \cite{PhysRevB.103.184429}, i.e., magnetism cancellation by Mn$_{\rm{Bi}}$ antisites. However, as we will show in this work, parallel alignment of the Mn$_{\rm{Bi}}$ moment also strongly suppresses the magnetic gap, contradicting the above scenario.

%%%%%%%%%%%%%%%%%%%%%%%%%%%%%%%%%%%%%%%%%%%%%%%%%%%%%%%%%%%%%%%%%%%%%%%%%%%%%
\begin{figure*}[tbp]
\includegraphics[width=1.95\columnwidth]{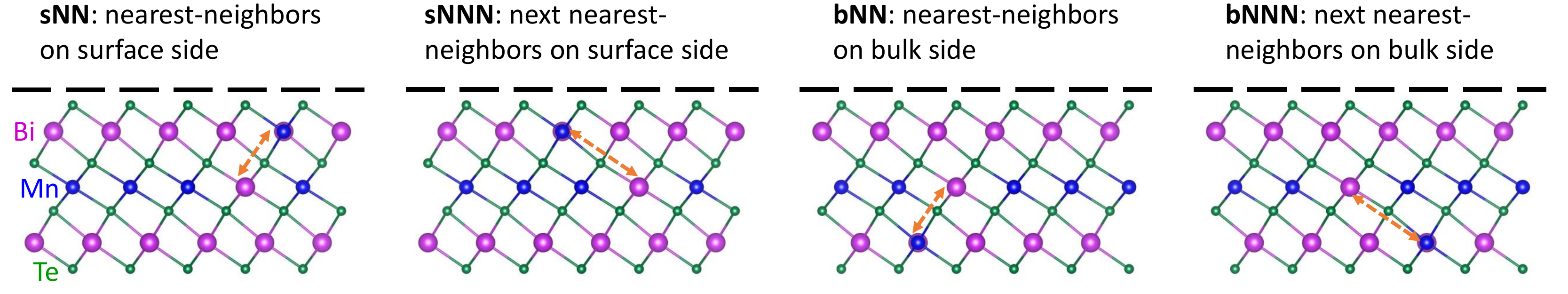}
\caption{\label{Fig1-antisites-33} Cation co-antisite configurations in the surface SL.
\textbf{sNN}: nearest-neighbouring Mn$_{\rm{Bi}}$ and Bi$_{\rm{Mn}}$ on the surface side;
\textbf{sNNN}: next nearest-neighbouring Mn$_{\rm{Bi}}$ and Bi$_{\rm{Mn}}$ on the surface side;
\textbf{bNN}: nearest-neighbouring Mn$_{\rm{Bi}}$ and Bi$_{\rm{Mn}}$ on the bulk side;
\textbf{bNNN}: next nearest-neighbouring Mn$_{\rm{Bi}}$ and Bi$_{\rm{Mn}}$ on the bulk side.
For simplicity, only the surface SL is shown where the bold dashed line indicates the surface.
The orange vector in each panel connects the antisites Mn$_{\rm{Bi}}$ and Bi$_{\rm{Mn}}$.
}
\end{figure*}
%%%%%%%%%%%%%%%%%%%%%%%%%%%%%%%%%%%%%%%%%%%%%%%%%%%%%%%%%%%%%%%%%%%%%%%%%%%%%

In this work, we find that cation co-antisites Mn$_{\rm{Bi}}$ and Bi$_{\rm{Mn}}$ (extra Bi replacing Mn) can resolve puzzles regarding the surface magnetic gap by satisfying all three conditions. These defects push topological surface states, which would distribute dominantly on the first vdW layer and show a sizable magnetic gap (60$\sim$70 meV) on the clean sample, partially into the second vdW layer. In the AFM phase, magnetic exchange interactions (with surface states) from the first and second layers cancel each other and significantly diminish the magnetic gap to several meV. In the FM phase, however, exchange interactions from the top two layers exhibit no cancellation and thus preserve the magnetic gap. 
This mechanism requires no reconstruction of surface magnetism at zero fields [condition (ii)]. In addition, we find that the resultant surface gap is insensitive to the orientation of Mn$_{\rm{Bi}}$ moment. Our results also provide insights to understand gapless surface states of MnBi$_2$Te$_4$(Bi$_2$Te$_3$)$_n$ ($n=1,2$) observed in the AFM phase \cite{hu2020natcom,klimovskikh2020tunable} and even in the FM phase \cite{klimovskikh2020tunable,hu2020sciadv}.

\textit{Method --}
In the following, each MnBi$_2$Te$_4$ vdW layer is called a septuple layer (SL).
All the calculations are performed within density functional theory as implemented in Vienna $ab$-$initio$ Simulation Package \cite{VASP,PRB54p11169}.
The generalized gradient approximation describes the exchange-correlation interaction as parameterized by Perdew-Burke-Ernzerhof \cite{PRL77p3865}.
The cutoff energy is 350 eV for the plane wave basis. The Mn $d$ electrons are treated with a Hubbard $U$ of 5 eV.
The surface band structures under different surface co-antisites are simulated with slab models of 5-SL thick.
The in-plane supercell is examined from $2\times2$ to $4\times4$ to investigate the antisite density effect.
Three different magnetic configurations are considered, i.e., FM where all magnetic moments are parallel, A-type AFM where the Mn$_{\rm{Bi}}$ moment is parallelly coupled to other magnetic moments in the same SL, and Ferri-AFM where the Mn$_{\rm{Bi}}$ moment is antiparallel to other moments in the same SL in the A-type AFM framework (see Supplemental Material\cite{SM} for details).

\textit{Results--}
We start from the configurations of antisites Bi$_{\rm{Mn}}$ and Mn$_{\rm{Bi}}$ on the surface of MnBi$_2$Te$_4$.
Previous works \cite{acsnano.0c03149,AFMp2006516} found that Bi$_{\rm{Mn}}$ and Mn$_{\rm{Bi}}$ have relatively low formation energies, confirming the common existence of these defects.
The single cation antisite dramatically changes the surface band dispersion due to charge doping (see Supplemental Material \cite{SM}), which deviates from experimentally observed band structures.
In this work, we consider neutral defect as formed by the combination of Bi$_{\rm{Mn}}$ and Mn$_{\rm{Bi}}$ antisites with close distance. If two antisites are far separated, the band structure, which can be the overlay of the band structure of single antisite cases, may still deviate from experiments.
We consider four co-antisite configurations of Bi$_{\rm{Mn}}$-Mn$_{\rm{Bi}}$ in the surface SL, i.e., sNN, sNNN, bNN, and bNNN as defined in Fig. \ref{Fig1-antisites-33}.

We find that the next nearest-neighboring Bi$_{\rm{Mn}}$-Mn$_{\rm{Bi}}$ (sNNN and bNNN in Fig. \ref{Fig1-antisites-33}) are more stable in energy (dozens of meV) than the corresponding nearest-neighboring Bi$_{\rm{Mn}}$-Mn$_{\rm{Bi}}$ (sNN and bNN in Fig. \ref{Fig1-antisites-33}) under the same magnetic configuration, and the sNNN configuration has the lowest total energy \cite{SM}.
Under high defect density (in-plane cell size 2$\times$2), total energies of different antisite configurations under Ferri-AFM are slightly smaller than those under A-type AFM (within 6 meV).
With decreasing the defect density (e.g., in-plane cell size 3$\times$3), total energies of different antisite configurations under Ferri-AFM are slightly larger than those under A-type AFM (within 7 meV).
The FM always has higher total energy.
As we will see \cite{SM}, band structures and charge densities at $\Gamma$ point under Ferri-AFM are very similar to those under A-type AFM.
Thus in the following, only results for FM and A-type AFM (which will be further abbreviated as AFM) are presented, and results for Ferri-AFM are presented in Supplemental Material \cite{SM}.

%%%%%%%%%%%%%%%%%%%%%%%%%%%%%%%%%%%%%%%%%%%%%%%%%%%%%%%%%%%%%%%%%%%%%%%%%%%%%
\begin{figure}[tbp]
\includegraphics[width=0.95\columnwidth]{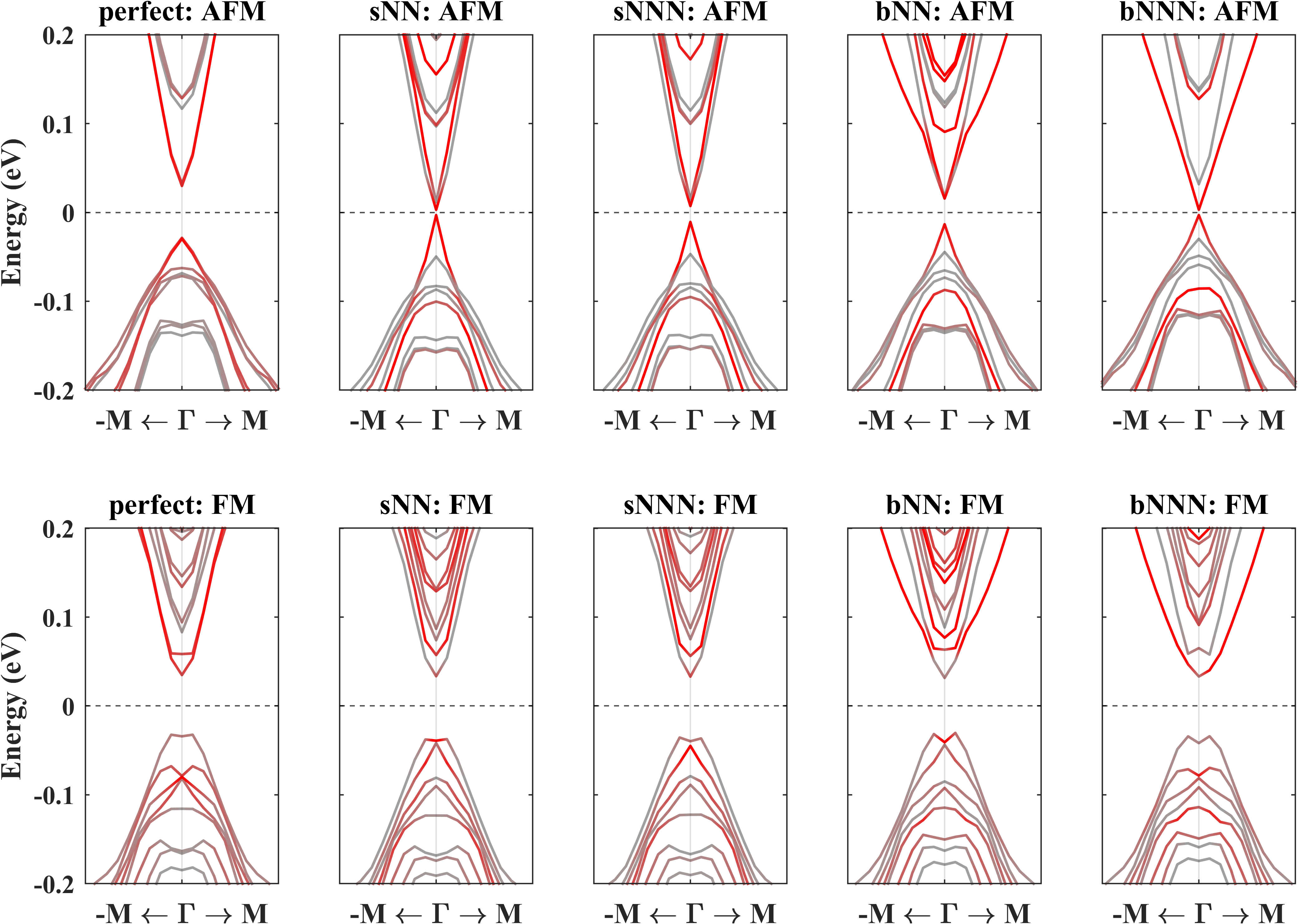}
\caption{\label{Fig2-super33-MGM_MBT1} Band structures of the different co-antisites as shown in Fig. \ref{Fig1-antisites-33}.
Upper panels: band structures under A-type AFM.
Lower panels: band structures under FM.
For better comparison, the band structures of the perfect surface are also shown (titled $perfect$).
The color red stands for the surface SL weight to the bands while the color grey stands for the weight from the rest part of the slab.
Here we still employ the $-M$-$\Gamma$-$M$ $k$-path of the hexagonal Brillouin zone of the perfect surface, though the surfaces with antisites may have no longer the hexagonal Brillouin zone.
The band structures along the $-K$-$\Gamma$-$K$ line of the hexagonal Brillouin zone share similar band characters (shown in the Supplemental Material).
Band gaps are found in Table \ref{Table_gap}.
}
\end{figure}
%%%%%%%%%%%%%%%%%%%%%%%%%%%%%%%%%%%%%%%%%%%%%%%%%%%%%%%%%%%%%%%%%%%%%%%%%%%%%

Figure \ref{Fig2-super33-MGM_MBT1} shows band structures of surfaces with different co-antisites under AFM and FM (in-plane cell size 3$\times$3).
For the perfect surface, band structures under AFM and FM show large gaps ($\sim$60 meV for AFM and $\sim$69 meV for FM, see Table \ref{Table_gap}).
However, when co-antisites appear on the surface SL, band structures (especially band gaps) under AFM become very different from those under FM.
The band gap of AFM or Ferri-AFM decreases to tens or even several meV, which depends on defect details. However, the band gap of FM remains above 55 meV, as summarized in Table \ref{Table_gap}.
FM band gaps are also robust to densities of the co-antisites (i.e., less sensitive to the in-plane cell size).
All these different co-antisite configurations can arise randomly with varied densities in real materials.
Thus, the experimental surface magnetic gap can be position- and field-dependent, i.e., the gap size depends on the magnetic field (magnetic configuration) and real space positions of measurements (different co-antisites), as evidenced by recent scanning tunneling microscopy/spectroscopy experiments \cite{arxiv.2205.09195}.
In ARPES, the overall surface gap can be reduced to the smallest one (or even smaller due to band misalignment) if different co-antisites coexist in the AFM or Ferri-AFM phase, rationalizing the diminishing gap in experiments.

%%%%%%%%%%%%%%%%%%%%%%%%%%%%%%%%%%%%%%%%%%%%%%%%%%%%%%%%%%%%%%%%%%%%%%%%%%%%%
\begin{table}
\centering
\caption{\label{Table_gap} Band gaps of MnBi$_2$Te$_4$ surfaces with different cation co-antisites under different magnetic configurations. The band gap is in units of millielectron volts (meV).}
\renewcommand\arraystretch{1.0}
\begin{ruledtabular}
\begin{tabular}{rccccc}
  surface & cell-size & FM & AFM & Ferri-AFM \\
  \hline
               perfect &   & 69 & 60 & \\
                       \specialrule{0em}{3pt}{3pt}
  \multirow{3}{*}{sNN} &2$\times$2& 55 & 13 & 1 \\
                       &3$\times$3& 72 &  6 & 4 \\
                       &4$\times$4& 74 & 31 & \\
                       \specialrule{0em}{3pt}{3pt}
  \multirow{3}{*}{sNNN}&2$\times$2& 58 & 23 & 15 \\
                       &3$\times$3& 73 & 18 & 20 \\
                       &4$\times$4& 72 & 40 & \\
                       \specialrule{0em}{3pt}{3pt}
  \multirow{3}{*}{bNN} &2$\times$2& 66 & 43 & 50 \\
                       &3$\times$3& 72 & 29 & 36 \\
                       &4$\times$4& 65 & 22 & \\
                       \specialrule{0em}{3pt}{3pt}
  \multirow{3}{*}{bNNN}&2$\times$2& 76 & 21 & 40 \\
                       &3$\times$3& 75 &  6 & 17 \\
                       &4$\times$4& 65 & 24 & \\
\end{tabular}
\end{ruledtabular}
\end{table}
%%%%%%%%%%%%%%%%%%%%%%%%%%%%%%%%%%%%%%%%%%%%%%%%%%%%%%%%%%%%%%%%%%%%%%%%%%%%%

\begin{figure*}[tbp]
\includegraphics[width=1.9\columnwidth]{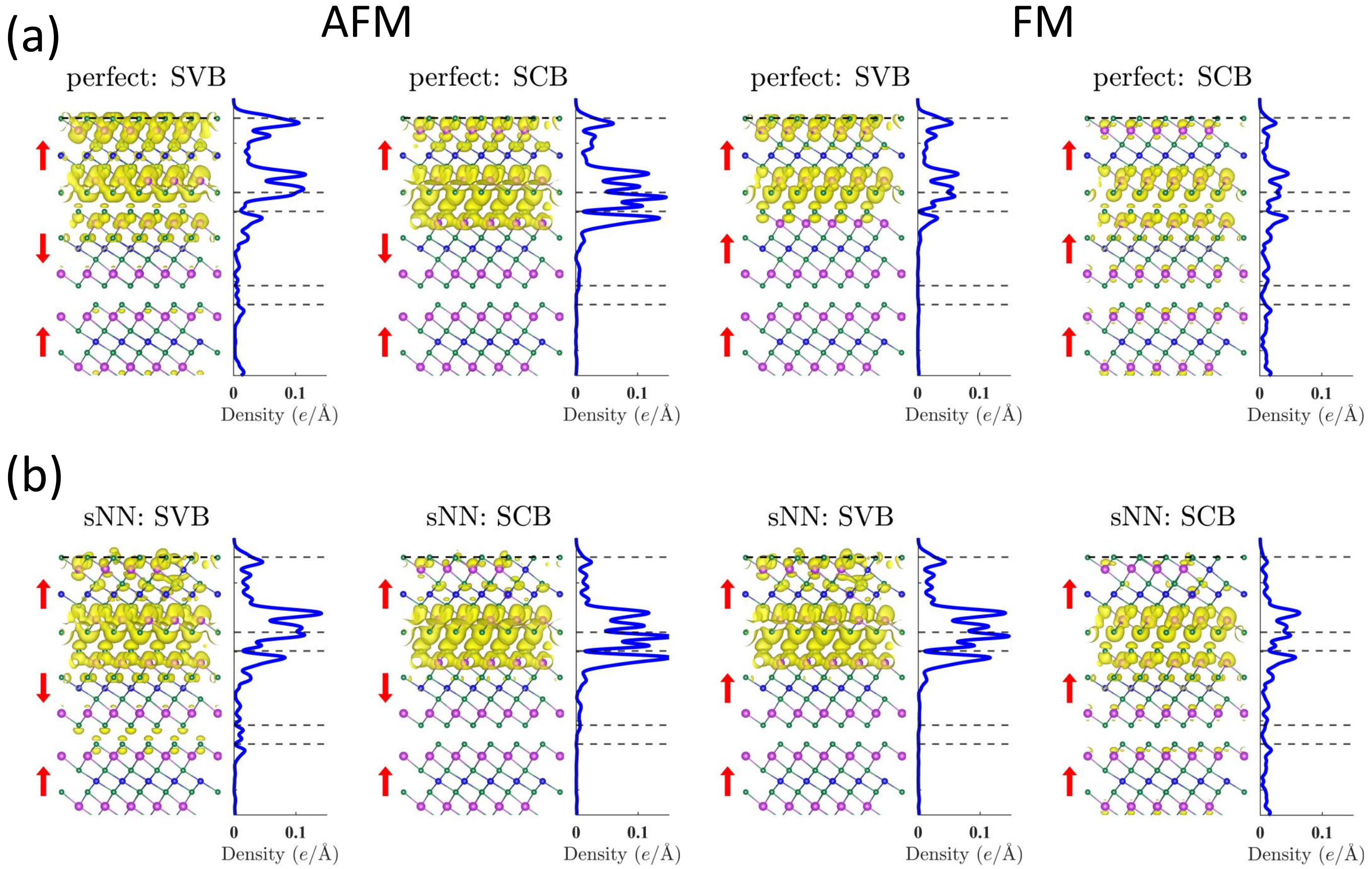}
\caption{\label{Fig3-super33-parchg-Gamma} Charge distribution of the top surface valence band (SVB) and bottom surface conduction band (SCB) at $\Gamma$ point of the perfect surface (a) and surface with sNN co-antisites (b) (see the configuration in Fig. \ref{Fig1-antisites-33}).
Band structures of the two surfaces under AFM and FM are shown in Fig. \ref{Fig2-super33-MGM_MBT1}.
The left part of each panel shows the charge distribution in the lattice (isosurface of 3.4$\times$10$^{-4}$ $e/\rm{\AA^3}$), and the right part shows the corresponding planar average along the out-of-plane direction \cite{notice}.
The red vectors indicate the magnetic configuration in each SL.
Only three SL layers are shown for simplicity, where the top dashed line of each panel indicates the surface.
Compared to the perfect surface, the charge density of sNN surface is distributed closer to the vdW gap (the centroid of charge density is moved 1$\sim$2 \AA~toward the vdW gap).
}
\end{figure*}

Band structures in Fig. \ref{Fig2-super33-MGM_MBT1} show that
the topological surface states near $\Gamma$ point change in dispersion
due to these co-antisites for both AFM and FM phases, especially the highest surface valence band (SVB) and lowest surface conduction band (SCB). 
The dispersion change originates in the charge redistribution of corresponding states. In Fig. \ref{Fig3-super33-parchg-Gamma}, we plot the charge distribution of SVB and SCB at $\Gamma$ and trace their evolution for the perfect surface and sNN case as an example.
For the perfect surface,  charge densities of SVB under both AFM and FM are mainly concentrated on the top SL with little distribution in the surface vdW gap and the second SL. For SCB, the charge density is relatively localized to the vdW gap under AFM and distributed in an extended way under FM.
When co-antisites appear, charge densities of both SVB and SCB are pushed toward the surface vdW gap under both AFM and FM (centroids of the surface states move about 1$\sim$2 \AA~toward the bulk).
Consequently, the top SL contribution to surface states is suppressed while the second SL contribution is enhanced.
Similar charge redistribution is also confirmed for the other three co-antisite configurations \cite{SM}.

%%%%%%%%%%%%%%%%%%%%%%%%%%%%%%%%%%%%%%%%%%%%%%%%%%%%%%%%%%%%%%%%%%%%%%%%%%%%%
%%%%%%%%%%%%%%%%%%%%%%%%%%%%%%%%%%%%%%%%%%%%%%%%%%%%%%%%%%%%%%%%%%%%%%%%%%%%%

%%%%%%%%%%%%%%%%%%%%%%%%%%%%%%%%%%%%%%%%%%%%%%%%%%%%%%%%%%%%%%%%%%%%%%%%%%%%%
\begin{figure}[tbp]
\includegraphics[width=0.95\columnwidth]{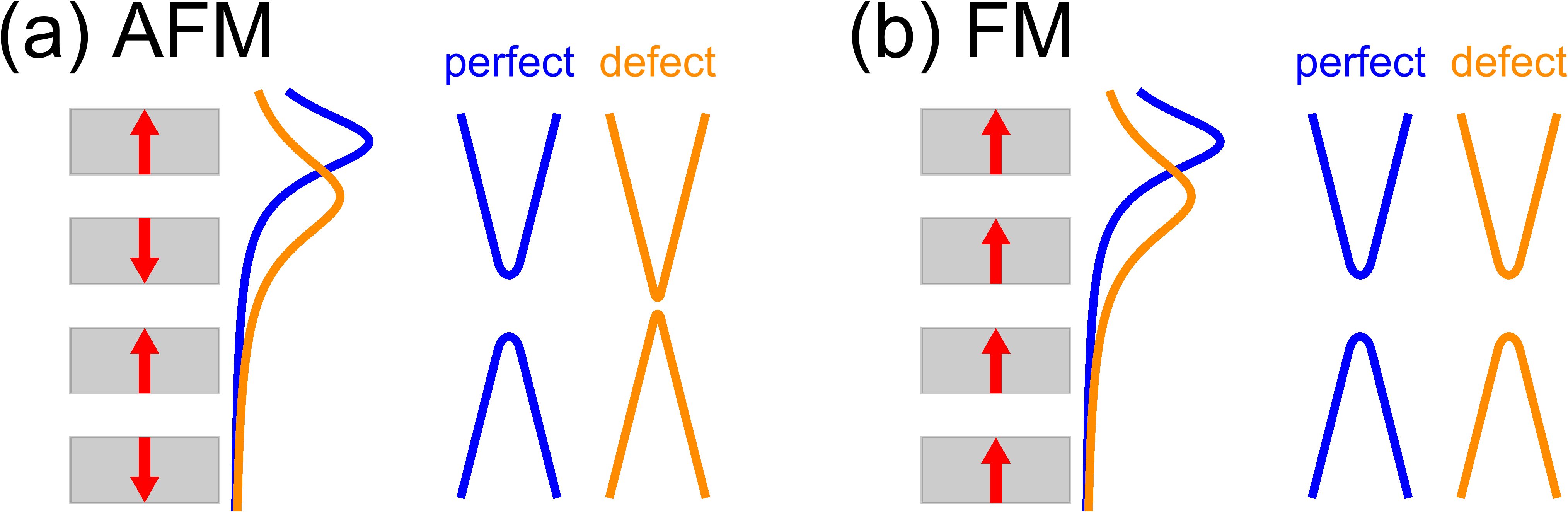}
\caption{\label{Fig4-super33}
Schematics for the relationship between surface charge distribution and magnetic gap under AFM (a) and FM (b).
The red vectors represent the magnetic configuration.
For a perfect surface where surface states (blue curves) are mainly distributed in the surface SL (grey rectangles), the topological surface states (blue linear crossing lines) under both AFM and FM open a magnetic gap.
For a defective surface where surface states (orange curves) are pushed down toward the second SL, the magnetic gap is diminished heavily under AFM while not changing under FM (orange crossing lines).
}
\end{figure}
%%%%%%%%%%%%%%%%%%%%%%%%%%%%%%%%%%%%%%%%%%%%%%%%%%%%%%%%%%%%%%%%%%%%%%%%%%%%%

On the perfect surface, where surface states are mainly distributed on the surface SL, the magnetism of the surface SL dominates the exchange interaction, which opens the magnetic gap, and the second SL has a weaker effect.
The major contribution of the top SL explains the comparable surface gap between FM and AFM phases. The extra contribution of the second SL induces a slightly larger gap in FM than in AFM (see Table \ref{Table_gap}).
On surfaces with co-antisites where surface charges are pushed toward the second SL, the top SL influence on the magnetic gap is reduced while the second SL influence is enhanced simultaneously.  
As illustrated in Fig. \ref{Fig4-super33}, the first and second SLs compete in magnetism, leading to a suppressed magnetic gap for AFM. In contrast, the top two SLs exhibit the same magnetization and thus will not necessarily reduce the magnetic gap for the FM phase.

This mechanism is different from the previously proposed one \cite{garnica2022native} based on the magnetism cancellation by the antiparallel alignment of the Mn$_{\rm{Bi}}$ moment to other magnetic moments in the same SL.
In our proposal, such antiparallel alignment of Mn$_{\rm{Bi}}$ moment has a minor effect on the surface magnetic gap in MnBi$_2$Te$_4$ \cite{SM}. We found that the magnetic effect from the second SL plays an essential role in the surface state redistribution. 

Recent theoretical works \cite{shikin2020nature,wang220508204} find that the magnetic gap can be reduced and even closed by surface vdW gap expansion.
At the same time, surface states are redistributed closer to the vdW gap with its expansion.
We point out that such a gap reduction can also be explained by our theory.
We also confirm that the FM magnetic gap is robust with the surface vdW gap expansion in MnBi$_2$Te$_4$, as shown in Supplemental Material \cite{SM}.
Thus, the cancellation of the magnetic interaction by the second SL under AFM can be a generic mechanism of the reduced surface magnetic gap for various surface defects or disorders.

\textit{Discussions --}
To realize the insulating topological states in MnBi$_2$Te$_4$, our work indicates practical strategies to optimize the material. 
One basic strategy is to improve the sample quality by reducing surface antisite defects to avoid surface state redistribution into the second SL.
The basic idea to preserve the surface magnetic gap in AFM  is that topological surface states should distribute mainly in the top SL. One may also proximity the MnBi$_2$Te$_4$ surface with another polar insulator, where topological surface charges are attracted mainly to the top SL by the modified surface potential. This may be a promising way that requires exploration in further work.

Another strategy is to realize intrinsic FM order in MnBi$_2$Te$_4$, where the surface gap may be more tolerant 
to the sample quality.
In recent experiments, FM was realized in superlattices (MnBi$_2$Te$_4$)(Bi$_2$Te$_3$)$_3$ \cite{klimovskikh2020tunable,hu2020sciadv} and Sb-doped (MnBi$_2$Te$_4$)(Bi$_2$Te$_3$) \cite{NiNi2021PRB,guan2022PRM}.
As proposed in Ref. \onlinecite{NiNi2021PRB}, the FM coupling between neighbouring SLs is mediated by the AFM coupling between SLs and the Mn$_{\rm{Bi}}$ antisites in Bi$_2$Te$_3$ layers.
Unfortunately, ARPES still observed gapless surface states in these FM samples. 
We point out that such Mn$_{\rm{Bi}}$ antisites in the Bi$_2$Te$_3$ layers may play the role of the second SL as in the AFM MnBi$_2$Te$_4$. Because they can cancel out the magnetism of the surface SL, Mn$_{\rm{Bi}}$ antisites in the Bi$_2$Te$_3$ layer may be less helpful for the magnetic gap than expected.
Besides, as we reported previously, such superlattices suffer from additional limitations (surface selection and film thickness) to ensure a surface band gap \cite{Tan2022mbt}.

\textit{Conclusion.}
We have studied the effect of co-antisites on the surface magnetic gap of MnBi$_2$Te$_4$ and found that the charge density of the topological surface states is pushed toward the surface van der Waals gap.
The increased interaction between the topological surface states and the second SL layer decreases the surface magnetic gap for the AFM phase but affects little the gap for the FM phase.
Our work reconciles contradicting observations between theory, ARPES and transport experiments and proposes material strategies to realize intriguing QAHE and AI in MnBi$_2$Te$_4$-family magnetic topological insulators.

\textbf{Acknowledgement.} 
We acknowledge helpful discussions with Igor Mazin and Chong Wang. 
B.Y. acknowledges the financial support by the European Research Council (ERC Consolidator Grant, No. 815869), the Israel Science Foundation (ISF No. 3520/20).

%\bibliography{Reference}

\begin{thebibliography}{68}%
\makeatletter
\providecommand \@ifxundefined [1]{%
 \@ifx{#1\undefined}
}%
\providecommand \@ifnum [1]{%
 \ifnum #1\expandafter \@firstoftwo
 \else \expandafter \@secondoftwo
 \fi
}%
\providecommand \@ifx [1]{%
 \ifx #1\expandafter \@firstoftwo
 \else \expandafter \@secondoftwo
 \fi
}%
\providecommand \natexlab [1]{#1}%
\providecommand \enquote  [1]{``#1''}%
\providecommand \bibnamefont  [1]{#1}%
\providecommand \bibfnamefont [1]{#1}%
\providecommand \citenamefont [1]{#1}%
\providecommand \href@noop [0]{\@secondoftwo}%
\providecommand \href [0]{\begingroup \@sanitize@url \@href}%
\providecommand \@href[1]{\@@startlink{#1}\@@href}%
\providecommand \@@href[1]{\endgroup#1\@@endlink}%
\providecommand \@sanitize@url [0]{\catcode `\\12\catcode `\$12\catcode
  `\&12\catcode `\#12\catcode `\^12\catcode `\_12\catcode `\%12\relax}%
\providecommand \@@startlink[1]{}%
\providecommand \@@endlink[0]{}%
\providecommand \url  [0]{\begingroup\@sanitize@url \@url }%
\providecommand \@url [1]{\endgroup\@href {#1}{\urlprefix }}%
\providecommand \urlprefix  [0]{URL }%
\providecommand \Eprint [0]{\href }%
\providecommand \doibase [0]{http://dx.doi.org/}%
\providecommand \selectlanguage [0]{\@gobble}%
\providecommand \bibinfo  [0]{\@secondoftwo}%
\providecommand \bibfield  [0]{\@secondoftwo}%
\providecommand \translation [1]{[#1]}%
\providecommand \BibitemOpen [0]{}%
\providecommand \bibitemStop [0]{}%
\providecommand \bibitemNoStop [0]{.\EOS\space}%
\providecommand \EOS [0]{\spacefactor3000\relax}%
\providecommand \BibitemShut  [1]{\csname bibitem#1\endcsname}%
\let\auto@bib@innerbib\@empty
%</preamble>
\bibitem [{\citenamefont {Chang}\ \emph {et~al.}(2013)\citenamefont {Chang},
  \citenamefont {Zhang}, \citenamefont {Feng}, \citenamefont {Shen},
  \citenamefont {Zhang}, \citenamefont {Guo}, \citenamefont {Li}, \citenamefont
  {Ou}, \citenamefont {Wei}, \citenamefont {Wang}, \citenamefont {Ji},
  \citenamefont {Feng}, \citenamefont {Ji}, \citenamefont {Chen}, \citenamefont
  {Jia}, \citenamefont {Dai}, \citenamefont {Fang}, \citenamefont {Zhang},
  \citenamefont {He}, \citenamefont {Wang}, \citenamefont {Lu}, \citenamefont
  {Ma},\ and\ \citenamefont {Xue}}]{chang2013}%
  \BibitemOpen
  \bibfield  {author} {\bibinfo {author} {\bibfnamefont {C.-Z.}\ \bibnamefont
  {Chang}}, \bibinfo {author} {\bibfnamefont {J.}~\bibnamefont {Zhang}},
  \bibinfo {author} {\bibfnamefont {X.}~\bibnamefont {Feng}}, \bibinfo {author}
  {\bibfnamefont {J.}~\bibnamefont {Shen}}, \bibinfo {author} {\bibfnamefont
  {Z.}~\bibnamefont {Zhang}}, \bibinfo {author} {\bibfnamefont
  {M.}~\bibnamefont {Guo}}, \bibinfo {author} {\bibfnamefont {K.}~\bibnamefont
  {Li}}, \bibinfo {author} {\bibfnamefont {Y.}~\bibnamefont {Ou}}, \bibinfo
  {author} {\bibfnamefont {P.}~\bibnamefont {Wei}}, \bibinfo {author}
  {\bibfnamefont {L.-L.}\ \bibnamefont {Wang}}, \bibinfo {author}
  {\bibfnamefont {Z.-Q.}\ \bibnamefont {Ji}}, \bibinfo {author} {\bibfnamefont
  {Y.}~\bibnamefont {Feng}}, \bibinfo {author} {\bibfnamefont {S.}~\bibnamefont
  {Ji}}, \bibinfo {author} {\bibfnamefont {X.}~\bibnamefont {Chen}}, \bibinfo
  {author} {\bibfnamefont {J.}~\bibnamefont {Jia}}, \bibinfo {author}
  {\bibfnamefont {X.}~\bibnamefont {Dai}}, \bibinfo {author} {\bibfnamefont
  {Z.}~\bibnamefont {Fang}}, \bibinfo {author} {\bibfnamefont {S.-C.}\
  \bibnamefont {Zhang}}, \bibinfo {author} {\bibfnamefont {K.}~\bibnamefont
  {He}}, \bibinfo {author} {\bibfnamefont {Y.}~\bibnamefont {Wang}}, \bibinfo
  {author} {\bibfnamefont {L.}~\bibnamefont {Lu}}, \bibinfo {author}
  {\bibfnamefont {X.-C.}\ \bibnamefont {Ma}}, \ and\ \bibinfo {author}
  {\bibfnamefont {Q.-K.}\ \bibnamefont {Xue}},\ }\href {\doibase
  10.1126/science.1234414} {\bibfield  {journal} {\bibinfo  {journal}
  {Science}\ }\textbf {\bibinfo {volume} {340}},\ \bibinfo {pages} {167}
  (\bibinfo {year} {2013})}\BibitemShut {NoStop}%
\bibitem [{\citenamefont {Mogi}\ \emph {et~al.}(2017)\citenamefont {Mogi},
  \citenamefont {Kawamura}, \citenamefont {Tsukazaki}, \citenamefont {Yoshimi},
  \citenamefont {Takahashi}, \citenamefont {Kawasaki},\ and\ \citenamefont
  {Tokura}}]{sciadv.aao1669}%
  \BibitemOpen
  \bibfield  {author} {\bibinfo {author} {\bibfnamefont {M.}~\bibnamefont
  {Mogi}}, \bibinfo {author} {\bibfnamefont {M.}~\bibnamefont {Kawamura}},
  \bibinfo {author} {\bibfnamefont {A.}~\bibnamefont {Tsukazaki}}, \bibinfo
  {author} {\bibfnamefont {R.}~\bibnamefont {Yoshimi}}, \bibinfo {author}
  {\bibfnamefont {K.~S.}\ \bibnamefont {Takahashi}}, \bibinfo {author}
  {\bibfnamefont {M.}~\bibnamefont {Kawasaki}}, \ and\ \bibinfo {author}
  {\bibfnamefont {Y.}~\bibnamefont {Tokura}},\ }\href {\doibase
  10.1126/sciadv.aao1669} {\bibfield  {journal} {\bibinfo  {journal} {Science
  Advances}\ }\textbf {\bibinfo {volume} {3}},\ \bibinfo {pages} {eaao1669}
  (\bibinfo {year} {2017})}\BibitemShut {NoStop}%
\bibitem [{\citenamefont {Xiao}\ \emph {et~al.}(2018)\citenamefont {Xiao},
  \citenamefont {Jiang}, \citenamefont {Shin}, \citenamefont {Wang},
  \citenamefont {Wang}, \citenamefont {Zhao}, \citenamefont {Liu},
  \citenamefont {Wu}, \citenamefont {Chan}, \citenamefont {Samarth},\ and\
  \citenamefont {Chang}}]{PhysRevLett.120.056801}%
  \BibitemOpen
  \bibfield  {author} {\bibinfo {author} {\bibfnamefont {D.}~\bibnamefont
  {Xiao}}, \bibinfo {author} {\bibfnamefont {J.}~\bibnamefont {Jiang}},
  \bibinfo {author} {\bibfnamefont {J.-H.}\ \bibnamefont {Shin}}, \bibinfo
  {author} {\bibfnamefont {W.}~\bibnamefont {Wang}}, \bibinfo {author}
  {\bibfnamefont {F.}~\bibnamefont {Wang}}, \bibinfo {author} {\bibfnamefont
  {Y.-F.}\ \bibnamefont {Zhao}}, \bibinfo {author} {\bibfnamefont
  {C.}~\bibnamefont {Liu}}, \bibinfo {author} {\bibfnamefont {W.}~\bibnamefont
  {Wu}}, \bibinfo {author} {\bibfnamefont {M.~H.~W.}\ \bibnamefont {Chan}},
  \bibinfo {author} {\bibfnamefont {N.}~\bibnamefont {Samarth}}, \ and\
  \bibinfo {author} {\bibfnamefont {C.-Z.}\ \bibnamefont {Chang}},\ }\href
  {\doibase 10.1103/PhysRevLett.120.056801} {\bibfield  {journal} {\bibinfo
  {journal} {Phys. Rev. Lett.}\ }\textbf {\bibinfo {volume} {120}},\ \bibinfo
  {pages} {056801} (\bibinfo {year} {2018})}\BibitemShut {NoStop}%
\bibitem [{\citenamefont {Gong}\ \emph {et~al.}(2019)\citenamefont {Gong},
  \citenamefont {Guo}, \citenamefont {Li}, \citenamefont {Zhu}, \citenamefont
  {Liao}, \citenamefont {Liu}, \citenamefont {Zhang}, \citenamefont {Gu},
  \citenamefont {Tang}, \citenamefont {Feng}, \citenamefont {Zhang},
  \citenamefont {Li}, \citenamefont {Song}, \citenamefont {Wang}, \citenamefont
  {Yu}, \citenamefont {Chen}, \citenamefont {Wang}, \citenamefont {Yao},
  \citenamefont {Duan}, \citenamefont {Xu}, \citenamefont {Zhang},
  \citenamefont {Ma}, \citenamefont {Xue},\ and\ \citenamefont
  {He}}]{gong2019experimental}%
  \BibitemOpen
  \bibfield  {author} {\bibinfo {author} {\bibfnamefont {Y.}~\bibnamefont
  {Gong}}, \bibinfo {author} {\bibfnamefont {J.}~\bibnamefont {Guo}}, \bibinfo
  {author} {\bibfnamefont {J.}~\bibnamefont {Li}}, \bibinfo {author}
  {\bibfnamefont {K.}~\bibnamefont {Zhu}}, \bibinfo {author} {\bibfnamefont
  {M.}~\bibnamefont {Liao}}, \bibinfo {author} {\bibfnamefont {X.}~\bibnamefont
  {Liu}}, \bibinfo {author} {\bibfnamefont {Q.}~\bibnamefont {Zhang}}, \bibinfo
  {author} {\bibfnamefont {L.}~\bibnamefont {Gu}}, \bibinfo {author}
  {\bibfnamefont {L.}~\bibnamefont {Tang}}, \bibinfo {author} {\bibfnamefont
  {X.}~\bibnamefont {Feng}}, \bibinfo {author} {\bibfnamefont {D.}~\bibnamefont
  {Zhang}}, \bibinfo {author} {\bibfnamefont {W.}~\bibnamefont {Li}}, \bibinfo
  {author} {\bibfnamefont {C.}~\bibnamefont {Song}}, \bibinfo {author}
  {\bibfnamefont {L.}~\bibnamefont {Wang}}, \bibinfo {author} {\bibfnamefont
  {P.}~\bibnamefont {Yu}}, \bibinfo {author} {\bibfnamefont {X.}~\bibnamefont
  {Chen}}, \bibinfo {author} {\bibfnamefont {Y.}~\bibnamefont {Wang}}, \bibinfo
  {author} {\bibfnamefont {H.}~\bibnamefont {Yao}}, \bibinfo {author}
  {\bibfnamefont {W.}~\bibnamefont {Duan}}, \bibinfo {author} {\bibfnamefont
  {Y.}~\bibnamefont {Xu}}, \bibinfo {author} {\bibfnamefont {S.-C.}\
  \bibnamefont {Zhang}}, \bibinfo {author} {\bibfnamefont {X.}~\bibnamefont
  {Ma}}, \bibinfo {author} {\bibfnamefont {Q.-K.}\ \bibnamefont {Xue}}, \ and\
  \bibinfo {author} {\bibfnamefont {K.}~\bibnamefont {He}},\ }\href {\doibase
  10.1088/0256-307x/36/7/076801} {\bibfield  {journal} {\bibinfo  {journal}
  {Chin. Phys. Lett.}\ }\textbf {\bibinfo {volume} {36}},\ \bibinfo {pages}
  {076801} (\bibinfo {year} {2019})}\BibitemShut {NoStop}%
\bibitem [{\citenamefont {Otrokov}\ \emph
  {et~al.}(2019{\natexlab{a}})\citenamefont {Otrokov}, \citenamefont
  {Klimovskikh}, \citenamefont {Bentmann}, \citenamefont {Estyunin},
  \citenamefont {Zeugner}, \citenamefont {Aliev}, \citenamefont {Ga{\ss}},
  \citenamefont {Wolter}, \citenamefont {Koroleva}, \citenamefont {Shikin},
  \citenamefont {Blanco-Rey}, \citenamefont {Hoffmann}, \citenamefont
  {Rusinov}, \citenamefont {Vyazovskaya}, \citenamefont {Eremeev},
  \citenamefont {Koroteev}, \citenamefont {Kuznetsov}, \citenamefont {Freyse},
  \citenamefont {S\'{a}nchez-Barriga}, \citenamefont {Amiraslanov},
  \citenamefont {Babanly}, \citenamefont {Mamedov}, \citenamefont {Abdullayev},
  \citenamefont {Zverev}, \citenamefont {Alfonsov}, \citenamefont {Kataev},
  \citenamefont {B{\"u}chner}, \citenamefont {Schwier}, \citenamefont {Kumar},
  \citenamefont {Kimura}, \citenamefont {Petaccia}, \citenamefont {Santo},
  \citenamefont {Vidal}, \citenamefont {Schatz}, \citenamefont {Ki{\ss}ner},
  \citenamefont {{\"U}nzelmann}, \citenamefont {Min}, \citenamefont {Moser},
  \citenamefont {Peixoto}, \citenamefont {Reinert}, \citenamefont {Ernst},
  \citenamefont {Echenique}, \citenamefont {Isaeva},\ and\ \citenamefont
  {Chulkov}}]{otrokov2019prediction}%
  \BibitemOpen
  \bibfield  {author} {\bibinfo {author} {\bibfnamefont {M.~M.}\ \bibnamefont
  {Otrokov}}, \bibinfo {author} {\bibfnamefont {I.~I.}\ \bibnamefont
  {Klimovskikh}}, \bibinfo {author} {\bibfnamefont {H.}~\bibnamefont
  {Bentmann}}, \bibinfo {author} {\bibfnamefont {D.}~\bibnamefont {Estyunin}},
  \bibinfo {author} {\bibfnamefont {A.}~\bibnamefont {Zeugner}}, \bibinfo
  {author} {\bibfnamefont {Z.~S.}\ \bibnamefont {Aliev}}, \bibinfo {author}
  {\bibfnamefont {S.}~\bibnamefont {Ga{\ss}}}, \bibinfo {author} {\bibfnamefont
  {A.~U.~B.}\ \bibnamefont {Wolter}}, \bibinfo {author} {\bibfnamefont {A.~V.}\
  \bibnamefont {Koroleva}}, \bibinfo {author} {\bibfnamefont {A.~M.}\
  \bibnamefont {Shikin}}, \bibinfo {author} {\bibfnamefont {M.}~\bibnamefont
  {Blanco-Rey}}, \bibinfo {author} {\bibfnamefont {M.}~\bibnamefont
  {Hoffmann}}, \bibinfo {author} {\bibfnamefont {I.~P.}\ \bibnamefont
  {Rusinov}}, \bibinfo {author} {\bibfnamefont {A.~Y.}\ \bibnamefont
  {Vyazovskaya}}, \bibinfo {author} {\bibfnamefont {S.~V.}\ \bibnamefont
  {Eremeev}}, \bibinfo {author} {\bibfnamefont {Y.~M.}\ \bibnamefont
  {Koroteev}}, \bibinfo {author} {\bibfnamefont {V.~M.}\ \bibnamefont
  {Kuznetsov}}, \bibinfo {author} {\bibfnamefont {F.}~\bibnamefont {Freyse}},
  \bibinfo {author} {\bibfnamefont {J.}~\bibnamefont {S\'{a}nchez-Barriga}},
  \bibinfo {author} {\bibfnamefont {I.~R.}\ \bibnamefont {Amiraslanov}},
  \bibinfo {author} {\bibfnamefont {M.~B.}\ \bibnamefont {Babanly}}, \bibinfo
  {author} {\bibfnamefont {N.~T.}\ \bibnamefont {Mamedov}}, \bibinfo {author}
  {\bibfnamefont {N.~A.}\ \bibnamefont {Abdullayev}}, \bibinfo {author}
  {\bibfnamefont {V.~N.}\ \bibnamefont {Zverev}}, \bibinfo {author}
  {\bibfnamefont {A.}~\bibnamefont {Alfonsov}}, \bibinfo {author}
  {\bibfnamefont {V.}~\bibnamefont {Kataev}}, \bibinfo {author} {\bibfnamefont
  {B.}~\bibnamefont {B{\"u}chner}}, \bibinfo {author} {\bibfnamefont {E.~F.}\
  \bibnamefont {Schwier}}, \bibinfo {author} {\bibfnamefont {S.}~\bibnamefont
  {Kumar}}, \bibinfo {author} {\bibfnamefont {A.}~\bibnamefont {Kimura}},
  \bibinfo {author} {\bibfnamefont {L.}~\bibnamefont {Petaccia}}, \bibinfo
  {author} {\bibfnamefont {G.~D.}\ \bibnamefont {Santo}}, \bibinfo {author}
  {\bibfnamefont {R.~C.}\ \bibnamefont {Vidal}}, \bibinfo {author}
  {\bibfnamefont {S.}~\bibnamefont {Schatz}}, \bibinfo {author} {\bibfnamefont
  {K.}~\bibnamefont {Ki{\ss}ner}}, \bibinfo {author} {\bibfnamefont
  {M.}~\bibnamefont {{\"U}nzelmann}}, \bibinfo {author} {\bibfnamefont {C.~H.}\
  \bibnamefont {Min}}, \bibinfo {author} {\bibfnamefont {S.}~\bibnamefont
  {Moser}}, \bibinfo {author} {\bibfnamefont {T.~R.~F.}\ \bibnamefont
  {Peixoto}}, \bibinfo {author} {\bibfnamefont {F.}~\bibnamefont {Reinert}},
  \bibinfo {author} {\bibfnamefont {A.}~\bibnamefont {Ernst}}, \bibinfo
  {author} {\bibfnamefont {P.~M.}\ \bibnamefont {Echenique}}, \bibinfo {author}
  {\bibfnamefont {A.}~\bibnamefont {Isaeva}}, \ and\ \bibinfo {author}
  {\bibfnamefont {E.~V.}\ \bibnamefont {Chulkov}},\ }\href {\doibase
  10.1038/s41586-019-1840-9} {\bibfield  {journal} {\bibinfo  {journal}
  {Nature}\ }\textbf {\bibinfo {volume} {576}},\ \bibinfo {pages} {416}
  (\bibinfo {year} {2019}{\natexlab{a}})}\BibitemShut {NoStop}%
\bibitem [{\citenamefont {Li}\ \emph {et~al.}(2019{\natexlab{a}})\citenamefont
  {Li}, \citenamefont {Li}, \citenamefont {Du}, \citenamefont {Wang},
  \citenamefont {Gu}, \citenamefont {Zhang}, \citenamefont {He}, \citenamefont
  {Duan},\ and\ \citenamefont {Xu}}]{li2019intrinsic}%
  \BibitemOpen
  \bibfield  {author} {\bibinfo {author} {\bibfnamefont {J.}~\bibnamefont
  {Li}}, \bibinfo {author} {\bibfnamefont {Y.}~\bibnamefont {Li}}, \bibinfo
  {author} {\bibfnamefont {S.}~\bibnamefont {Du}}, \bibinfo {author}
  {\bibfnamefont {Z.}~\bibnamefont {Wang}}, \bibinfo {author} {\bibfnamefont
  {B.-L.}\ \bibnamefont {Gu}}, \bibinfo {author} {\bibfnamefont {S.-C.}\
  \bibnamefont {Zhang}}, \bibinfo {author} {\bibfnamefont {K.}~\bibnamefont
  {He}}, \bibinfo {author} {\bibfnamefont {W.}~\bibnamefont {Duan}}, \ and\
  \bibinfo {author} {\bibfnamefont {Y.}~\bibnamefont {Xu}},\ }\href {\doibase
  10.1126/sciadv.aaw5685} {\bibfield  {journal} {\bibinfo  {journal} {Sci.
  Adv.}\ }\textbf {\bibinfo {volume} {5}},\ \bibinfo {pages} {eaaw5685}
  (\bibinfo {year} {2019}{\natexlab{a}})}\BibitemShut {NoStop}%
\bibitem [{\citenamefont {Zhang}\ \emph {et~al.}(2019)\citenamefont {Zhang},
  \citenamefont {Shi}, \citenamefont {Zhu}, \citenamefont {Xing}, \citenamefont
  {Zhang},\ and\ \citenamefont {Wang}}]{PhysRevLett.122.206401}%
  \BibitemOpen
  \bibfield  {author} {\bibinfo {author} {\bibfnamefont {D.}~\bibnamefont
  {Zhang}}, \bibinfo {author} {\bibfnamefont {M.}~\bibnamefont {Shi}}, \bibinfo
  {author} {\bibfnamefont {T.}~\bibnamefont {Zhu}}, \bibinfo {author}
  {\bibfnamefont {D.}~\bibnamefont {Xing}}, \bibinfo {author} {\bibfnamefont
  {H.}~\bibnamefont {Zhang}}, \ and\ \bibinfo {author} {\bibfnamefont
  {J.}~\bibnamefont {Wang}},\ }\href {\doibase 10.1103/PhysRevLett.122.206401}
  {\bibfield  {journal} {\bibinfo  {journal} {Phys. Rev. Lett.}\ }\textbf
  {\bibinfo {volume} {122}},\ \bibinfo {pages} {206401} (\bibinfo {year}
  {2019})}\BibitemShut {NoStop}%
\bibitem [{\citenamefont {Deng}\ \emph {et~al.}(2020)\citenamefont {Deng},
  \citenamefont {Yu}, \citenamefont {Shi}, \citenamefont {Guo}, \citenamefont
  {Xu}, \citenamefont {Wang}, \citenamefont {Chen},\ and\ \citenamefont
  {Zhang}}]{deng2020science}%
  \BibitemOpen
  \bibfield  {author} {\bibinfo {author} {\bibfnamefont {Y.}~\bibnamefont
  {Deng}}, \bibinfo {author} {\bibfnamefont {Y.}~\bibnamefont {Yu}}, \bibinfo
  {author} {\bibfnamefont {M.~Z.}\ \bibnamefont {Shi}}, \bibinfo {author}
  {\bibfnamefont {Z.}~\bibnamefont {Guo}}, \bibinfo {author} {\bibfnamefont
  {Z.}~\bibnamefont {Xu}}, \bibinfo {author} {\bibfnamefont {J.}~\bibnamefont
  {Wang}}, \bibinfo {author} {\bibfnamefont {X.~H.}\ \bibnamefont {Chen}}, \
  and\ \bibinfo {author} {\bibfnamefont {Y.}~\bibnamefont {Zhang}},\ }\href
  {\doibase 10.1126/science.aax8156} {\bibfield  {journal} {\bibinfo  {journal}
  {Science}\ }\textbf {\bibinfo {volume} {367}},\ \bibinfo {pages} {895}
  (\bibinfo {year} {2020})}\BibitemShut {NoStop}%
\bibitem [{\citenamefont {Ge}\ \emph {et~al.}(2020)\citenamefont {Ge},
  \citenamefont {Liu}, \citenamefont {Li}, \citenamefont {Li}, \citenamefont
  {Luo}, \citenamefont {Wu}, \citenamefont {Xu},\ and\ \citenamefont
  {Wang}}]{nwaa089}%
  \BibitemOpen
  \bibfield  {author} {\bibinfo {author} {\bibfnamefont {J.}~\bibnamefont
  {Ge}}, \bibinfo {author} {\bibfnamefont {Y.}~\bibnamefont {Liu}}, \bibinfo
  {author} {\bibfnamefont {J.}~\bibnamefont {Li}}, \bibinfo {author}
  {\bibfnamefont {H.}~\bibnamefont {Li}}, \bibinfo {author} {\bibfnamefont
  {T.}~\bibnamefont {Luo}}, \bibinfo {author} {\bibfnamefont {Y.}~\bibnamefont
  {Wu}}, \bibinfo {author} {\bibfnamefont {Y.}~\bibnamefont {Xu}}, \ and\
  \bibinfo {author} {\bibfnamefont {J.}~\bibnamefont {Wang}},\ }\href {\doibase
  10.1093/nsr/nwaa089} {\bibfield  {journal} {\bibinfo  {journal} {National
  Science Review}\ }\textbf {\bibinfo {volume} {7}},\ \bibinfo {pages} {1280}
  (\bibinfo {year} {2020})}\BibitemShut {NoStop}%
\bibitem [{\citenamefont {Liu}\ \emph {et~al.}(2020)\citenamefont {Liu},
  \citenamefont {Wang}, \citenamefont {Li}, \citenamefont {Wu}, \citenamefont
  {Li}, \citenamefont {Li}, \citenamefont {He}, \citenamefont {Xu},
  \citenamefont {Zhang},\ and\ \citenamefont {Wang}}]{liu2020robust}%
  \BibitemOpen
  \bibfield  {author} {\bibinfo {author} {\bibfnamefont {C.}~\bibnamefont
  {Liu}}, \bibinfo {author} {\bibfnamefont {Y.}~\bibnamefont {Wang}}, \bibinfo
  {author} {\bibfnamefont {H.}~\bibnamefont {Li}}, \bibinfo {author}
  {\bibfnamefont {Y.}~\bibnamefont {Wu}}, \bibinfo {author} {\bibfnamefont
  {Y.}~\bibnamefont {Li}}, \bibinfo {author} {\bibfnamefont {J.}~\bibnamefont
  {Li}}, \bibinfo {author} {\bibfnamefont {K.}~\bibnamefont {He}}, \bibinfo
  {author} {\bibfnamefont {Y.}~\bibnamefont {Xu}}, \bibinfo {author}
  {\bibfnamefont {J.}~\bibnamefont {Zhang}}, \ and\ \bibinfo {author}
  {\bibfnamefont {Y.}~\bibnamefont {Wang}},\ }\href {\doibase
  10.1038/s41563-019-0573-3} {\bibfield  {journal} {\bibinfo  {journal} {Nature
  materials}\ }\textbf {\bibinfo {volume} {19}},\ \bibinfo {pages} {522}
  (\bibinfo {year} {2020})}\BibitemShut {NoStop}%
\bibitem [{\citenamefont {Deng}\ \emph {et~al.}(2021)\citenamefont {Deng},
  \citenamefont {Chen}, \citenamefont {Wo{\l}o{\'s}}, \citenamefont
  {Konczykowski}, \citenamefont {Sobczak}, \citenamefont {Sitnicka},
  \citenamefont {Fedorchenko}, \citenamefont {Borysiuk}, \citenamefont
  {Heider}, \citenamefont {Pluci{\'n}ski}, \citenamefont {Park}, \citenamefont
  {Georgescu}, \citenamefont {Cano},\ and\ \citenamefont
  {Krusin-Elbaum}}]{deng2021high}%
  \BibitemOpen
  \bibfield  {author} {\bibinfo {author} {\bibfnamefont {H.}~\bibnamefont
  {Deng}}, \bibinfo {author} {\bibfnamefont {Z.}~\bibnamefont {Chen}}, \bibinfo
  {author} {\bibfnamefont {A.}~\bibnamefont {Wo{\l}o{\'s}}}, \bibinfo {author}
  {\bibfnamefont {M.}~\bibnamefont {Konczykowski}}, \bibinfo {author}
  {\bibfnamefont {K.}~\bibnamefont {Sobczak}}, \bibinfo {author} {\bibfnamefont
  {J.}~\bibnamefont {Sitnicka}}, \bibinfo {author} {\bibfnamefont {I.~V.}\
  \bibnamefont {Fedorchenko}}, \bibinfo {author} {\bibfnamefont
  {J.}~\bibnamefont {Borysiuk}}, \bibinfo {author} {\bibfnamefont
  {T.}~\bibnamefont {Heider}}, \bibinfo {author} {\bibfnamefont
  {{\L}.}~\bibnamefont {Pluci{\'n}ski}}, \bibinfo {author} {\bibfnamefont
  {K.}~\bibnamefont {Park}}, \bibinfo {author} {\bibfnamefont {A.~B.}\
  \bibnamefont {Georgescu}}, \bibinfo {author} {\bibfnamefont {J.}~\bibnamefont
  {Cano}}, \ and\ \bibinfo {author} {\bibfnamefont {L.}~\bibnamefont
  {Krusin-Elbaum}},\ }\href {\doibase 10.1038/s41567-020-0998-2} {\bibfield
  {journal} {\bibinfo  {journal} {Nature Physics}\ }\textbf {\bibinfo {volume}
  {17}},\ \bibinfo {pages} {36} (\bibinfo {year} {2021})}\BibitemShut {NoStop}%
\bibitem [{\citenamefont {Yan}(2022)}]{yan2021elusive}%
  \BibitemOpen
  \bibfield  {author} {\bibinfo {author} {\bibfnamefont {J.-Q.}\ \bibnamefont
  {Yan}},\ }\href {\doibase 10.1149/2162-8777/ac70fc} {\bibfield  {journal}
  {\bibinfo  {journal} {{ECS} Journal of Solid State Science and Technology}\
  }\textbf {\bibinfo {volume} {11}},\ \bibinfo {pages} {063007} (\bibinfo
  {year} {2022})}\BibitemShut {NoStop}%
\bibitem [{\citenamefont {Wu}\ \emph {et~al.}(2019)\citenamefont {Wu},
  \citenamefont {Liu}, \citenamefont {Sasase}, \citenamefont {Ienaga},
  \citenamefont {Obata}, \citenamefont {Yukawa}, \citenamefont {Horiba},
  \citenamefont {Kumigashira}, \citenamefont {Okuma}, \citenamefont
  {Inoshita},\ and\ \citenamefont {Hosono}}]{wu2019natural}%
  \BibitemOpen
  \bibfield  {author} {\bibinfo {author} {\bibfnamefont {J.}~\bibnamefont
  {Wu}}, \bibinfo {author} {\bibfnamefont {F.}~\bibnamefont {Liu}}, \bibinfo
  {author} {\bibfnamefont {M.}~\bibnamefont {Sasase}}, \bibinfo {author}
  {\bibfnamefont {K.}~\bibnamefont {Ienaga}}, \bibinfo {author} {\bibfnamefont
  {Y.}~\bibnamefont {Obata}}, \bibinfo {author} {\bibfnamefont
  {R.}~\bibnamefont {Yukawa}}, \bibinfo {author} {\bibfnamefont
  {K.}~\bibnamefont {Horiba}}, \bibinfo {author} {\bibfnamefont
  {H.}~\bibnamefont {Kumigashira}}, \bibinfo {author} {\bibfnamefont
  {S.}~\bibnamefont {Okuma}}, \bibinfo {author} {\bibfnamefont
  {T.}~\bibnamefont {Inoshita}}, \ and\ \bibinfo {author} {\bibfnamefont
  {H.}~\bibnamefont {Hosono}},\ }\href {\doibase 10.1126/sciadv.aax9989}
  {\bibfield  {journal} {\bibinfo  {journal} {Science Advances}\ }\textbf
  {\bibinfo {volume} {5}},\ \bibinfo {pages} {eaax9989} (\bibinfo {year}
  {2019})}\BibitemShut {NoStop}%
\bibitem [{\citenamefont {Lee}\ \emph {et~al.}(2021)\citenamefont {Lee},
  \citenamefont {Graf}, \citenamefont {Min}, \citenamefont {Zhu}, \citenamefont
  {Yi}, \citenamefont {Ciocys}, \citenamefont {Wang}, \citenamefont {Choi},
  \citenamefont {Basnet}, \citenamefont {Fereidouni}, \citenamefont {Wegner},
  \citenamefont {Zhao}, \citenamefont {Verlinde}, \citenamefont {He},
  \citenamefont {Redwing}, \citenamefont {Gopalan}, \citenamefont {Churchill},
  \citenamefont {Lanzara}, \citenamefont {Samarth}, \citenamefont {Chang},
  \citenamefont {Hu},\ and\ \citenamefont {Mao}}]{PhysRevX.11.031032}%
  \BibitemOpen
  \bibfield  {author} {\bibinfo {author} {\bibfnamefont {S.~H.}\ \bibnamefont
  {Lee}}, \bibinfo {author} {\bibfnamefont {D.}~\bibnamefont {Graf}}, \bibinfo
  {author} {\bibfnamefont {L.}~\bibnamefont {Min}}, \bibinfo {author}
  {\bibfnamefont {Y.}~\bibnamefont {Zhu}}, \bibinfo {author} {\bibfnamefont
  {H.}~\bibnamefont {Yi}}, \bibinfo {author} {\bibfnamefont {S.}~\bibnamefont
  {Ciocys}}, \bibinfo {author} {\bibfnamefont {Y.}~\bibnamefont {Wang}},
  \bibinfo {author} {\bibfnamefont {E.~S.}\ \bibnamefont {Choi}}, \bibinfo
  {author} {\bibfnamefont {R.}~\bibnamefont {Basnet}}, \bibinfo {author}
  {\bibfnamefont {A.}~\bibnamefont {Fereidouni}}, \bibinfo {author}
  {\bibfnamefont {A.}~\bibnamefont {Wegner}}, \bibinfo {author} {\bibfnamefont
  {Y.-F.}\ \bibnamefont {Zhao}}, \bibinfo {author} {\bibfnamefont
  {K.}~\bibnamefont {Verlinde}}, \bibinfo {author} {\bibfnamefont
  {J.}~\bibnamefont {He}}, \bibinfo {author} {\bibfnamefont {R.}~\bibnamefont
  {Redwing}}, \bibinfo {author} {\bibfnamefont {V.}~\bibnamefont {Gopalan}},
  \bibinfo {author} {\bibfnamefont {H.~O.~H.}\ \bibnamefont {Churchill}},
  \bibinfo {author} {\bibfnamefont {A.}~\bibnamefont {Lanzara}}, \bibinfo
  {author} {\bibfnamefont {N.}~\bibnamefont {Samarth}}, \bibinfo {author}
  {\bibfnamefont {C.-Z.}\ \bibnamefont {Chang}}, \bibinfo {author}
  {\bibfnamefont {J.}~\bibnamefont {Hu}}, \ and\ \bibinfo {author}
  {\bibfnamefont {Z.~Q.}\ \bibnamefont {Mao}},\ }\href {\doibase
  10.1103/PhysRevX.11.031032} {\bibfield  {journal} {\bibinfo  {journal} {Phys.
  Rev. X}\ }\textbf {\bibinfo {volume} {11}},\ \bibinfo {pages} {031032}
  (\bibinfo {year} {2021})}\BibitemShut {NoStop}%
\bibitem [{\citenamefont {Lei}\ \emph {et~al.}(2020)\citenamefont {Lei},
  \citenamefont {Chen},\ and\ \citenamefont {MacDonald}}]{lei2020magnetized}%
  \BibitemOpen
  \bibfield  {author} {\bibinfo {author} {\bibfnamefont {C.}~\bibnamefont
  {Lei}}, \bibinfo {author} {\bibfnamefont {S.}~\bibnamefont {Chen}}, \ and\
  \bibinfo {author} {\bibfnamefont {A.~H.}\ \bibnamefont {MacDonald}},\ }\href
  {\doibase 10.1073/pnas.2014004117} {\bibfield  {journal} {\bibinfo  {journal}
  {Proc. Natl. Acad. Sci.}\ }\textbf {\bibinfo {volume} {117}},\ \bibinfo
  {pages} {27224} (\bibinfo {year} {2020})}\BibitemShut {NoStop}%
\bibitem [{\citenamefont {Li}\ \emph {et~al.}(2021{\natexlab{a}})\citenamefont
  {Li}, \citenamefont {Chen}, \citenamefont {Jiang},\ and\ \citenamefont
  {Xie}}]{PhysRevLett.127.236402}%
  \BibitemOpen
  \bibfield  {author} {\bibinfo {author} {\bibfnamefont {H.}~\bibnamefont
  {Li}}, \bibinfo {author} {\bibfnamefont {C.-Z.}\ \bibnamefont {Chen}},
  \bibinfo {author} {\bibfnamefont {H.}~\bibnamefont {Jiang}}, \ and\ \bibinfo
  {author} {\bibfnamefont {X.~C.}\ \bibnamefont {Xie}},\ }\href {\doibase
  10.1103/PhysRevLett.127.236402} {\bibfield  {journal} {\bibinfo  {journal}
  {Phys. Rev. Lett.}\ }\textbf {\bibinfo {volume} {127}},\ \bibinfo {pages}
  {236402} (\bibinfo {year} {2021}{\natexlab{a}})}\BibitemShut {NoStop}%
\bibitem [{\citenamefont {Wang}\ \emph {et~al.}(2021)\citenamefont {Wang},
  \citenamefont {Ge}, \citenamefont {Li}, \citenamefont {Liu}, \citenamefont
  {Xu},\ and\ \citenamefont {Wang}}]{WANG2021100098}%
  \BibitemOpen
  \bibfield  {author} {\bibinfo {author} {\bibfnamefont {P.}~\bibnamefont
  {Wang}}, \bibinfo {author} {\bibfnamefont {J.}~\bibnamefont {Ge}}, \bibinfo
  {author} {\bibfnamefont {J.}~\bibnamefont {Li}}, \bibinfo {author}
  {\bibfnamefont {Y.}~\bibnamefont {Liu}}, \bibinfo {author} {\bibfnamefont
  {Y.}~\bibnamefont {Xu}}, \ and\ \bibinfo {author} {\bibfnamefont
  {J.}~\bibnamefont {Wang}},\ }\href {\doibase
  https://doi.org/10.1016/j.xinn.2021.100098} {\bibfield  {journal} {\bibinfo
  {journal} {The Innovation}\ }\textbf {\bibinfo {volume} {2}},\ \bibinfo
  {pages} {100098} (\bibinfo {year} {2021})}\BibitemShut {NoStop}%
\bibitem [{\citenamefont {Zhao}\ and\ \citenamefont
  {Liu}(2021{\natexlab{a}})}]{Zhao2021}%
  \BibitemOpen
  \bibfield  {author} {\bibinfo {author} {\bibfnamefont {Y.}~\bibnamefont
  {Zhao}}\ and\ \bibinfo {author} {\bibfnamefont {Q.}~\bibnamefont {Liu}},\
  }\href {\doibase 10.1063/5.0059447} {\bibfield  {journal} {\bibinfo
  {journal} {Applied Physics Letters}\ }\textbf {\bibinfo {volume} {119}},\
  \bibinfo {pages} {060502} (\bibinfo {year} {2021}{\natexlab{a}})}\BibitemShut
  {NoStop}%
\bibitem [{\citenamefont {Li}\ \emph {et~al.}(2021{\natexlab{b}})\citenamefont
  {Li}, \citenamefont {Trang}, \citenamefont {Wu}, \citenamefont {Hwang},
  \citenamefont {Medhekar}, \citenamefont {Mo}, \citenamefont {Yang},\ and\
  \citenamefont {Edmonds}}]{li2021large}%
  \BibitemOpen
  \bibfield  {author} {\bibinfo {author} {\bibfnamefont {Q.}~\bibnamefont
  {Li}}, \bibinfo {author} {\bibfnamefont {C.~X.}\ \bibnamefont {Trang}},
  \bibinfo {author} {\bibfnamefont {W.}~\bibnamefont {Wu}}, \bibinfo {author}
  {\bibfnamefont {J.}~\bibnamefont {Hwang}}, \bibinfo {author} {\bibfnamefont
  {N.}~\bibnamefont {Medhekar}}, \bibinfo {author} {\bibfnamefont {S.-K.}\
  \bibnamefont {Mo}}, \bibinfo {author} {\bibfnamefont {S.~A.}\ \bibnamefont
  {Yang}}, \ and\ \bibinfo {author} {\bibfnamefont {M.~T.}\ \bibnamefont
  {Edmonds}},\ }\href {https://arxiv.org/ftp/arxiv/papers/2105/2105.08298.pdf}
  {\bibfield  {journal} {\bibinfo  {journal} {arXiv:2105.08298}\ } (\bibinfo
  {year} {2021}{\natexlab{b}})}\BibitemShut {NoStop}%
\bibitem [{\citenamefont {Klimovskikh}\ \emph {et~al.}(2020)\citenamefont
  {Klimovskikh}, \citenamefont {Otrokov}, \citenamefont {Estyunin},
  \citenamefont {Eremeev}, \citenamefont {Filnov}, \citenamefont {Koroleva},
  \citenamefont {Shevchenko}, \citenamefont {Voroshnin}, \citenamefont
  {Rybkin}, \citenamefont {Rusinov}, \citenamefont {Blanco-Rey}, \citenamefont
  {Hoffmann}, \citenamefont {Aliev}, \citenamefont {Babanly}, \citenamefont
  {Amiraslanov}, \citenamefont {Abdullayev}, \citenamefont {Zverev},
  \citenamefont {Kimura}, \citenamefont {Tereshchenko}, \citenamefont {Kokh},
  \citenamefont {Petaccia}, \citenamefont {Santo}, \citenamefont {Ernst},
  \citenamefont {Echenique}, \citenamefont {Mamedov}, \citenamefont {Shikin},\
  and\ \citenamefont {Chulkov}}]{klimovskikh2020tunable}%
  \BibitemOpen
  \bibfield  {author} {\bibinfo {author} {\bibfnamefont {I.~I.}\ \bibnamefont
  {Klimovskikh}}, \bibinfo {author} {\bibfnamefont {M.~M.}\ \bibnamefont
  {Otrokov}}, \bibinfo {author} {\bibfnamefont {D.}~\bibnamefont {Estyunin}},
  \bibinfo {author} {\bibfnamefont {S.~V.}\ \bibnamefont {Eremeev}}, \bibinfo
  {author} {\bibfnamefont {S.~O.}\ \bibnamefont {Filnov}}, \bibinfo {author}
  {\bibfnamefont {A.}~\bibnamefont {Koroleva}}, \bibinfo {author}
  {\bibfnamefont {E.}~\bibnamefont {Shevchenko}}, \bibinfo {author}
  {\bibfnamefont {V.}~\bibnamefont {Voroshnin}}, \bibinfo {author}
  {\bibfnamefont {A.~G.}\ \bibnamefont {Rybkin}}, \bibinfo {author}
  {\bibfnamefont {I.~P.}\ \bibnamefont {Rusinov}}, \bibinfo {author}
  {\bibfnamefont {M.}~\bibnamefont {Blanco-Rey}}, \bibinfo {author}
  {\bibfnamefont {M.}~\bibnamefont {Hoffmann}}, \bibinfo {author}
  {\bibfnamefont {Z.~S.}\ \bibnamefont {Aliev}}, \bibinfo {author}
  {\bibfnamefont {M.~B.}\ \bibnamefont {Babanly}}, \bibinfo {author}
  {\bibfnamefont {I.~R.}\ \bibnamefont {Amiraslanov}}, \bibinfo {author}
  {\bibfnamefont {N.~A.}\ \bibnamefont {Abdullayev}}, \bibinfo {author}
  {\bibfnamefont {V.~N.}\ \bibnamefont {Zverev}}, \bibinfo {author}
  {\bibfnamefont {A.}~\bibnamefont {Kimura}}, \bibinfo {author} {\bibfnamefont
  {O.~E.}\ \bibnamefont {Tereshchenko}}, \bibinfo {author} {\bibfnamefont
  {K.~A.}\ \bibnamefont {Kokh}}, \bibinfo {author} {\bibfnamefont
  {L.}~\bibnamefont {Petaccia}}, \bibinfo {author} {\bibfnamefont {G.~D.}\
  \bibnamefont {Santo}}, \bibinfo {author} {\bibfnamefont {A.}~\bibnamefont
  {Ernst}}, \bibinfo {author} {\bibfnamefont {P.~M.}\ \bibnamefont
  {Echenique}}, \bibinfo {author} {\bibfnamefont {N.~T.}\ \bibnamefont
  {Mamedov}}, \bibinfo {author} {\bibfnamefont {A.~M.}\ \bibnamefont {Shikin}},
  \ and\ \bibinfo {author} {\bibfnamefont {E.~V.}\ \bibnamefont {Chulkov}},\
  }\href {\doibase 10.1038/s41535-020-00255-9} {\bibfield  {journal} {\bibinfo
  {journal} {npj Quantum Materials}\ }\textbf {\bibinfo {volume} {5}},\
  \bibinfo {pages} {54} (\bibinfo {year} {2020})}\BibitemShut {NoStop}%
\bibitem [{\citenamefont {Hu}\ \emph {et~al.}(2020{\natexlab{a}})\citenamefont
  {Hu}, \citenamefont {Ding}, \citenamefont {Gordon}, \citenamefont {Ghosh},
  \citenamefont {Tien}, \citenamefont {Li}, \citenamefont {Linn}, \citenamefont
  {Lien}, \citenamefont {Huang}, \citenamefont {Mackey}, \citenamefont {Liu},
  \citenamefont {Reddy}, \citenamefont {Singh}, \citenamefont {Agarwal},
  \citenamefont {Bansil}, \citenamefont {Song}, \citenamefont {Li},
  \citenamefont {Xu}, \citenamefont {Lin}, \citenamefont {Cao}, \citenamefont
  {Chang}, \citenamefont {Dessau},\ and\ \citenamefont {Ni}}]{hu2020sciadv}%
  \BibitemOpen
  \bibfield  {author} {\bibinfo {author} {\bibfnamefont {C.}~\bibnamefont
  {Hu}}, \bibinfo {author} {\bibfnamefont {L.}~\bibnamefont {Ding}}, \bibinfo
  {author} {\bibfnamefont {K.~N.}\ \bibnamefont {Gordon}}, \bibinfo {author}
  {\bibfnamefont {B.}~\bibnamefont {Ghosh}}, \bibinfo {author} {\bibfnamefont
  {H.-J.}\ \bibnamefont {Tien}}, \bibinfo {author} {\bibfnamefont
  {H.}~\bibnamefont {Li}}, \bibinfo {author} {\bibfnamefont {A.~G.}\
  \bibnamefont {Linn}}, \bibinfo {author} {\bibfnamefont {S.-W.}\ \bibnamefont
  {Lien}}, \bibinfo {author} {\bibfnamefont {C.-Y.}\ \bibnamefont {Huang}},
  \bibinfo {author} {\bibfnamefont {S.}~\bibnamefont {Mackey}}, \bibinfo
  {author} {\bibfnamefont {J.}~\bibnamefont {Liu}}, \bibinfo {author}
  {\bibfnamefont {P.~V.~S.}\ \bibnamefont {Reddy}}, \bibinfo {author}
  {\bibfnamefont {B.}~\bibnamefont {Singh}}, \bibinfo {author} {\bibfnamefont
  {A.}~\bibnamefont {Agarwal}}, \bibinfo {author} {\bibfnamefont
  {A.}~\bibnamefont {Bansil}}, \bibinfo {author} {\bibfnamefont
  {M.}~\bibnamefont {Song}}, \bibinfo {author} {\bibfnamefont {D.}~\bibnamefont
  {Li}}, \bibinfo {author} {\bibfnamefont {S.-Y.}\ \bibnamefont {Xu}}, \bibinfo
  {author} {\bibfnamefont {H.}~\bibnamefont {Lin}}, \bibinfo {author}
  {\bibfnamefont {H.}~\bibnamefont {Cao}}, \bibinfo {author} {\bibfnamefont
  {T.-R.}\ \bibnamefont {Chang}}, \bibinfo {author} {\bibfnamefont
  {D.}~\bibnamefont {Dessau}}, \ and\ \bibinfo {author} {\bibfnamefont
  {N.}~\bibnamefont {Ni}},\ }\href {\doibase 10.1126/sciadv.aba4275} {\bibfield
   {journal} {\bibinfo  {journal} {Science Advances}\ }\textbf {\bibinfo
  {volume} {6}},\ \bibinfo {pages} {eaba4275} (\bibinfo {year}
  {2020}{\natexlab{a}})}\BibitemShut {NoStop}%
\bibitem [{\citenamefont {Wu}\ \emph {et~al.}(2020{\natexlab{a}})\citenamefont
  {Wu}, \citenamefont {Liu}, \citenamefont {Liu}, \citenamefont {Wang},
  \citenamefont {Li}, \citenamefont {Lu}, \citenamefont {Matsuishi},\ and\
  \citenamefont {Hosono}}]{wu2020toward}%
  \BibitemOpen
  \bibfield  {author} {\bibinfo {author} {\bibfnamefont {J.}~\bibnamefont
  {Wu}}, \bibinfo {author} {\bibfnamefont {F.}~\bibnamefont {Liu}}, \bibinfo
  {author} {\bibfnamefont {C.}~\bibnamefont {Liu}}, \bibinfo {author}
  {\bibfnamefont {Y.}~\bibnamefont {Wang}}, \bibinfo {author} {\bibfnamefont
  {C.}~\bibnamefont {Li}}, \bibinfo {author} {\bibfnamefont {Y.}~\bibnamefont
  {Lu}}, \bibinfo {author} {\bibfnamefont {S.}~\bibnamefont {Matsuishi}}, \
  and\ \bibinfo {author} {\bibfnamefont {H.}~\bibnamefont {Hosono}},\ }\href
  {\doibase 10.1002/adma.202001815} {\bibfield  {journal} {\bibinfo  {journal}
  {Advanced Materials}\ }\textbf {\bibinfo {volume} {32}},\ \bibinfo {pages}
  {2001815} (\bibinfo {year} {2020}{\natexlab{a}})}\BibitemShut {NoStop}%
\bibitem [{\citenamefont {Otrokov}\ \emph
  {et~al.}(2019{\natexlab{b}})\citenamefont {Otrokov}, \citenamefont {Rusinov},
  \citenamefont {Blanco-Rey}, \citenamefont {Hoffmann}, \citenamefont
  {Vyazovskaya}, \citenamefont {Eremeev}, \citenamefont {Ernst}, \citenamefont
  {Echenique}, \citenamefont {Arnau},\ and\ \citenamefont
  {Chulkov}}]{PhysRevLett.122.107202}%
  \BibitemOpen
  \bibfield  {author} {\bibinfo {author} {\bibfnamefont {M.~M.}\ \bibnamefont
  {Otrokov}}, \bibinfo {author} {\bibfnamefont {I.~P.}\ \bibnamefont
  {Rusinov}}, \bibinfo {author} {\bibfnamefont {M.}~\bibnamefont {Blanco-Rey}},
  \bibinfo {author} {\bibfnamefont {M.}~\bibnamefont {Hoffmann}}, \bibinfo
  {author} {\bibfnamefont {A.~Y.}\ \bibnamefont {Vyazovskaya}}, \bibinfo
  {author} {\bibfnamefont {S.~V.}\ \bibnamefont {Eremeev}}, \bibinfo {author}
  {\bibfnamefont {A.}~\bibnamefont {Ernst}}, \bibinfo {author} {\bibfnamefont
  {P.~M.}\ \bibnamefont {Echenique}}, \bibinfo {author} {\bibfnamefont
  {A.}~\bibnamefont {Arnau}}, \ and\ \bibinfo {author} {\bibfnamefont {E.~V.}\
  \bibnamefont {Chulkov}},\ }\href {\doibase 10.1103/PhysRevLett.122.107202}
  {\bibfield  {journal} {\bibinfo  {journal} {Phys. Rev. Lett.}\ }\textbf
  {\bibinfo {volume} {122}},\ \bibinfo {pages} {107202} (\bibinfo {year}
  {2019}{\natexlab{b}})}\BibitemShut {NoStop}%
\bibitem [{\citenamefont {Tian}\ \emph {et~al.}(2020)\citenamefont {Tian},
  \citenamefont {Gao}, \citenamefont {Nie}, \citenamefont {Qian}, \citenamefont
  {Gong}, \citenamefont {Fu}, \citenamefont {Li}, \citenamefont {Fan},
  \citenamefont {Zhang}, \citenamefont {Kondo}, \citenamefont {Shin},
  \citenamefont {Adell}, \citenamefont {Fedderwitz}, \citenamefont {Ding},
  \citenamefont {Wang}, \citenamefont {Qian},\ and\ \citenamefont
  {Lei}}]{PhysRevB.102.035144}%
  \BibitemOpen
  \bibfield  {author} {\bibinfo {author} {\bibfnamefont {S.}~\bibnamefont
  {Tian}}, \bibinfo {author} {\bibfnamefont {S.}~\bibnamefont {Gao}}, \bibinfo
  {author} {\bibfnamefont {S.}~\bibnamefont {Nie}}, \bibinfo {author}
  {\bibfnamefont {Y.}~\bibnamefont {Qian}}, \bibinfo {author} {\bibfnamefont
  {C.}~\bibnamefont {Gong}}, \bibinfo {author} {\bibfnamefont {Y.}~\bibnamefont
  {Fu}}, \bibinfo {author} {\bibfnamefont {H.}~\bibnamefont {Li}}, \bibinfo
  {author} {\bibfnamefont {W.}~\bibnamefont {Fan}}, \bibinfo {author}
  {\bibfnamefont {P.}~\bibnamefont {Zhang}}, \bibinfo {author} {\bibfnamefont
  {T.}~\bibnamefont {Kondo}}, \bibinfo {author} {\bibfnamefont
  {S.}~\bibnamefont {Shin}}, \bibinfo {author} {\bibfnamefont {J.}~\bibnamefont
  {Adell}}, \bibinfo {author} {\bibfnamefont {H.}~\bibnamefont {Fedderwitz}},
  \bibinfo {author} {\bibfnamefont {H.}~\bibnamefont {Ding}}, \bibinfo {author}
  {\bibfnamefont {Z.}~\bibnamefont {Wang}}, \bibinfo {author} {\bibfnamefont
  {T.}~\bibnamefont {Qian}}, \ and\ \bibinfo {author} {\bibfnamefont
  {H.}~\bibnamefont {Lei}},\ }\href {\doibase 10.1103/PhysRevB.102.035144}
  {\bibfield  {journal} {\bibinfo  {journal} {Phys. Rev. B}\ }\textbf {\bibinfo
  {volume} {102}},\ \bibinfo {pages} {035144} (\bibinfo {year}
  {2020})}\BibitemShut {NoStop}%
\bibitem [{\citenamefont {Ma}\ \emph {et~al.}(2020)\citenamefont {Ma},
  \citenamefont {Chen}, \citenamefont {Schwier}, \citenamefont {Zhang},
  \citenamefont {Hao}, \citenamefont {Kumar}, \citenamefont {Lu} \emph
  {et~al.}}]{PhysRevB.102.245136}%
  \BibitemOpen
  \bibfield  {author} {\bibinfo {author} {\bibfnamefont {X.-M.}\ \bibnamefont
  {Ma}}, \bibinfo {author} {\bibfnamefont {Z.}~\bibnamefont {Chen}}, \bibinfo
  {author} {\bibfnamefont {E.~F.}\ \bibnamefont {Schwier}}, \bibinfo {author}
  {\bibfnamefont {Y.}~\bibnamefont {Zhang}}, \bibinfo {author} {\bibfnamefont
  {Y.-J.}\ \bibnamefont {Hao}}, \bibinfo {author} {\bibfnamefont
  {S.}~\bibnamefont {Kumar}}, \bibinfo {author} {\bibfnamefont
  {R.}~\bibnamefont {Lu}},  \emph {et~al.},\ }\href {\doibase
  10.1103/PhysRevB.102.245136} {\bibfield  {journal} {\bibinfo  {journal}
  {Phys. Rev. B}\ }\textbf {\bibinfo {volume} {102}},\ \bibinfo {pages}
  {245136} (\bibinfo {year} {2020})}\BibitemShut {NoStop}%
\bibitem [{\citenamefont {Zhong}\ \emph {et~al.}(2021)\citenamefont {Zhong},
  \citenamefont {Bao}, \citenamefont {Wang}, \citenamefont {Li}, \citenamefont
  {Yin}, \citenamefont {Xu}, \citenamefont {Duan}, \citenamefont {Xia},\ and\
  \citenamefont {Zhou}}]{zhong2021light}%
  \BibitemOpen
  \bibfield  {author} {\bibinfo {author} {\bibfnamefont {H.}~\bibnamefont
  {Zhong}}, \bibinfo {author} {\bibfnamefont {C.}~\bibnamefont {Bao}}, \bibinfo
  {author} {\bibfnamefont {H.}~\bibnamefont {Wang}}, \bibinfo {author}
  {\bibfnamefont {J.}~\bibnamefont {Li}}, \bibinfo {author} {\bibfnamefont
  {Z.}~\bibnamefont {Yin}}, \bibinfo {author} {\bibfnamefont {Y.}~\bibnamefont
  {Xu}}, \bibinfo {author} {\bibfnamefont {W.}~\bibnamefont {Duan}}, \bibinfo
  {author} {\bibfnamefont {T.-L.}\ \bibnamefont {Xia}}, \ and\ \bibinfo
  {author} {\bibfnamefont {S.}~\bibnamefont {Zhou}},\ }\href {\doibase
  10.1021/acs.nanolett.1c01448} {\bibfield  {journal} {\bibinfo  {journal}
  {Nano Letters}\ }\textbf {\bibinfo {volume} {21}},\ \bibinfo {pages} {6080}
  (\bibinfo {year} {2021})}\BibitemShut {NoStop}%
\bibitem [{\citenamefont {Wu}\ \emph {et~al.}(2020{\natexlab{b}})\citenamefont
  {Wu}, \citenamefont {Li}, \citenamefont {Ma}, \citenamefont {Zhang},
  \citenamefont {Liu}, \citenamefont {Zhou}, \citenamefont {Shao},
  \citenamefont {Wang}, \citenamefont {Hao}, \citenamefont {Feng},
  \citenamefont {Schwier}, \citenamefont {Kumar}, \citenamefont {Sun},
  \citenamefont {Liu}, \citenamefont {Shimada}, \citenamefont {Miyamoto},
  \citenamefont {Okuda}, \citenamefont {Wang}, \citenamefont {Xie},
  \citenamefont {Chen}, \citenamefont {Liu}, \citenamefont {Liu},\ and\
  \citenamefont {Zhao}}]{PhysRevX.10.031013}%
  \BibitemOpen
  \bibfield  {author} {\bibinfo {author} {\bibfnamefont {X.}~\bibnamefont
  {Wu}}, \bibinfo {author} {\bibfnamefont {J.}~\bibnamefont {Li}}, \bibinfo
  {author} {\bibfnamefont {X.-M.}\ \bibnamefont {Ma}}, \bibinfo {author}
  {\bibfnamefont {Y.}~\bibnamefont {Zhang}}, \bibinfo {author} {\bibfnamefont
  {Y.}~\bibnamefont {Liu}}, \bibinfo {author} {\bibfnamefont {C.-S.}\
  \bibnamefont {Zhou}}, \bibinfo {author} {\bibfnamefont {J.}~\bibnamefont
  {Shao}}, \bibinfo {author} {\bibfnamefont {Q.}~\bibnamefont {Wang}}, \bibinfo
  {author} {\bibfnamefont {Y.-J.}\ \bibnamefont {Hao}}, \bibinfo {author}
  {\bibfnamefont {Y.}~\bibnamefont {Feng}}, \bibinfo {author} {\bibfnamefont
  {E.~F.}\ \bibnamefont {Schwier}}, \bibinfo {author} {\bibfnamefont
  {S.}~\bibnamefont {Kumar}}, \bibinfo {author} {\bibfnamefont
  {H.}~\bibnamefont {Sun}}, \bibinfo {author} {\bibfnamefont {P.}~\bibnamefont
  {Liu}}, \bibinfo {author} {\bibfnamefont {K.}~\bibnamefont {Shimada}},
  \bibinfo {author} {\bibfnamefont {K.}~\bibnamefont {Miyamoto}}, \bibinfo
  {author} {\bibfnamefont {T.}~\bibnamefont {Okuda}}, \bibinfo {author}
  {\bibfnamefont {K.}~\bibnamefont {Wang}}, \bibinfo {author} {\bibfnamefont
  {M.}~\bibnamefont {Xie}}, \bibinfo {author} {\bibfnamefont {C.}~\bibnamefont
  {Chen}}, \bibinfo {author} {\bibfnamefont {Q.}~\bibnamefont {Liu}}, \bibinfo
  {author} {\bibfnamefont {C.}~\bibnamefont {Liu}}, \ and\ \bibinfo {author}
  {\bibfnamefont {Y.}~\bibnamefont {Zhao}},\ }\href {\doibase
  10.1103/PhysRevX.10.031013} {\bibfield  {journal} {\bibinfo  {journal} {Phys.
  Rev. X}\ }\textbf {\bibinfo {volume} {10}},\ \bibinfo {pages} {031013}
  (\bibinfo {year} {2020}{\natexlab{b}})}\BibitemShut {NoStop}%
\bibitem [{\citenamefont {Lee}\ \emph {et~al.}(2019{\natexlab{a}})\citenamefont
  {Lee}, \citenamefont {Zhu}, \citenamefont {Wang}, \citenamefont {Miao},
  \citenamefont {Pillsbury}, \citenamefont {Yi}, \citenamefont {Kempinger},
  \citenamefont {Hu}, \citenamefont {Heikes}, \citenamefont {Quarterman},
  \citenamefont {Ratcliff}, \citenamefont {Borchers}, \citenamefont {Zhang},
  \citenamefont {Ke}, \citenamefont {Graf}, \citenamefont {Alem}, \citenamefont
  {Chang}, \citenamefont {Samarth},\ and\ \citenamefont
  {Mao}}]{PhysRevResearch.1.012011}%
  \BibitemOpen
  \bibfield  {author} {\bibinfo {author} {\bibfnamefont {S.~H.}\ \bibnamefont
  {Lee}}, \bibinfo {author} {\bibfnamefont {Y.}~\bibnamefont {Zhu}}, \bibinfo
  {author} {\bibfnamefont {Y.}~\bibnamefont {Wang}}, \bibinfo {author}
  {\bibfnamefont {L.}~\bibnamefont {Miao}}, \bibinfo {author} {\bibfnamefont
  {T.}~\bibnamefont {Pillsbury}}, \bibinfo {author} {\bibfnamefont
  {H.}~\bibnamefont {Yi}}, \bibinfo {author} {\bibfnamefont {S.}~\bibnamefont
  {Kempinger}}, \bibinfo {author} {\bibfnamefont {J.}~\bibnamefont {Hu}},
  \bibinfo {author} {\bibfnamefont {C.~A.}\ \bibnamefont {Heikes}}, \bibinfo
  {author} {\bibfnamefont {P.}~\bibnamefont {Quarterman}}, \bibinfo {author}
  {\bibfnamefont {W.}~\bibnamefont {Ratcliff}}, \bibinfo {author}
  {\bibfnamefont {J.~A.}\ \bibnamefont {Borchers}}, \bibinfo {author}
  {\bibfnamefont {H.}~\bibnamefont {Zhang}}, \bibinfo {author} {\bibfnamefont
  {X.}~\bibnamefont {Ke}}, \bibinfo {author} {\bibfnamefont {D.}~\bibnamefont
  {Graf}}, \bibinfo {author} {\bibfnamefont {N.}~\bibnamefont {Alem}}, \bibinfo
  {author} {\bibfnamefont {C.-Z.}\ \bibnamefont {Chang}}, \bibinfo {author}
  {\bibfnamefont {N.}~\bibnamefont {Samarth}}, \ and\ \bibinfo {author}
  {\bibfnamefont {Z.}~\bibnamefont {Mao}},\ }\href {\doibase
  10.1103/PhysRevResearch.1.012011} {\bibfield  {journal} {\bibinfo  {journal}
  {Phys. Rev. Research}\ }\textbf {\bibinfo {volume} {1}},\ \bibinfo {pages}
  {012011} (\bibinfo {year} {2019}{\natexlab{a}})}\BibitemShut {NoStop}%
\bibitem [{\citenamefont {Shikin}\ \emph {et~al.}(2021)\citenamefont {Shikin},
  \citenamefont {Estyunin}, \citenamefont {Zaitsev}, \citenamefont {Glazkova},
  \citenamefont {Klimovskikh} \emph {et~al.}}]{shikin2021sample}%
  \BibitemOpen
  \bibfield  {author} {\bibinfo {author} {\bibfnamefont {A.~M.}\ \bibnamefont
  {Shikin}}, \bibinfo {author} {\bibfnamefont {D.~A.}\ \bibnamefont
  {Estyunin}}, \bibinfo {author} {\bibfnamefont {N.~L.}\ \bibnamefont
  {Zaitsev}}, \bibinfo {author} {\bibfnamefont {D.}~\bibnamefont {Glazkova}},
  \bibinfo {author} {\bibfnamefont {I.~I.}\ \bibnamefont {Klimovskikh}},  \emph
  {et~al.},\ }\href {\doibase 10.1103/PhysRevB.104.115168} {\bibfield
  {journal} {\bibinfo  {journal} {Phys. Rev. B}\ }\textbf {\bibinfo {volume}
  {104}},\ \bibinfo {pages} {115168} (\bibinfo {year} {2021})}\BibitemShut
  {NoStop}%
\bibitem [{\citenamefont {Ji}\ \emph {et~al.}(2021)\citenamefont {Ji},
  \citenamefont {Liu}, \citenamefont {Wang}, \citenamefont {Luo}, \citenamefont
  {Li}, \citenamefont {Li}, \citenamefont {Wu}, \citenamefont {Xu},\ and\
  \citenamefont {Wang}}]{Ji2021Detection}%
  \BibitemOpen
  \bibfield  {author} {\bibinfo {author} {\bibfnamefont {H.-R.}\ \bibnamefont
  {Ji}}, \bibinfo {author} {\bibfnamefont {Y.-Z.}\ \bibnamefont {Liu}},
  \bibinfo {author} {\bibfnamefont {H.}~\bibnamefont {Wang}}, \bibinfo {author}
  {\bibfnamefont {J.-W.}\ \bibnamefont {Luo}}, \bibinfo {author} {\bibfnamefont
  {J.-H.}\ \bibnamefont {Li}}, \bibinfo {author} {\bibfnamefont
  {H.}~\bibnamefont {Li}}, \bibinfo {author} {\bibfnamefont {Y.}~\bibnamefont
  {Wu}}, \bibinfo {author} {\bibfnamefont {Y.}~\bibnamefont {Xu}}, \ and\
  \bibinfo {author} {\bibfnamefont {J.}~\bibnamefont {Wang}},\ }\href {\doibase
  10.1088/0256-307x/38/10/107404} {\bibfield  {journal} {\bibinfo  {journal}
  {Chinese Physics Letters}\ }\textbf {\bibinfo {volume} {38}},\ \bibinfo
  {pages} {107404} (\bibinfo {year} {2021})}\BibitemShut {NoStop}%
\bibitem [{\citenamefont {Hao}\ \emph {et~al.}(2019)\citenamefont {Hao},
  \citenamefont {Liu}, \citenamefont {Feng}, \citenamefont {Ma}, \citenamefont
  {Schwier}, \citenamefont {Arita}, \citenamefont {Kumar}, \citenamefont {Hu},
  \citenamefont {Lu}, \citenamefont {Zeng}, \citenamefont {Wang}, \citenamefont
  {Hao}, \citenamefont {Sun}, \citenamefont {Zhang}, \citenamefont {Mei},
  \citenamefont {Ni}, \citenamefont {Wu}, \citenamefont {Shimada},
  \citenamefont {Chen}, \citenamefont {Liu},\ and\ \citenamefont
  {Liu}}]{PhysRevX.9.041038}%
  \BibitemOpen
  \bibfield  {author} {\bibinfo {author} {\bibfnamefont {Y.-J.}\ \bibnamefont
  {Hao}}, \bibinfo {author} {\bibfnamefont {P.}~\bibnamefont {Liu}}, \bibinfo
  {author} {\bibfnamefont {Y.}~\bibnamefont {Feng}}, \bibinfo {author}
  {\bibfnamefont {X.-M.}\ \bibnamefont {Ma}}, \bibinfo {author} {\bibfnamefont
  {E.~F.}\ \bibnamefont {Schwier}}, \bibinfo {author} {\bibfnamefont
  {M.}~\bibnamefont {Arita}}, \bibinfo {author} {\bibfnamefont
  {S.}~\bibnamefont {Kumar}}, \bibinfo {author} {\bibfnamefont
  {C.}~\bibnamefont {Hu}}, \bibinfo {author} {\bibfnamefont {R.}~\bibnamefont
  {Lu}}, \bibinfo {author} {\bibfnamefont {M.}~\bibnamefont {Zeng}}, \bibinfo
  {author} {\bibfnamefont {Y.}~\bibnamefont {Wang}}, \bibinfo {author}
  {\bibfnamefont {Z.}~\bibnamefont {Hao}}, \bibinfo {author} {\bibfnamefont
  {H.-Y.}\ \bibnamefont {Sun}}, \bibinfo {author} {\bibfnamefont
  {K.}~\bibnamefont {Zhang}}, \bibinfo {author} {\bibfnamefont
  {J.}~\bibnamefont {Mei}}, \bibinfo {author} {\bibfnamefont {N.}~\bibnamefont
  {Ni}}, \bibinfo {author} {\bibfnamefont {L.}~\bibnamefont {Wu}}, \bibinfo
  {author} {\bibfnamefont {K.}~\bibnamefont {Shimada}}, \bibinfo {author}
  {\bibfnamefont {C.}~\bibnamefont {Chen}}, \bibinfo {author} {\bibfnamefont
  {Q.}~\bibnamefont {Liu}}, \ and\ \bibinfo {author} {\bibfnamefont
  {C.}~\bibnamefont {Liu}},\ }\href {\doibase 10.1103/PhysRevX.9.041038}
  {\bibfield  {journal} {\bibinfo  {journal} {Phys. Rev. X}\ }\textbf {\bibinfo
  {volume} {9}},\ \bibinfo {pages} {041038} (\bibinfo {year}
  {2019})}\BibitemShut {NoStop}%
\bibitem [{\citenamefont {Li}\ \emph {et~al.}(2019{\natexlab{b}})\citenamefont
  {Li}, \citenamefont {Gao}, \citenamefont {Duan}, \citenamefont {Xu},
  \citenamefont {Zhu}, \citenamefont {Tian}, \citenamefont {Gao}, \citenamefont
  {Fan}, \citenamefont {Rao}, \citenamefont {Huang}, \citenamefont {Li},
  \citenamefont {Yan}, \citenamefont {Liu}, \citenamefont {Liu}, \citenamefont
  {Huang}, \citenamefont {Li}, \citenamefont {Liu}, \citenamefont {Zhang},
  \citenamefont {Zhang}, \citenamefont {Kondo}, \citenamefont {Shin},
  \citenamefont {Lei}, \citenamefont {Shi}, \citenamefont {Zhang},
  \citenamefont {Weng}, \citenamefont {Qian},\ and\ \citenamefont
  {Ding}}]{PhysRevX.9.041039}%
  \BibitemOpen
  \bibfield  {author} {\bibinfo {author} {\bibfnamefont {H.}~\bibnamefont
  {Li}}, \bibinfo {author} {\bibfnamefont {S.-Y.}\ \bibnamefont {Gao}},
  \bibinfo {author} {\bibfnamefont {S.-F.}\ \bibnamefont {Duan}}, \bibinfo
  {author} {\bibfnamefont {Y.-F.}\ \bibnamefont {Xu}}, \bibinfo {author}
  {\bibfnamefont {K.-J.}\ \bibnamefont {Zhu}}, \bibinfo {author} {\bibfnamefont
  {S.-J.}\ \bibnamefont {Tian}}, \bibinfo {author} {\bibfnamefont {J.-C.}\
  \bibnamefont {Gao}}, \bibinfo {author} {\bibfnamefont {W.-H.}\ \bibnamefont
  {Fan}}, \bibinfo {author} {\bibfnamefont {Z.-C.}\ \bibnamefont {Rao}},
  \bibinfo {author} {\bibfnamefont {J.-R.}\ \bibnamefont {Huang}}, \bibinfo
  {author} {\bibfnamefont {J.-J.}\ \bibnamefont {Li}}, \bibinfo {author}
  {\bibfnamefont {D.-Y.}\ \bibnamefont {Yan}}, \bibinfo {author} {\bibfnamefont
  {Z.-T.}\ \bibnamefont {Liu}}, \bibinfo {author} {\bibfnamefont {W.-L.}\
  \bibnamefont {Liu}}, \bibinfo {author} {\bibfnamefont {Y.-B.}\ \bibnamefont
  {Huang}}, \bibinfo {author} {\bibfnamefont {Y.-L.}\ \bibnamefont {Li}},
  \bibinfo {author} {\bibfnamefont {Y.}~\bibnamefont {Liu}}, \bibinfo {author}
  {\bibfnamefont {G.-B.}\ \bibnamefont {Zhang}}, \bibinfo {author}
  {\bibfnamefont {P.}~\bibnamefont {Zhang}}, \bibinfo {author} {\bibfnamefont
  {T.}~\bibnamefont {Kondo}}, \bibinfo {author} {\bibfnamefont
  {S.}~\bibnamefont {Shin}}, \bibinfo {author} {\bibfnamefont {H.-C.}\
  \bibnamefont {Lei}}, \bibinfo {author} {\bibfnamefont {Y.-G.}\ \bibnamefont
  {Shi}}, \bibinfo {author} {\bibfnamefont {W.-T.}\ \bibnamefont {Zhang}},
  \bibinfo {author} {\bibfnamefont {H.-M.}\ \bibnamefont {Weng}}, \bibinfo
  {author} {\bibfnamefont {T.}~\bibnamefont {Qian}}, \ and\ \bibinfo {author}
  {\bibfnamefont {H.}~\bibnamefont {Ding}},\ }\href {\doibase
  10.1103/PhysRevX.9.041039} {\bibfield  {journal} {\bibinfo  {journal} {Phys.
  Rev. X}\ }\textbf {\bibinfo {volume} {9}},\ \bibinfo {pages} {041039}
  (\bibinfo {year} {2019}{\natexlab{b}})}\BibitemShut {NoStop}%
\bibitem [{\citenamefont {Chen}\ \emph {et~al.}(2019)\citenamefont {Chen},
  \citenamefont {Xu}, \citenamefont {Li}, \citenamefont {Li}, \citenamefont
  {Wang}, \citenamefont {Zhang}, \citenamefont {Li}, \citenamefont {Wu},
  \citenamefont {Liang}, \citenamefont {Chen}, \citenamefont {Jung},
  \citenamefont {Cacho}, \citenamefont {Mao}, \citenamefont {Liu},
  \citenamefont {Wang}, \citenamefont {Guo}, \citenamefont {Xu}, \citenamefont
  {Liu}, \citenamefont {Yang},\ and\ \citenamefont {Chen}}]{PhysRevX.9.041040}%
  \BibitemOpen
  \bibfield  {author} {\bibinfo {author} {\bibfnamefont {Y.~J.}\ \bibnamefont
  {Chen}}, \bibinfo {author} {\bibfnamefont {L.~X.}\ \bibnamefont {Xu}},
  \bibinfo {author} {\bibfnamefont {J.~H.}\ \bibnamefont {Li}}, \bibinfo
  {author} {\bibfnamefont {Y.~W.}\ \bibnamefont {Li}}, \bibinfo {author}
  {\bibfnamefont {H.~Y.}\ \bibnamefont {Wang}}, \bibinfo {author}
  {\bibfnamefont {C.~F.}\ \bibnamefont {Zhang}}, \bibinfo {author}
  {\bibfnamefont {H.}~\bibnamefont {Li}}, \bibinfo {author} {\bibfnamefont
  {Y.}~\bibnamefont {Wu}}, \bibinfo {author} {\bibfnamefont {A.~J.}\
  \bibnamefont {Liang}}, \bibinfo {author} {\bibfnamefont {C.}~\bibnamefont
  {Chen}}, \bibinfo {author} {\bibfnamefont {S.~W.}\ \bibnamefont {Jung}},
  \bibinfo {author} {\bibfnamefont {C.}~\bibnamefont {Cacho}}, \bibinfo
  {author} {\bibfnamefont {Y.~H.}\ \bibnamefont {Mao}}, \bibinfo {author}
  {\bibfnamefont {S.}~\bibnamefont {Liu}}, \bibinfo {author} {\bibfnamefont
  {M.~X.}\ \bibnamefont {Wang}}, \bibinfo {author} {\bibfnamefont {Y.~F.}\
  \bibnamefont {Guo}}, \bibinfo {author} {\bibfnamefont {Y.}~\bibnamefont
  {Xu}}, \bibinfo {author} {\bibfnamefont {Z.~K.}\ \bibnamefont {Liu}},
  \bibinfo {author} {\bibfnamefont {L.~X.}\ \bibnamefont {Yang}}, \ and\
  \bibinfo {author} {\bibfnamefont {Y.~L.}\ \bibnamefont {Chen}},\ }\href
  {\doibase 10.1103/PhysRevX.9.041040} {\bibfield  {journal} {\bibinfo
  {journal} {Phys. Rev. X}\ }\textbf {\bibinfo {volume} {9}},\ \bibinfo {pages}
  {041040} (\bibinfo {year} {2019})}\BibitemShut {NoStop}%
\bibitem [{\citenamefont {Vidal}\ \emph {et~al.}(2019)\citenamefont {Vidal},
  \citenamefont {Zeugner}, \citenamefont {Facio}, \citenamefont {Ray},
  \citenamefont {Haghighi}, \citenamefont {Wolter}, \citenamefont
  {Corredor~Bohorquez}, \citenamefont {Caglieris}, \citenamefont {Moser},
  \citenamefont {Figgemeier}, \citenamefont {Peixoto}, \citenamefont {Vasili},
  \citenamefont {Valvidares}, \citenamefont {Jung}, \citenamefont {Cacho},
  \citenamefont {Alfonsov}, \citenamefont {Mehlawat}, \citenamefont {Kataev},
  \citenamefont {Hess}, \citenamefont {Richter}, \citenamefont {B\"uchner},
  \citenamefont {van~den Brink}, \citenamefont {Ruck}, \citenamefont {Reinert},
  \citenamefont {Bentmann},\ and\ \citenamefont {Isaeva}}]{PhysRevX.9.041065}%
  \BibitemOpen
  \bibfield  {author} {\bibinfo {author} {\bibfnamefont {R.~C.}\ \bibnamefont
  {Vidal}}, \bibinfo {author} {\bibfnamefont {A.}~\bibnamefont {Zeugner}},
  \bibinfo {author} {\bibfnamefont {J.~I.}\ \bibnamefont {Facio}}, \bibinfo
  {author} {\bibfnamefont {R.}~\bibnamefont {Ray}}, \bibinfo {author}
  {\bibfnamefont {M.~H.}\ \bibnamefont {Haghighi}}, \bibinfo {author}
  {\bibfnamefont {A.~U.~B.}\ \bibnamefont {Wolter}}, \bibinfo {author}
  {\bibfnamefont {L.~T.}\ \bibnamefont {Corredor~Bohorquez}}, \bibinfo {author}
  {\bibfnamefont {F.}~\bibnamefont {Caglieris}}, \bibinfo {author}
  {\bibfnamefont {S.}~\bibnamefont {Moser}}, \bibinfo {author} {\bibfnamefont
  {T.}~\bibnamefont {Figgemeier}}, \bibinfo {author} {\bibfnamefont {T.~R.~F.}\
  \bibnamefont {Peixoto}}, \bibinfo {author} {\bibfnamefont {H.~B.}\
  \bibnamefont {Vasili}}, \bibinfo {author} {\bibfnamefont {M.}~\bibnamefont
  {Valvidares}}, \bibinfo {author} {\bibfnamefont {S.}~\bibnamefont {Jung}},
  \bibinfo {author} {\bibfnamefont {C.}~\bibnamefont {Cacho}}, \bibinfo
  {author} {\bibfnamefont {A.}~\bibnamefont {Alfonsov}}, \bibinfo {author}
  {\bibfnamefont {K.}~\bibnamefont {Mehlawat}}, \bibinfo {author}
  {\bibfnamefont {V.}~\bibnamefont {Kataev}}, \bibinfo {author} {\bibfnamefont
  {C.}~\bibnamefont {Hess}}, \bibinfo {author} {\bibfnamefont {M.}~\bibnamefont
  {Richter}}, \bibinfo {author} {\bibfnamefont {B.}~\bibnamefont {B\"uchner}},
  \bibinfo {author} {\bibfnamefont {J.}~\bibnamefont {van~den Brink}}, \bibinfo
  {author} {\bibfnamefont {M.}~\bibnamefont {Ruck}}, \bibinfo {author}
  {\bibfnamefont {F.}~\bibnamefont {Reinert}}, \bibinfo {author} {\bibfnamefont
  {H.}~\bibnamefont {Bentmann}}, \ and\ \bibinfo {author} {\bibfnamefont
  {A.}~\bibnamefont {Isaeva}},\ }\href {\doibase 10.1103/PhysRevX.9.041065}
  {\bibfield  {journal} {\bibinfo  {journal} {Phys. Rev. X}\ }\textbf {\bibinfo
  {volume} {9}},\ \bibinfo {pages} {041065} (\bibinfo {year}
  {2019})}\BibitemShut {NoStop}%
\bibitem [{\citenamefont {Swatek}\ \emph {et~al.}(2020)\citenamefont {Swatek},
  \citenamefont {Wu}, \citenamefont {Wang}, \citenamefont {Lee}, \citenamefont
  {Schrunk}, \citenamefont {Yan},\ and\ \citenamefont
  {Kaminski}}]{PhysRevB.101.161109}%
  \BibitemOpen
  \bibfield  {author} {\bibinfo {author} {\bibfnamefont {P.}~\bibnamefont
  {Swatek}}, \bibinfo {author} {\bibfnamefont {Y.}~\bibnamefont {Wu}}, \bibinfo
  {author} {\bibfnamefont {L.-L.}\ \bibnamefont {Wang}}, \bibinfo {author}
  {\bibfnamefont {K.}~\bibnamefont {Lee}}, \bibinfo {author} {\bibfnamefont
  {B.}~\bibnamefont {Schrunk}}, \bibinfo {author} {\bibfnamefont
  {J.}~\bibnamefont {Yan}}, \ and\ \bibinfo {author} {\bibfnamefont
  {A.}~\bibnamefont {Kaminski}},\ }\href {\doibase 10.1103/PhysRevB.101.161109}
  {\bibfield  {journal} {\bibinfo  {journal} {Phys. Rev. B}\ }\textbf {\bibinfo
  {volume} {101}},\ \bibinfo {pages} {161109} (\bibinfo {year}
  {2020})}\BibitemShut {NoStop}%
\bibitem [{\citenamefont {Xu}\ \emph {et~al.}(2020)\citenamefont {Xu},
  \citenamefont {Mao}, \citenamefont {Wang}, \citenamefont {Li}, \citenamefont
  {Chen}, \citenamefont {Xia}, \citenamefont {Li}, \citenamefont {Pei},
  \citenamefont {Zhang}, \citenamefont {Zheng}, \citenamefont {Huang},
  \citenamefont {Zhang}, \citenamefont {Cui}, \citenamefont {Liang},
  \citenamefont {Xia}, \citenamefont {Su}, \citenamefont {Jung}, \citenamefont
  {Cacho}, \citenamefont {Wang}, \citenamefont {Li}, \citenamefont {Xu},
  \citenamefont {Guo}, \citenamefont {Yang}, \citenamefont {Liu}, \citenamefont
  {Chen},\ and\ \citenamefont {Jiang}}]{XU20202086}%
  \BibitemOpen
  \bibfield  {author} {\bibinfo {author} {\bibfnamefont {L.}~\bibnamefont
  {Xu}}, \bibinfo {author} {\bibfnamefont {Y.}~\bibnamefont {Mao}}, \bibinfo
  {author} {\bibfnamefont {H.}~\bibnamefont {Wang}}, \bibinfo {author}
  {\bibfnamefont {J.}~\bibnamefont {Li}}, \bibinfo {author} {\bibfnamefont
  {Y.}~\bibnamefont {Chen}}, \bibinfo {author} {\bibfnamefont {Y.}~\bibnamefont
  {Xia}}, \bibinfo {author} {\bibfnamefont {Y.}~\bibnamefont {Li}}, \bibinfo
  {author} {\bibfnamefont {D.}~\bibnamefont {Pei}}, \bibinfo {author}
  {\bibfnamefont {J.}~\bibnamefont {Zhang}}, \bibinfo {author} {\bibfnamefont
  {H.}~\bibnamefont {Zheng}}, \bibinfo {author} {\bibfnamefont
  {K.}~\bibnamefont {Huang}}, \bibinfo {author} {\bibfnamefont
  {C.}~\bibnamefont {Zhang}}, \bibinfo {author} {\bibfnamefont
  {S.}~\bibnamefont {Cui}}, \bibinfo {author} {\bibfnamefont {A.}~\bibnamefont
  {Liang}}, \bibinfo {author} {\bibfnamefont {W.}~\bibnamefont {Xia}}, \bibinfo
  {author} {\bibfnamefont {H.}~\bibnamefont {Su}}, \bibinfo {author}
  {\bibfnamefont {S.}~\bibnamefont {Jung}}, \bibinfo {author} {\bibfnamefont
  {C.}~\bibnamefont {Cacho}}, \bibinfo {author} {\bibfnamefont
  {M.}~\bibnamefont {Wang}}, \bibinfo {author} {\bibfnamefont {G.}~\bibnamefont
  {Li}}, \bibinfo {author} {\bibfnamefont {Y.}~\bibnamefont {Xu}}, \bibinfo
  {author} {\bibfnamefont {Y.}~\bibnamefont {Guo}}, \bibinfo {author}
  {\bibfnamefont {L.}~\bibnamefont {Yang}}, \bibinfo {author} {\bibfnamefont
  {Z.}~\bibnamefont {Liu}}, \bibinfo {author} {\bibfnamefont {Y.}~\bibnamefont
  {Chen}}, \ and\ \bibinfo {author} {\bibfnamefont {M.}~\bibnamefont {Jiang}},\
  }\href {\doibase https://doi.org/10.1016/j.scib.2020.07.032} {\bibfield
  {journal} {\bibinfo  {journal} {Science Bulletin}\ }\textbf {\bibinfo
  {volume} {65}},\ \bibinfo {pages} {2086} (\bibinfo {year}
  {2020})}\BibitemShut {NoStop}%
\bibitem [{\citenamefont {Hu}\ \emph {et~al.}(2020{\natexlab{b}})\citenamefont
  {Hu}, \citenamefont {Gordon}, \citenamefont {Liu}, \citenamefont {Liu},
  \citenamefont {Zhou}, \citenamefont {Hao}, \citenamefont {Narayan},
  \citenamefont {Emmanouilidou}, \citenamefont {Sun}, \citenamefont {Liu},
  \citenamefont {Brawer}, \citenamefont {Ramirez}, \citenamefont {Ding},
  \citenamefont {Cao}, \citenamefont {Liu}, \citenamefont {Dessau},\ and\
  \citenamefont {Ni}}]{hu2020natcom}%
  \BibitemOpen
  \bibfield  {author} {\bibinfo {author} {\bibfnamefont {C.}~\bibnamefont
  {Hu}}, \bibinfo {author} {\bibfnamefont {K.~N.}\ \bibnamefont {Gordon}},
  \bibinfo {author} {\bibfnamefont {P.}~\bibnamefont {Liu}}, \bibinfo {author}
  {\bibfnamefont {J.}~\bibnamefont {Liu}}, \bibinfo {author} {\bibfnamefont
  {X.}~\bibnamefont {Zhou}}, \bibinfo {author} {\bibfnamefont {P.}~\bibnamefont
  {Hao}}, \bibinfo {author} {\bibfnamefont {D.}~\bibnamefont {Narayan}},
  \bibinfo {author} {\bibfnamefont {E.}~\bibnamefont {Emmanouilidou}}, \bibinfo
  {author} {\bibfnamefont {H.}~\bibnamefont {Sun}}, \bibinfo {author}
  {\bibfnamefont {Y.}~\bibnamefont {Liu}}, \bibinfo {author} {\bibfnamefont
  {H.}~\bibnamefont {Brawer}}, \bibinfo {author} {\bibfnamefont {A.~P.}\
  \bibnamefont {Ramirez}}, \bibinfo {author} {\bibfnamefont {L.}~\bibnamefont
  {Ding}}, \bibinfo {author} {\bibfnamefont {H.}~\bibnamefont {Cao}}, \bibinfo
  {author} {\bibfnamefont {Q.}~\bibnamefont {Liu}}, \bibinfo {author}
  {\bibfnamefont {D.}~\bibnamefont {Dessau}}, \ and\ \bibinfo {author}
  {\bibfnamefont {N.}~\bibnamefont {Ni}},\ }\href {\doibase
  10.1038/s41467-019-13814-x} {\bibfield  {journal} {\bibinfo  {journal}
  {Nature communications}\ }\textbf {\bibinfo {volume} {11}},\ \bibinfo {pages}
  {97} (\bibinfo {year} {2020}{\natexlab{b}})}\BibitemShut {NoStop}%
\bibitem [{\citenamefont {Hu}\ \emph {et~al.}(2020{\natexlab{c}})\citenamefont
  {Hu}, \citenamefont {Xu}, \citenamefont {Shi}, \citenamefont {Luo},
  \citenamefont {Peng}, \citenamefont {Wang}, \citenamefont {Ying},
  \citenamefont {Wu}, \citenamefont {Liu}, \citenamefont {Zhang}, \citenamefont
  {Chen}, \citenamefont {Xu}, \citenamefont {Chen},\ and\ \citenamefont
  {He}}]{PhysRevB.101.161113}%
  \BibitemOpen
  \bibfield  {author} {\bibinfo {author} {\bibfnamefont {Y.}~\bibnamefont
  {Hu}}, \bibinfo {author} {\bibfnamefont {L.}~\bibnamefont {Xu}}, \bibinfo
  {author} {\bibfnamefont {M.}~\bibnamefont {Shi}}, \bibinfo {author}
  {\bibfnamefont {A.}~\bibnamefont {Luo}}, \bibinfo {author} {\bibfnamefont
  {S.}~\bibnamefont {Peng}}, \bibinfo {author} {\bibfnamefont {Z.~Y.}\
  \bibnamefont {Wang}}, \bibinfo {author} {\bibfnamefont {J.~J.}\ \bibnamefont
  {Ying}}, \bibinfo {author} {\bibfnamefont {T.}~\bibnamefont {Wu}}, \bibinfo
  {author} {\bibfnamefont {Z.~K.}\ \bibnamefont {Liu}}, \bibinfo {author}
  {\bibfnamefont {C.~F.}\ \bibnamefont {Zhang}}, \bibinfo {author}
  {\bibfnamefont {Y.~L.}\ \bibnamefont {Chen}}, \bibinfo {author}
  {\bibfnamefont {G.}~\bibnamefont {Xu}}, \bibinfo {author} {\bibfnamefont
  {X.-H.}\ \bibnamefont {Chen}}, \ and\ \bibinfo {author} {\bibfnamefont
  {J.-F.}\ \bibnamefont {He}},\ }\href {\doibase 10.1103/PhysRevB.101.161113}
  {\bibfield  {journal} {\bibinfo  {journal} {Phys. Rev. B}\ }\textbf {\bibinfo
  {volume} {101}},\ \bibinfo {pages} {161113} (\bibinfo {year}
  {2020}{\natexlab{c}})}\BibitemShut {NoStop}%
\bibitem [{\citenamefont {Jo}\ \emph {et~al.}(2020)\citenamefont {Jo},
  \citenamefont {Wang}, \citenamefont {Slager}, \citenamefont {Yan},
  \citenamefont {Wu}, \citenamefont {Lee}, \citenamefont {Schrunk},
  \citenamefont {Vishwanath},\ and\ \citenamefont
  {Kaminski}}]{PhysRevB.102.045130}%
  \BibitemOpen
  \bibfield  {author} {\bibinfo {author} {\bibfnamefont {N.~H.}\ \bibnamefont
  {Jo}}, \bibinfo {author} {\bibfnamefont {L.-L.}\ \bibnamefont {Wang}},
  \bibinfo {author} {\bibfnamefont {R.-J.}\ \bibnamefont {Slager}}, \bibinfo
  {author} {\bibfnamefont {J.}~\bibnamefont {Yan}}, \bibinfo {author}
  {\bibfnamefont {Y.}~\bibnamefont {Wu}}, \bibinfo {author} {\bibfnamefont
  {K.}~\bibnamefont {Lee}}, \bibinfo {author} {\bibfnamefont {B.}~\bibnamefont
  {Schrunk}}, \bibinfo {author} {\bibfnamefont {A.}~\bibnamefont {Vishwanath}},
  \ and\ \bibinfo {author} {\bibfnamefont {A.}~\bibnamefont {Kaminski}},\
  }\href {\doibase 10.1103/PhysRevB.102.045130} {\bibfield  {journal} {\bibinfo
   {journal} {Phys. Rev. B}\ }\textbf {\bibinfo {volume} {102}},\ \bibinfo
  {pages} {045130} (\bibinfo {year} {2020})}\BibitemShut {NoStop}%
\bibitem [{\citenamefont {Vidal}\ \emph {et~al.}(2021)\citenamefont {Vidal},
  \citenamefont {Bentmann}, \citenamefont {Facio}, \citenamefont {Heider},
  \citenamefont {Kagerer}, \citenamefont {Fornari}, \citenamefont {Peixoto},
  \citenamefont {Figgemeier}, \citenamefont {Jung}, \citenamefont {Cacho},
  \citenamefont {B\"uchner}, \citenamefont {van~den Brink}, \citenamefont
  {Schneider}, \citenamefont {Plucinski}, \citenamefont {Schwier},
  \citenamefont {Shimada}, \citenamefont {Richter}, \citenamefont {Isaeva},\
  and\ \citenamefont {Reinert}}]{PhysRevLett.126.176403}%
  \BibitemOpen
  \bibfield  {author} {\bibinfo {author} {\bibfnamefont {R.~C.}\ \bibnamefont
  {Vidal}}, \bibinfo {author} {\bibfnamefont {H.}~\bibnamefont {Bentmann}},
  \bibinfo {author} {\bibfnamefont {J.~I.}\ \bibnamefont {Facio}}, \bibinfo
  {author} {\bibfnamefont {T.}~\bibnamefont {Heider}}, \bibinfo {author}
  {\bibfnamefont {P.}~\bibnamefont {Kagerer}}, \bibinfo {author} {\bibfnamefont
  {C.~I.}\ \bibnamefont {Fornari}}, \bibinfo {author} {\bibfnamefont
  {T.~R.~F.}\ \bibnamefont {Peixoto}}, \bibinfo {author} {\bibfnamefont
  {T.}~\bibnamefont {Figgemeier}}, \bibinfo {author} {\bibfnamefont
  {S.}~\bibnamefont {Jung}}, \bibinfo {author} {\bibfnamefont {C.}~\bibnamefont
  {Cacho}}, \bibinfo {author} {\bibfnamefont {B.}~\bibnamefont {B\"uchner}},
  \bibinfo {author} {\bibfnamefont {J.}~\bibnamefont {van~den Brink}}, \bibinfo
  {author} {\bibfnamefont {C.~M.}\ \bibnamefont {Schneider}}, \bibinfo {author}
  {\bibfnamefont {L.}~\bibnamefont {Plucinski}}, \bibinfo {author}
  {\bibfnamefont {E.~F.}\ \bibnamefont {Schwier}}, \bibinfo {author}
  {\bibfnamefont {K.}~\bibnamefont {Shimada}}, \bibinfo {author} {\bibfnamefont
  {M.}~\bibnamefont {Richter}}, \bibinfo {author} {\bibfnamefont
  {A.}~\bibnamefont {Isaeva}}, \ and\ \bibinfo {author} {\bibfnamefont
  {F.}~\bibnamefont {Reinert}},\ }\href {\doibase
  10.1103/PhysRevLett.126.176403} {\bibfield  {journal} {\bibinfo  {journal}
  {Phys. Rev. Lett.}\ }\textbf {\bibinfo {volume} {126}},\ \bibinfo {pages}
  {176403} (\bibinfo {year} {2021})}\BibitemShut {NoStop}%
\bibitem [{\citenamefont {Sass}\ \emph {et~al.}(2020)\citenamefont {Sass},
  \citenamefont {Kim}, \citenamefont {Vanderbilt}, \citenamefont {Yan},\ and\
  \citenamefont {Wu}}]{PhysRevLett.125.037201}%
  \BibitemOpen
  \bibfield  {author} {\bibinfo {author} {\bibfnamefont {P.~M.}\ \bibnamefont
  {Sass}}, \bibinfo {author} {\bibfnamefont {J.}~\bibnamefont {Kim}}, \bibinfo
  {author} {\bibfnamefont {D.}~\bibnamefont {Vanderbilt}}, \bibinfo {author}
  {\bibfnamefont {J.}~\bibnamefont {Yan}}, \ and\ \bibinfo {author}
  {\bibfnamefont {W.}~\bibnamefont {Wu}},\ }\href {\doibase
  10.1103/PhysRevLett.125.037201} {\bibfield  {journal} {\bibinfo  {journal}
  {Phys. Rev. Lett.}\ }\textbf {\bibinfo {volume} {125}},\ \bibinfo {pages}
  {037201} (\bibinfo {year} {2020})}\BibitemShut {NoStop}%
\bibitem [{\citenamefont {Nevola}\ \emph {et~al.}(2020)\citenamefont {Nevola},
  \citenamefont {Li}, \citenamefont {Yan}, \citenamefont {Moore}, \citenamefont
  {Lee}, \citenamefont {Miao},\ and\ \citenamefont
  {Johnson}}]{PhysRevLett.125.117205}%
  \BibitemOpen
  \bibfield  {author} {\bibinfo {author} {\bibfnamefont {D.}~\bibnamefont
  {Nevola}}, \bibinfo {author} {\bibfnamefont {H.~X.}\ \bibnamefont {Li}},
  \bibinfo {author} {\bibfnamefont {J.-Q.}\ \bibnamefont {Yan}}, \bibinfo
  {author} {\bibfnamefont {R.~G.}\ \bibnamefont {Moore}}, \bibinfo {author}
  {\bibfnamefont {H.-N.}\ \bibnamefont {Lee}}, \bibinfo {author} {\bibfnamefont
  {H.}~\bibnamefont {Miao}}, \ and\ \bibinfo {author} {\bibfnamefont {P.~D.}\
  \bibnamefont {Johnson}},\ }\href {\doibase 10.1103/PhysRevLett.125.117205}
  {\bibfield  {journal} {\bibinfo  {journal} {Phys. Rev. Lett.}\ }\textbf
  {\bibinfo {volume} {125}},\ \bibinfo {pages} {117205} (\bibinfo {year}
  {2020})}\BibitemShut {NoStop}%
\bibitem [{\citenamefont {Zhao}\ and\ \citenamefont
  {Liu}(2021{\natexlab{b}})}]{zhao2021routes}%
  \BibitemOpen
  \bibfield  {author} {\bibinfo {author} {\bibfnamefont {Y.}~\bibnamefont
  {Zhao}}\ and\ \bibinfo {author} {\bibfnamefont {Q.}~\bibnamefont {Liu}},\
  }\href {\doibase 10.1063/5.0059447} {\bibfield  {journal} {\bibinfo
  {journal} {Applied Physics Letters}\ }\textbf {\bibinfo {volume} {119}},\
  \bibinfo {pages} {060502} (\bibinfo {year} {2021}{\natexlab{b}})}\BibitemShut
  {NoStop}%
\bibitem [{\citenamefont {Lee}\ \emph {et~al.}(2019{\natexlab{b}})\citenamefont
  {Lee}, \citenamefont {Zhu}, \citenamefont {Wang}, \citenamefont {Miao},
  \citenamefont {Pillsbury}, \citenamefont {Yi}, \citenamefont {Kempinger},
  \citenamefont {Hu}, \citenamefont {Heikes}, \citenamefont {Quarterman},
  \citenamefont {Ratcliff}, \citenamefont {Borchers}, \citenamefont {Zhang},
  \citenamefont {Ke}, \citenamefont {Graf}, \citenamefont {Alem}, \citenamefont
  {Chang}, \citenamefont {Samarth},\ and\ \citenamefont {Mao}}]{Lee2019}%
  \BibitemOpen
  \bibfield  {author} {\bibinfo {author} {\bibfnamefont {S.~H.}\ \bibnamefont
  {Lee}}, \bibinfo {author} {\bibfnamefont {Y.}~\bibnamefont {Zhu}}, \bibinfo
  {author} {\bibfnamefont {Y.}~\bibnamefont {Wang}}, \bibinfo {author}
  {\bibfnamefont {L.}~\bibnamefont {Miao}}, \bibinfo {author} {\bibfnamefont
  {T.}~\bibnamefont {Pillsbury}}, \bibinfo {author} {\bibfnamefont
  {H.}~\bibnamefont {Yi}}, \bibinfo {author} {\bibfnamefont {S.}~\bibnamefont
  {Kempinger}}, \bibinfo {author} {\bibfnamefont {J.}~\bibnamefont {Hu}},
  \bibinfo {author} {\bibfnamefont {C.~A.}\ \bibnamefont {Heikes}}, \bibinfo
  {author} {\bibfnamefont {P.}~\bibnamefont {Quarterman}}, \bibinfo {author}
  {\bibfnamefont {W.}~\bibnamefont {Ratcliff}}, \bibinfo {author}
  {\bibfnamefont {J.~A.}\ \bibnamefont {Borchers}}, \bibinfo {author}
  {\bibfnamefont {H.}~\bibnamefont {Zhang}}, \bibinfo {author} {\bibfnamefont
  {X.}~\bibnamefont {Ke}}, \bibinfo {author} {\bibfnamefont {D.}~\bibnamefont
  {Graf}}, \bibinfo {author} {\bibfnamefont {N.}~\bibnamefont {Alem}}, \bibinfo
  {author} {\bibfnamefont {C.-Z.}\ \bibnamefont {Chang}}, \bibinfo {author}
  {\bibfnamefont {N.}~\bibnamefont {Samarth}}, \ and\ \bibinfo {author}
  {\bibfnamefont {Z.}~\bibnamefont {Mao}},\ }\href {\doibase
  10.1103/PhysRevResearch.1.012011} {\bibfield  {journal} {\bibinfo  {journal}
  {Phys. Rev. Research}\ }\textbf {\bibinfo {volume} {1}},\ \bibinfo {pages}
  {012011} (\bibinfo {year} {2019}{\natexlab{b}})}\BibitemShut {NoStop}%
\bibitem [{\citenamefont {Ovchinnikov}\ \emph {et~al.}(2021)\citenamefont
  {Ovchinnikov}, \citenamefont {Huang}, \citenamefont {Lin}, \citenamefont
  {Fei}, \citenamefont {Cai}, \citenamefont {Song}, \citenamefont {He},
  \citenamefont {Jiang}, \citenamefont {Wang}, \citenamefont {Li},
  \citenamefont {Wang}, \citenamefont {Wu}, \citenamefont {Xiao}, \citenamefont
  {Chu}, \citenamefont {Yan}, \citenamefont {Chang}, \citenamefont {Cui},\ and\
  \citenamefont {Xu}}]{nanolett.0c05117}%
  \BibitemOpen
  \bibfield  {author} {\bibinfo {author} {\bibfnamefont {D.}~\bibnamefont
  {Ovchinnikov}}, \bibinfo {author} {\bibfnamefont {X.}~\bibnamefont {Huang}},
  \bibinfo {author} {\bibfnamefont {Z.}~\bibnamefont {Lin}}, \bibinfo {author}
  {\bibfnamefont {Z.}~\bibnamefont {Fei}}, \bibinfo {author} {\bibfnamefont
  {J.}~\bibnamefont {Cai}}, \bibinfo {author} {\bibfnamefont {T.}~\bibnamefont
  {Song}}, \bibinfo {author} {\bibfnamefont {M.}~\bibnamefont {He}}, \bibinfo
  {author} {\bibfnamefont {Q.}~\bibnamefont {Jiang}}, \bibinfo {author}
  {\bibfnamefont {C.}~\bibnamefont {Wang}}, \bibinfo {author} {\bibfnamefont
  {H.}~\bibnamefont {Li}}, \bibinfo {author} {\bibfnamefont {Y.}~\bibnamefont
  {Wang}}, \bibinfo {author} {\bibfnamefont {Y.}~\bibnamefont {Wu}}, \bibinfo
  {author} {\bibfnamefont {D.}~\bibnamefont {Xiao}}, \bibinfo {author}
  {\bibfnamefont {J.-H.}\ \bibnamefont {Chu}}, \bibinfo {author} {\bibfnamefont
  {J.}~\bibnamefont {Yan}}, \bibinfo {author} {\bibfnamefont {C.-Z.}\
  \bibnamefont {Chang}}, \bibinfo {author} {\bibfnamefont {Y.-T.}\ \bibnamefont
  {Cui}}, \ and\ \bibinfo {author} {\bibfnamefont {X.}~\bibnamefont {Xu}},\
  }\href {\doibase 10.1021/acs.nanolett.0c05117} {\bibfield  {journal}
  {\bibinfo  {journal} {Nano Letters}\ }\textbf {\bibinfo {volume} {21}},\
  \bibinfo {pages} {2544} (\bibinfo {year} {2021})}\BibitemShut {NoStop}%
\bibitem [{\citenamefont {Liu}\ \emph {et~al.}(2021{\natexlab{a}})\citenamefont
  {Liu}, \citenamefont {Wang}, \citenamefont {Yang}, \citenamefont {Mao},
  \citenamefont {Li}, \citenamefont {Li}, \citenamefont {Li}, \citenamefont
  {Zhu}, \citenamefont {Wang}, \citenamefont {Li}, \citenamefont {Wu},
  \citenamefont {Xu}, \citenamefont {Zhang},\ and\ \citenamefont
  {Wang}}]{liu2021magnetic}%
  \BibitemOpen
  \bibfield  {author} {\bibinfo {author} {\bibfnamefont {C.}~\bibnamefont
  {Liu}}, \bibinfo {author} {\bibfnamefont {Y.}~\bibnamefont {Wang}}, \bibinfo
  {author} {\bibfnamefont {M.}~\bibnamefont {Yang}}, \bibinfo {author}
  {\bibfnamefont {J.}~\bibnamefont {Mao}}, \bibinfo {author} {\bibfnamefont
  {H.}~\bibnamefont {Li}}, \bibinfo {author} {\bibfnamefont {Y.}~\bibnamefont
  {Li}}, \bibinfo {author} {\bibfnamefont {J.}~\bibnamefont {Li}}, \bibinfo
  {author} {\bibfnamefont {H.}~\bibnamefont {Zhu}}, \bibinfo {author}
  {\bibfnamefont {J.}~\bibnamefont {Wang}}, \bibinfo {author} {\bibfnamefont
  {L.}~\bibnamefont {Li}}, \bibinfo {author} {\bibfnamefont {Y.}~\bibnamefont
  {Wu}}, \bibinfo {author} {\bibfnamefont {Y.}~\bibnamefont {Xu}}, \bibinfo
  {author} {\bibfnamefont {J.}~\bibnamefont {Zhang}}, \ and\ \bibinfo {author}
  {\bibfnamefont {Y.}~\bibnamefont {Wang}},\ }\href {\doibase
  10.1038/s41467-021-25002-x} {\bibfield  {journal} {\bibinfo  {journal}
  {Nature Communications}\ }\textbf {\bibinfo {volume} {12}},\ \bibinfo {pages}
  {4647} (\bibinfo {year} {2021}{\natexlab{a}})}\BibitemShut {NoStop}%
\bibitem [{\citenamefont {Ying}\ \emph {et~al.}(2022)\citenamefont {Ying},
  \citenamefont {Zhang}, \citenamefont {Chen}, \citenamefont {Jia},
  \citenamefont {Fei}, \citenamefont {Zhang}, \citenamefont {Zhang},
  \citenamefont {Wang},\ and\ \citenamefont {Song}}]{ying2021experimental}%
  \BibitemOpen
  \bibfield  {author} {\bibinfo {author} {\bibfnamefont {Z.}~\bibnamefont
  {Ying}}, \bibinfo {author} {\bibfnamefont {S.}~\bibnamefont {Zhang}},
  \bibinfo {author} {\bibfnamefont {B.}~\bibnamefont {Chen}}, \bibinfo {author}
  {\bibfnamefont {B.}~\bibnamefont {Jia}}, \bibinfo {author} {\bibfnamefont
  {F.}~\bibnamefont {Fei}}, \bibinfo {author} {\bibfnamefont {M.}~\bibnamefont
  {Zhang}}, \bibinfo {author} {\bibfnamefont {H.}~\bibnamefont {Zhang}},
  \bibinfo {author} {\bibfnamefont {X.}~\bibnamefont {Wang}}, \ and\ \bibinfo
  {author} {\bibfnamefont {F.}~\bibnamefont {Song}},\ }\href {\doibase
  10.1103/PhysRevB.105.085412} {\bibfield  {journal} {\bibinfo  {journal}
  {Phys. Rev. B}\ }\textbf {\bibinfo {volume} {105}},\ \bibinfo {pages}
  {085412} (\bibinfo {year} {2022})}\BibitemShut {NoStop}%
\bibitem [{\citenamefont {Cai}\ \emph {et~al.}(2022)\citenamefont {Cai},
  \citenamefont {Ovchinnikov}, \citenamefont {Fei}, \citenamefont {He},
  \citenamefont {Song}, \citenamefont {Lin}, \citenamefont {Wang},
  \citenamefont {Cobden}, \citenamefont {Chu}, \citenamefont {Cui} \emph
  {et~al.}}]{cai2022electric}%
  \BibitemOpen
  \bibfield  {author} {\bibinfo {author} {\bibfnamefont {J.}~\bibnamefont
  {Cai}}, \bibinfo {author} {\bibfnamefont {D.}~\bibnamefont {Ovchinnikov}},
  \bibinfo {author} {\bibfnamefont {Z.}~\bibnamefont {Fei}}, \bibinfo {author}
  {\bibfnamefont {M.}~\bibnamefont {He}}, \bibinfo {author} {\bibfnamefont
  {T.}~\bibnamefont {Song}}, \bibinfo {author} {\bibfnamefont {Z.}~\bibnamefont
  {Lin}}, \bibinfo {author} {\bibfnamefont {C.}~\bibnamefont {Wang}}, \bibinfo
  {author} {\bibfnamefont {D.}~\bibnamefont {Cobden}}, \bibinfo {author}
  {\bibfnamefont {J.-H.}\ \bibnamefont {Chu}}, \bibinfo {author} {\bibfnamefont
  {Y.-T.}\ \bibnamefont {Cui}},  \emph {et~al.},\ }\href {\doibase
  10.1038/s41467-022-29259-8} {\bibfield  {journal} {\bibinfo  {journal}
  {Nature communications}\ }\textbf {\bibinfo {volume} {13}},\ \bibinfo {pages}
  {1} (\bibinfo {year} {2022})}\BibitemShut {NoStop}%
\bibitem [{\citenamefont {Lai}\ \emph {et~al.}(2021)\citenamefont {Lai},
  \citenamefont {Ke}, \citenamefont {Yan}, \citenamefont {McDonald},\ and\
  \citenamefont {McQueeney}}]{PhysRevB.103.184429}%
  \BibitemOpen
  \bibfield  {author} {\bibinfo {author} {\bibfnamefont {Y.}~\bibnamefont
  {Lai}}, \bibinfo {author} {\bibfnamefont {L.}~\bibnamefont {Ke}}, \bibinfo
  {author} {\bibfnamefont {J.}~\bibnamefont {Yan}}, \bibinfo {author}
  {\bibfnamefont {R.~D.}\ \bibnamefont {McDonald}}, \ and\ \bibinfo {author}
  {\bibfnamefont {R.~J.}\ \bibnamefont {McQueeney}},\ }\href {\doibase
  10.1103/PhysRevB.103.184429} {\bibfield  {journal} {\bibinfo  {journal}
  {Phys. Rev. B}\ }\textbf {\bibinfo {volume} {103}},\ \bibinfo {pages}
  {184429} (\bibinfo {year} {2021})}\BibitemShut {NoStop}%
\bibitem [{\citenamefont {Huang}\ \emph {et~al.}(2020)\citenamefont {Huang},
  \citenamefont {Du}, \citenamefont {Yan},\ and\ \citenamefont
  {Wu}}]{PhysRevMaterials.4.121202}%
  \BibitemOpen
  \bibfield  {author} {\bibinfo {author} {\bibfnamefont {Z.}~\bibnamefont
  {Huang}}, \bibinfo {author} {\bibfnamefont {M.-H.}\ \bibnamefont {Du}},
  \bibinfo {author} {\bibfnamefont {J.}~\bibnamefont {Yan}}, \ and\ \bibinfo
  {author} {\bibfnamefont {W.}~\bibnamefont {Wu}},\ }\href {\doibase
  10.1103/PhysRevMaterials.4.121202} {\bibfield  {journal} {\bibinfo  {journal}
  {Phys. Rev. Materials}\ }\textbf {\bibinfo {volume} {4}},\ \bibinfo {pages}
  {121202} (\bibinfo {year} {2020})}\BibitemShut {NoStop}%
\bibitem [{\citenamefont {Liu}\ \emph {et~al.}(2021{\natexlab{b}})\citenamefont
  {Liu}, \citenamefont {Wang}, \citenamefont {Zheng}, \citenamefont {Huang},
  \citenamefont {Wang}, \citenamefont {Chi}, \citenamefont {Wu}, \citenamefont
  {Chakoumakos}, \citenamefont {McGuire}, \citenamefont {Sales}, \citenamefont
  {Wu},\ and\ \citenamefont {Yan}}]{PhysRevX.11.021033}%
  \BibitemOpen
  \bibfield  {author} {\bibinfo {author} {\bibfnamefont {Y.}~\bibnamefont
  {Liu}}, \bibinfo {author} {\bibfnamefont {L.-L.}\ \bibnamefont {Wang}},
  \bibinfo {author} {\bibfnamefont {Q.}~\bibnamefont {Zheng}}, \bibinfo
  {author} {\bibfnamefont {Z.}~\bibnamefont {Huang}}, \bibinfo {author}
  {\bibfnamefont {X.}~\bibnamefont {Wang}}, \bibinfo {author} {\bibfnamefont
  {M.}~\bibnamefont {Chi}}, \bibinfo {author} {\bibfnamefont {Y.}~\bibnamefont
  {Wu}}, \bibinfo {author} {\bibfnamefont {B.~C.}\ \bibnamefont {Chakoumakos}},
  \bibinfo {author} {\bibfnamefont {M.~A.}\ \bibnamefont {McGuire}}, \bibinfo
  {author} {\bibfnamefont {B.~C.}\ \bibnamefont {Sales}}, \bibinfo {author}
  {\bibfnamefont {W.}~\bibnamefont {Wu}}, \ and\ \bibinfo {author}
  {\bibfnamefont {J.}~\bibnamefont {Yan}},\ }\href {\doibase
  10.1103/PhysRevX.11.021033} {\bibfield  {journal} {\bibinfo  {journal} {Phys.
  Rev. X}\ }\textbf {\bibinfo {volume} {11}},\ \bibinfo {pages} {021033}
  (\bibinfo {year} {2021}{\natexlab{b}})}\BibitemShut {NoStop}%
\bibitem [{\citenamefont {Hou}\ \emph {et~al.}(2020)\citenamefont {Hou},
  \citenamefont {Yao}, \citenamefont {Zhou}, \citenamefont {Ma}, \citenamefont
  {Han}, \citenamefont {Hao}, \citenamefont {Wu}, \citenamefont {Zhang},
  \citenamefont {Sun}, \citenamefont {Liu}, \citenamefont {Zhao}, \citenamefont
  {Liu},\ and\ \citenamefont {Lin}}]{acsnano.0c03149}%
  \BibitemOpen
  \bibfield  {author} {\bibinfo {author} {\bibfnamefont {F.}~\bibnamefont
  {Hou}}, \bibinfo {author} {\bibfnamefont {Q.}~\bibnamefont {Yao}}, \bibinfo
  {author} {\bibfnamefont {C.-S.}\ \bibnamefont {Zhou}}, \bibinfo {author}
  {\bibfnamefont {X.-M.}\ \bibnamefont {Ma}}, \bibinfo {author} {\bibfnamefont
  {M.}~\bibnamefont {Han}}, \bibinfo {author} {\bibfnamefont {Y.-J.}\
  \bibnamefont {Hao}}, \bibinfo {author} {\bibfnamefont {X.}~\bibnamefont
  {Wu}}, \bibinfo {author} {\bibfnamefont {Y.}~\bibnamefont {Zhang}}, \bibinfo
  {author} {\bibfnamefont {H.}~\bibnamefont {Sun}}, \bibinfo {author}
  {\bibfnamefont {C.}~\bibnamefont {Liu}}, \bibinfo {author} {\bibfnamefont
  {Y.}~\bibnamefont {Zhao}}, \bibinfo {author} {\bibfnamefont {Q.}~\bibnamefont
  {Liu}}, \ and\ \bibinfo {author} {\bibfnamefont {J.}~\bibnamefont {Lin}},\
  }\href {\doibase 10.1021/acsnano.0c03149} {\bibfield  {journal} {\bibinfo
  {journal} {ACS Nano}\ }\textbf {\bibinfo {volume} {14}},\ \bibinfo {pages}
  {11262} (\bibinfo {year} {2020})}\BibitemShut {NoStop}%
\bibitem [{\citenamefont {Yuan}\ \emph {et~al.}(2020)\citenamefont {Yuan},
  \citenamefont {Wang}, \citenamefont {Li}, \citenamefont {Li}, \citenamefont
  {Ji}, \citenamefont {Hao}, \citenamefont {Wu}, \citenamefont {He},
  \citenamefont {Wang}, \citenamefont {Xu}, \citenamefont {Duan}, \citenamefont
  {Li},\ and\ \citenamefont {Xue}}]{acs.nanolett.0c00031}%
  \BibitemOpen
  \bibfield  {author} {\bibinfo {author} {\bibfnamefont {Y.}~\bibnamefont
  {Yuan}}, \bibinfo {author} {\bibfnamefont {X.}~\bibnamefont {Wang}}, \bibinfo
  {author} {\bibfnamefont {H.}~\bibnamefont {Li}}, \bibinfo {author}
  {\bibfnamefont {J.}~\bibnamefont {Li}}, \bibinfo {author} {\bibfnamefont
  {Y.}~\bibnamefont {Ji}}, \bibinfo {author} {\bibfnamefont {Z.}~\bibnamefont
  {Hao}}, \bibinfo {author} {\bibfnamefont {Y.}~\bibnamefont {Wu}}, \bibinfo
  {author} {\bibfnamefont {K.}~\bibnamefont {He}}, \bibinfo {author}
  {\bibfnamefont {Y.}~\bibnamefont {Wang}}, \bibinfo {author} {\bibfnamefont
  {Y.}~\bibnamefont {Xu}}, \bibinfo {author} {\bibfnamefont {W.}~\bibnamefont
  {Duan}}, \bibinfo {author} {\bibfnamefont {W.}~\bibnamefont {Li}}, \ and\
  \bibinfo {author} {\bibfnamefont {Q.-K.}\ \bibnamefont {Xue}},\ }\href
  {\doibase 10.1021/acs.nanolett.0c00031} {\bibfield  {journal} {\bibinfo
  {journal} {Nano Letters}\ }\textbf {\bibinfo {volume} {20}},\ \bibinfo
  {pages} {3271} (\bibinfo {year} {2020})}\BibitemShut {NoStop}%
\bibitem [{\citenamefont {Sun}\ \emph {et~al.}(2020)\citenamefont {Sun},
  \citenamefont {Wang}, \citenamefont {Zhang}, \citenamefont {Chen},
  \citenamefont {Zhao}, \citenamefont {Liu}, \citenamefont {Liu}, \citenamefont
  {Chen}, \citenamefont {Lu},\ and\ \citenamefont {Xie}}]{PhysRevB.102.241406}%
  \BibitemOpen
  \bibfield  {author} {\bibinfo {author} {\bibfnamefont {H.-P.}\ \bibnamefont
  {Sun}}, \bibinfo {author} {\bibfnamefont {C.~M.}\ \bibnamefont {Wang}},
  \bibinfo {author} {\bibfnamefont {S.-B.}\ \bibnamefont {Zhang}}, \bibinfo
  {author} {\bibfnamefont {R.}~\bibnamefont {Chen}}, \bibinfo {author}
  {\bibfnamefont {Y.}~\bibnamefont {Zhao}}, \bibinfo {author} {\bibfnamefont
  {C.}~\bibnamefont {Liu}}, \bibinfo {author} {\bibfnamefont {Q.}~\bibnamefont
  {Liu}}, \bibinfo {author} {\bibfnamefont {C.}~\bibnamefont {Chen}}, \bibinfo
  {author} {\bibfnamefont {H.-Z.}\ \bibnamefont {Lu}}, \ and\ \bibinfo {author}
  {\bibfnamefont {X.~C.}\ \bibnamefont {Xie}},\ }\href {\doibase
  10.1103/PhysRevB.102.241406} {\bibfield  {journal} {\bibinfo  {journal}
  {Phys. Rev. B}\ }\textbf {\bibinfo {volume} {102}},\ \bibinfo {pages}
  {241406} (\bibinfo {year} {2020})}\BibitemShut {NoStop}%
\bibitem [{\citenamefont {Shikin}\ \emph {et~al.}(2020)\citenamefont {Shikin},
  \citenamefont {Estyunin}, \citenamefont {Klimovskikh}, \citenamefont
  {Filnov}, \citenamefont {Schwier}, \citenamefont {Kumar}, \citenamefont
  {Miyamoto}, \citenamefont {Okuda}, \citenamefont {Kimura}, \citenamefont
  {Kuroda} \emph {et~al.}}]{shikin2020nature}%
  \BibitemOpen
  \bibfield  {author} {\bibinfo {author} {\bibfnamefont {A.~M.}\ \bibnamefont
  {Shikin}}, \bibinfo {author} {\bibfnamefont {D.}~\bibnamefont {Estyunin}},
  \bibinfo {author} {\bibfnamefont {I.~I.}\ \bibnamefont {Klimovskikh}},
  \bibinfo {author} {\bibfnamefont {S.}~\bibnamefont {Filnov}}, \bibinfo
  {author} {\bibfnamefont {E.}~\bibnamefont {Schwier}}, \bibinfo {author}
  {\bibfnamefont {S.}~\bibnamefont {Kumar}}, \bibinfo {author} {\bibfnamefont
  {K.}~\bibnamefont {Miyamoto}}, \bibinfo {author} {\bibfnamefont
  {T.}~\bibnamefont {Okuda}}, \bibinfo {author} {\bibfnamefont
  {A.}~\bibnamefont {Kimura}}, \bibinfo {author} {\bibfnamefont
  {K.}~\bibnamefont {Kuroda}},  \emph {et~al.},\ }\href {\doibase
  10.1038/s41598-020-70089-9} {\bibfield  {journal} {\bibinfo  {journal}
  {Scientific Reports}\ }\textbf {\bibinfo {volume} {10}},\ \bibinfo {pages}
  {1} (\bibinfo {year} {2020})}\BibitemShut {NoStop}%
\bibitem [{\citenamefont {Wang}\ \emph {et~al.}(2022)\citenamefont {Wang},
  \citenamefont {Wang}, \citenamefont {Xing},\ and\ \citenamefont
  {Zhang}}]{wang220508204}%
  \BibitemOpen
  \bibfield  {author} {\bibinfo {author} {\bibfnamefont {D.}~\bibnamefont
  {Wang}}, \bibinfo {author} {\bibfnamefont {H.}~\bibnamefont {Wang}}, \bibinfo
  {author} {\bibfnamefont {D.}~\bibnamefont {Xing}}, \ and\ \bibinfo {author}
  {\bibfnamefont {H.}~\bibnamefont {Zhang}},\ }\href
  {https://arxiv.org/pdf/2205.08204v1.pdf} {\bibfield  {journal} {\bibinfo
  {journal} {arXiv:2205.08204}\ } (\bibinfo {year} {2022})}\BibitemShut
  {NoStop}%
\bibitem [{\citenamefont {Shikin}\ \emph {et~al.}(2022)\citenamefont {Shikin},
  \citenamefont {Makarova}, \citenamefont {Eryzhenkov}, \citenamefont
  {Usachov}, \citenamefont {Estyunin}, \citenamefont {Glazkova}, \citenamefont
  {Klimovskikh}, \citenamefont {Rybkin},\ and\ \citenamefont
  {Tarasov}}]{Shikin2022factors}%
  \BibitemOpen
  \bibfield  {author} {\bibinfo {author} {\bibfnamefont {A.~M.}\ \bibnamefont
  {Shikin}}, \bibinfo {author} {\bibfnamefont {T.~P.}\ \bibnamefont
  {Makarova}}, \bibinfo {author} {\bibfnamefont {A.~V.}\ \bibnamefont
  {Eryzhenkov}}, \bibinfo {author} {\bibfnamefont {D.~Y.}\ \bibnamefont
  {Usachov}}, \bibinfo {author} {\bibfnamefont {D.~A.}\ \bibnamefont
  {Estyunin}}, \bibinfo {author} {\bibfnamefont {D.~A.}\ \bibnamefont
  {Glazkova}}, \bibinfo {author} {\bibfnamefont {I.~I.}\ \bibnamefont
  {Klimovskikh}}, \bibinfo {author} {\bibfnamefont {A.~G.}\ \bibnamefont
  {Rybkin}}, \ and\ \bibinfo {author} {\bibfnamefont {A.~V.}\ \bibnamefont
  {Tarasov}},\ }\href {https://arxiv.org/pdf/2205.07501v1.pdf} {\bibfield
  {journal} {\bibinfo  {journal} {arXiv:2205.07501}\ } (\bibinfo {year}
  {2022})}\BibitemShut {NoStop}%
\bibitem [{\citenamefont {Garnica}\ \emph {et~al.}(2022)\citenamefont
  {Garnica}, \citenamefont {Otrokov}, \citenamefont {Aguilar}, \citenamefont
  {Klimovskikh}, \citenamefont {Estyunin}, \citenamefont {Aliev}, \citenamefont
  {Amiraslanov}, \citenamefont {Abdullayev}, \citenamefont {Zverev},
  \citenamefont {Babanly} \emph {et~al.}}]{garnica2022native}%
  \BibitemOpen
  \bibfield  {author} {\bibinfo {author} {\bibfnamefont {M.}~\bibnamefont
  {Garnica}}, \bibinfo {author} {\bibfnamefont {M.}~\bibnamefont {Otrokov}},
  \bibinfo {author} {\bibfnamefont {P.~C.}\ \bibnamefont {Aguilar}}, \bibinfo
  {author} {\bibfnamefont {I.}~\bibnamefont {Klimovskikh}}, \bibinfo {author}
  {\bibfnamefont {D.}~\bibnamefont {Estyunin}}, \bibinfo {author}
  {\bibfnamefont {Z.}~\bibnamefont {Aliev}}, \bibinfo {author} {\bibfnamefont
  {I.}~\bibnamefont {Amiraslanov}}, \bibinfo {author} {\bibfnamefont
  {N.}~\bibnamefont {Abdullayev}}, \bibinfo {author} {\bibfnamefont
  {V.}~\bibnamefont {Zverev}}, \bibinfo {author} {\bibfnamefont
  {M.}~\bibnamefont {Babanly}},  \emph {et~al.},\ }\href {\doibase
  10.1038/s41535-021-00414-6} {\bibfield  {journal} {\bibinfo  {journal} {npj
  Quantum Materials}\ }\textbf {\bibinfo {volume} {7}},\ \bibinfo {pages} {1}
  (\bibinfo {year} {2022})}\BibitemShut {NoStop}%
\bibitem [{\citenamefont {Kresse}\ and\ \citenamefont
  {Furthm{\"u}ller}(1996{\natexlab{a}})}]{VASP}%
  \BibitemOpen
  \bibfield  {author} {\bibinfo {author} {\bibfnamefont {G.}~\bibnamefont
  {Kresse}}\ and\ \bibinfo {author} {\bibfnamefont {J.}~\bibnamefont
  {Furthm{\"u}ller}},\ }\href {\doibase
  https://doi.org/10.1016/0927-0256(96)00008-0} {\bibfield  {journal} {\bibinfo
   {journal} {Comput. Mater. Sci.}\ }\textbf {\bibinfo {volume} {6}},\ \bibinfo
  {pages} {15} (\bibinfo {year} {1996}{\natexlab{a}})}\BibitemShut {NoStop}%
\bibitem [{\citenamefont {Kresse}\ and\ \citenamefont
  {Furthm{\"u}ller}(1996{\natexlab{b}})}]{PRB54p11169}%
  \BibitemOpen
  \bibfield  {author} {\bibinfo {author} {\bibfnamefont {G.}~\bibnamefont
  {Kresse}}\ and\ \bibinfo {author} {\bibfnamefont {J.}~\bibnamefont
  {Furthm{\"u}ller}},\ }\href {\doibase 10.1103/PhysRevB.54.11169} {\bibfield
  {journal} {\bibinfo  {journal} {Phys. Rev. B}\ }\textbf {\bibinfo {volume}
  {54}},\ \bibinfo {pages} {11169} (\bibinfo {year}
  {1996}{\natexlab{b}})}\BibitemShut {NoStop}%
\bibitem [{\citenamefont {Perdew}\ \emph {et~al.}(1996)\citenamefont {Perdew},
  \citenamefont {Burke},\ and\ \citenamefont {Ernzerhof}}]{PRL77p3865}%
  \BibitemOpen
  \bibfield  {author} {\bibinfo {author} {\bibfnamefont {J.~P.}\ \bibnamefont
  {Perdew}}, \bibinfo {author} {\bibfnamefont {K.}~\bibnamefont {Burke}}, \
  and\ \bibinfo {author} {\bibfnamefont {M.}~\bibnamefont {Ernzerhof}},\ }\href
  {\doibase 10.1103/PhysRevLett.77.3865} {\bibfield  {journal} {\bibinfo
  {journal} {Phys. Rev. Lett.}\ }\textbf {\bibinfo {volume} {77}},\ \bibinfo
  {pages} {3865} (\bibinfo {year} {1996})}\BibitemShut {NoStop}%
\bibitem [{SM()}]{SM}%
  \BibitemOpen
  \href@noop {} {}\bibinfo {note} {See the Supplemental Material for magnetic
  configurations as well as total energies of different surfaces, band
  structures of MnBi$_2$Te$_4$ surfaces with different single antisites and
  co-antisites under different densities, charge distributions of SVB and SCB
  of defects not shown in the main text, band structure evolution of
  MnBi$_2$Te$_4$ with surface vdW gap expansion (under AFM and FM), as well as
  band structures of MnBi$_2$Te$_4$-terminations with different co-antisites in
  MnBi$_4$Te$_7$ and MnBi$_6$Te$_{10}$.}\BibitemShut {Stop}%
\bibitem [{\citenamefont {Du}\ \emph {et~al.}(2021)\citenamefont {Du},
  \citenamefont {Yan}, \citenamefont {Cooper},\ and\ \citenamefont
  {Eisenbach}}]{AFMp2006516}%
  \BibitemOpen
  \bibfield  {author} {\bibinfo {author} {\bibfnamefont {M.-H.}\ \bibnamefont
  {Du}}, \bibinfo {author} {\bibfnamefont {J.}~\bibnamefont {Yan}}, \bibinfo
  {author} {\bibfnamefont {V.~R.}\ \bibnamefont {Cooper}}, \ and\ \bibinfo
  {author} {\bibfnamefont {M.}~\bibnamefont {Eisenbach}},\ }\href {\doibase
  https://doi.org/10.1002/adfm.202006516} {\bibfield  {journal} {\bibinfo
  {journal} {Advanced Functional Materials}\ }\textbf {\bibinfo {volume}
  {31}},\ \bibinfo {pages} {2006516} (\bibinfo {year} {2021})}\BibitemShut
  {NoStop}%
\bibitem [{\citenamefont {Liu}\ \emph {et~al.}(2022)\citenamefont {Liu},
  \citenamefont {Lei}, \citenamefont {Kim}, \citenamefont {Li}, \citenamefont
  {Frammolino}, \citenamefont {Yan}, \citenamefont {Macdonald},\ and\
  \citenamefont {Shih}}]{arxiv.2205.09195}%
  \BibitemOpen
  \bibfield  {author} {\bibinfo {author} {\bibfnamefont {M.}~\bibnamefont
  {Liu}}, \bibinfo {author} {\bibfnamefont {C.}~\bibnamefont {Lei}}, \bibinfo
  {author} {\bibfnamefont {H.}~\bibnamefont {Kim}}, \bibinfo {author}
  {\bibfnamefont {Y.}~\bibnamefont {Li}}, \bibinfo {author} {\bibfnamefont
  {L.}~\bibnamefont {Frammolino}}, \bibinfo {author} {\bibfnamefont
  {J.}~\bibnamefont {Yan}}, \bibinfo {author} {\bibfnamefont {A.~H.}\
  \bibnamefont {Macdonald}}, \ and\ \bibinfo {author} {\bibfnamefont {C.-K.}\
  \bibnamefont {Shih}},\ }\href {https://arxiv.org/abs/2205.09195} {\bibfield
  {journal} {\bibinfo  {journal} {arxiv.2205.09195}\ } (\bibinfo {year}
  {2022})}\BibitemShut {NoStop}%
\bibitem [{not()}]{notice}%
  \BibitemOpen
  \href@noop {} {}\bibinfo {note} {For the perfect surface under AFM, the
  charge density of SVB (SCB) is distributed in two bands with nearly the same
  energies. Thus the charge densities of the two band are added together for
  SVB (SCB) charge densities of the perfect surface under AFM.}\BibitemShut
  {Stop}%
\bibitem [{\citenamefont {Hu}\ \emph {et~al.}(2021)\citenamefont {Hu},
  \citenamefont {Lien}, \citenamefont {Feng}, \citenamefont {Mackey},
  \citenamefont {Tien}, \citenamefont {Mazin}, \citenamefont {Cao},
  \citenamefont {Chang},\ and\ \citenamefont {Ni}}]{NiNi2021PRB}%
  \BibitemOpen
  \bibfield  {author} {\bibinfo {author} {\bibfnamefont {C.}~\bibnamefont
  {Hu}}, \bibinfo {author} {\bibfnamefont {S.-W.}\ \bibnamefont {Lien}},
  \bibinfo {author} {\bibfnamefont {E.}~\bibnamefont {Feng}}, \bibinfo {author}
  {\bibfnamefont {S.}~\bibnamefont {Mackey}}, \bibinfo {author} {\bibfnamefont
  {H.-J.}\ \bibnamefont {Tien}}, \bibinfo {author} {\bibfnamefont {I.~I.}\
  \bibnamefont {Mazin}}, \bibinfo {author} {\bibfnamefont {H.}~\bibnamefont
  {Cao}}, \bibinfo {author} {\bibfnamefont {T.-R.}\ \bibnamefont {Chang}}, \
  and\ \bibinfo {author} {\bibfnamefont {N.}~\bibnamefont {Ni}},\ }\href
  {\doibase 10.1103/PhysRevB.104.054422} {\bibfield  {journal} {\bibinfo
  {journal} {Phys. Rev. B}\ }\textbf {\bibinfo {volume} {104}},\ \bibinfo
  {pages} {054422} (\bibinfo {year} {2021})}\BibitemShut {NoStop}%
\bibitem [{\citenamefont {Guan}\ \emph {et~al.}(2022)\citenamefont {Guan},
  \citenamefont {Yan}, \citenamefont {Lee}, \citenamefont {Gui}, \citenamefont
  {Ning}, \citenamefont {Ning}, \citenamefont {Zhu}, \citenamefont
  {Kothakonda}, \citenamefont {Xu}, \citenamefont {Ke}, \citenamefont {Sun},
  \citenamefont {Xie}, \citenamefont {Yang},\ and\ \citenamefont
  {Mao}}]{guan2022PRM}%
  \BibitemOpen
  \bibfield  {author} {\bibinfo {author} {\bibfnamefont {Y.~D.}\ \bibnamefont
  {Guan}}, \bibinfo {author} {\bibfnamefont {C.~H.}\ \bibnamefont {Yan}},
  \bibinfo {author} {\bibfnamefont {S.~H.}\ \bibnamefont {Lee}}, \bibinfo
  {author} {\bibfnamefont {X.}~\bibnamefont {Gui}}, \bibinfo {author}
  {\bibfnamefont {W.}~\bibnamefont {Ning}}, \bibinfo {author} {\bibfnamefont
  {J.~L.}\ \bibnamefont {Ning}}, \bibinfo {author} {\bibfnamefont {Y.~L.}\
  \bibnamefont {Zhu}}, \bibinfo {author} {\bibfnamefont {M.}~\bibnamefont
  {Kothakonda}}, \bibinfo {author} {\bibfnamefont {C.~Q.}\ \bibnamefont {Xu}},
  \bibinfo {author} {\bibfnamefont {X.~L.}\ \bibnamefont {Ke}}, \bibinfo
  {author} {\bibfnamefont {J.~W.}\ \bibnamefont {Sun}}, \bibinfo {author}
  {\bibfnamefont {W.~W.}\ \bibnamefont {Xie}}, \bibinfo {author} {\bibfnamefont
  {S.~L.}\ \bibnamefont {Yang}}, \ and\ \bibinfo {author} {\bibfnamefont
  {Z.~Q.}\ \bibnamefont {Mao}},\ }\href {\doibase
  10.1103/PhysRevMaterials.6.054203} {\bibfield  {journal} {\bibinfo  {journal}
  {Phys. Rev. Materials}\ }\textbf {\bibinfo {volume} {6}},\ \bibinfo {pages}
  {054203} (\bibinfo {year} {2022})}\BibitemShut {NoStop}%
\bibitem [{\citenamefont {Tan}\ and\ \citenamefont {Yan}(2022)}]{Tan2022mbt}%
  \BibitemOpen
  \bibfield  {author} {\bibinfo {author} {\bibfnamefont {H.}~\bibnamefont
  {Tan}}\ and\ \bibinfo {author} {\bibfnamefont {B.}~\bibnamefont {Yan}},\
  }\href {\doibase 10.1103/PhysRevB.105.165130} {\bibfield  {journal} {\bibinfo
   {journal} {Phys. Rev. B}\ }\textbf {\bibinfo {volume} {105}},\ \bibinfo
  {pages} {165130} (\bibinfo {year} {2022})}\BibitemShut {NoStop}%
\end{thebibliography}
%merlin.mbs apsrev4-1.bst 2010-07-25 4.21a (PWD, AO, DPC) hacked
%Control: key (0)
%Control: author (8) initials jnrlst
%Control: editor formatted (1) identically to author
%Control: production of article title (-1) disabled
%Control: page (0) single
%Control: year (1) truncated
%Control: production of eprint (0) enabled
%

\end{document}

\title{Supplemental Material for ``Distinct magnetic gaps between antiferromagnetic and ferromagnetic orders driven by surface defects in the topological magnet MnBi$_2$Te$_4$''}

\author{Hengxin Tan}
\author{Binghai Yan}
\affiliation{Department of Condensed Matter Physics, Weizmann Institute of Science, Rehovot 7610001, Israel}

%\date{\today}

\maketitle

\begin{figure}
\centering
\includegraphics[width=0.8\linewidth]{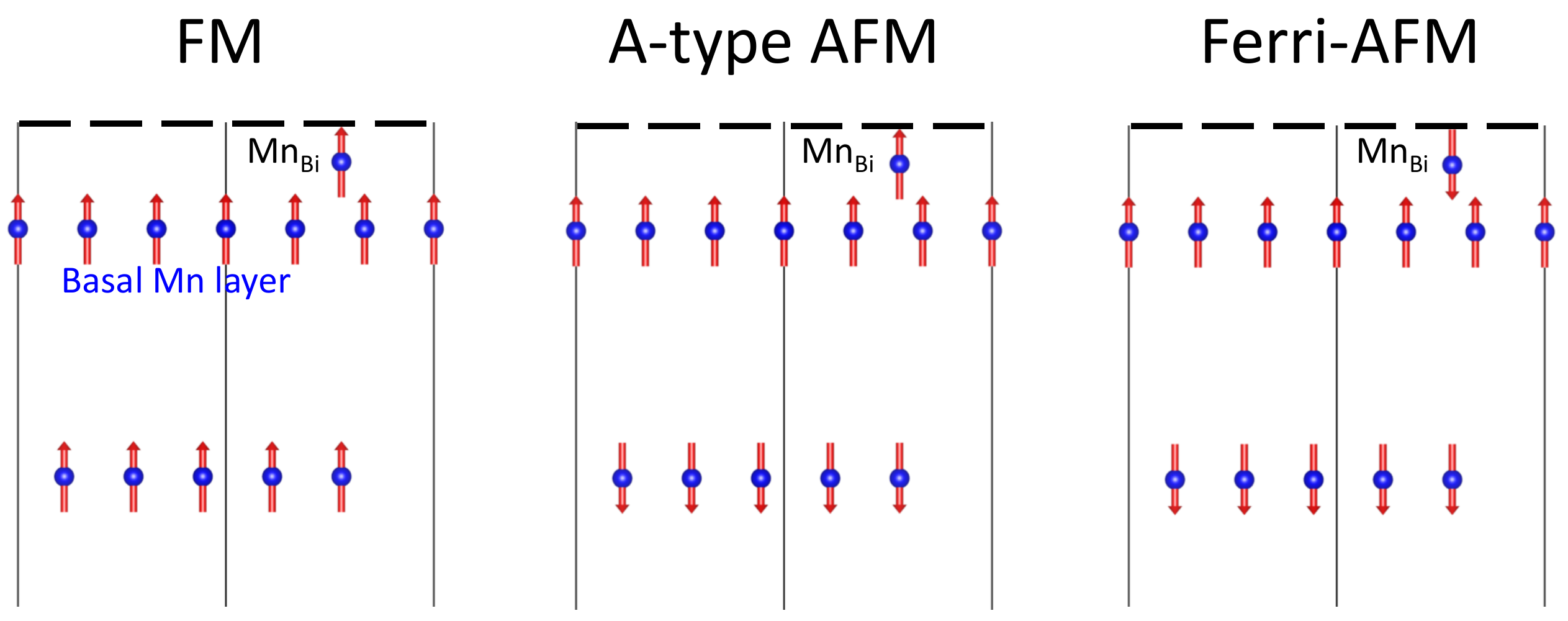}
\caption{\label{FigS1}
Magnetic structures of MnBi$_2$Te$_4$ surfaces with antisites. The sNN structure (in-plane cell size 3$\times$3) is taken as an example to show the three magnetic configurations (FM, AFM, Ferri-AFM), as shown by the red vectors.
For simplicity, only two Mn sublayers (blue spheres) nearest to the surface (bold dashed line) are shown.
Notice that the magnetic moment of Mn$_{\rm{Bi}}$ is parallel to the basal Mn sublayer in A-type AFM while it is antiparallel to the basal Mn sublayer in Ferri-AFM.
In the main text and the following, we do not distinguish between A-type AFM (abbreviated further AFM) and Ferri-AFM if further stated.
}
\end{figure}

\begin{figure}
\centering
\includegraphics[width=0.75\linewidth]{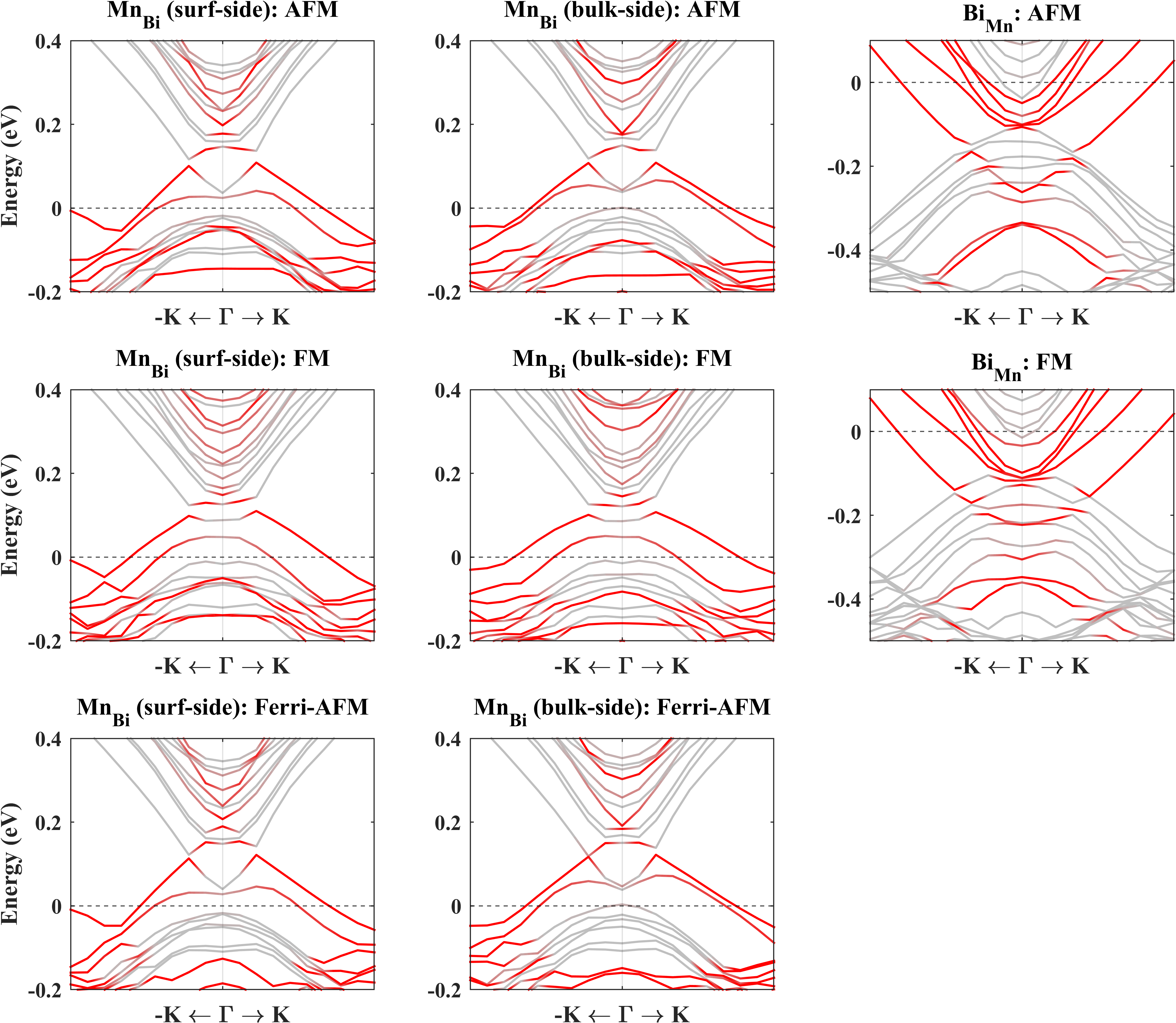}
\caption{\label{FigS2}
band structures of MnBi$_2$Te$_4$ surfaces with single antisite defects.
The slabs have 5 septuple layers with in-plane supercell size 3$\times$3.
The Mn$_{\rm{Bi}}$ antisite has two configurations, i.e., the Mn can replace Bi on the surface side or bulk side (see figure 1 in the main text).
The color red reflects the band weight of the surface van der Waals layer.
Due to the charged defect, the band structures do not show a simple gapped Dirac corn at the bulk band gap.
The band structures are shown along the $-K$-$\Gamma$-$K$ line of the hexagonal Brillouin zone of the perfect surface.
}
\end{figure}

\begin{figure}
\centering
\includegraphics[width=0.75\linewidth]{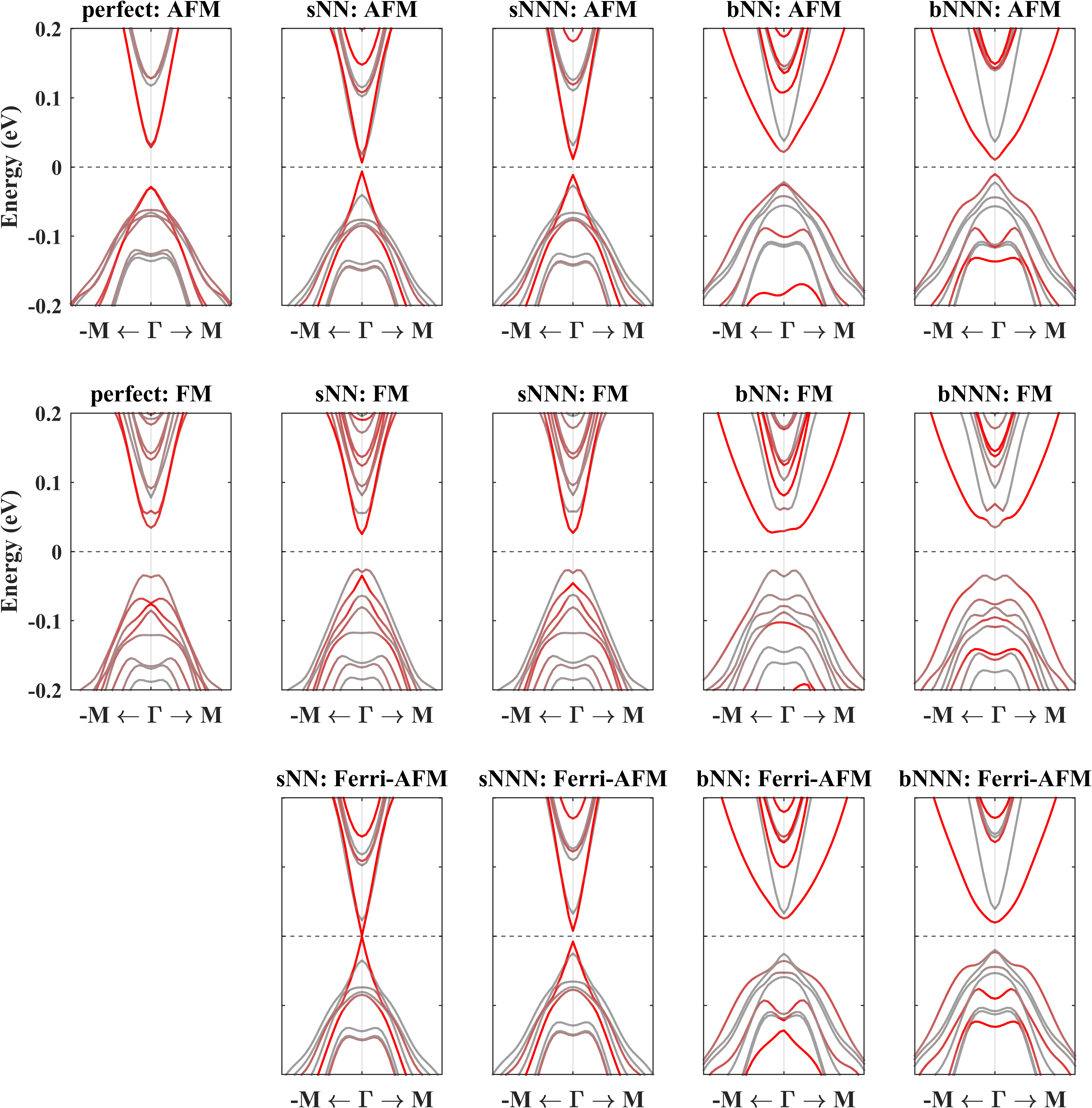}
\caption{\label{FigS3}
band structure of MnBi$_2$Te$_4$ surfaces with cation co-antisites.
The slabs have 5 septuple layers with an in-plane cell size 2$\times$2.
This figure is similar to figure 2 in the main text but for a smaller in-plane supercell (higher antisite density).
}
\end{figure}

\begin{figure}
\centering
\includegraphics[width=0.75\linewidth]{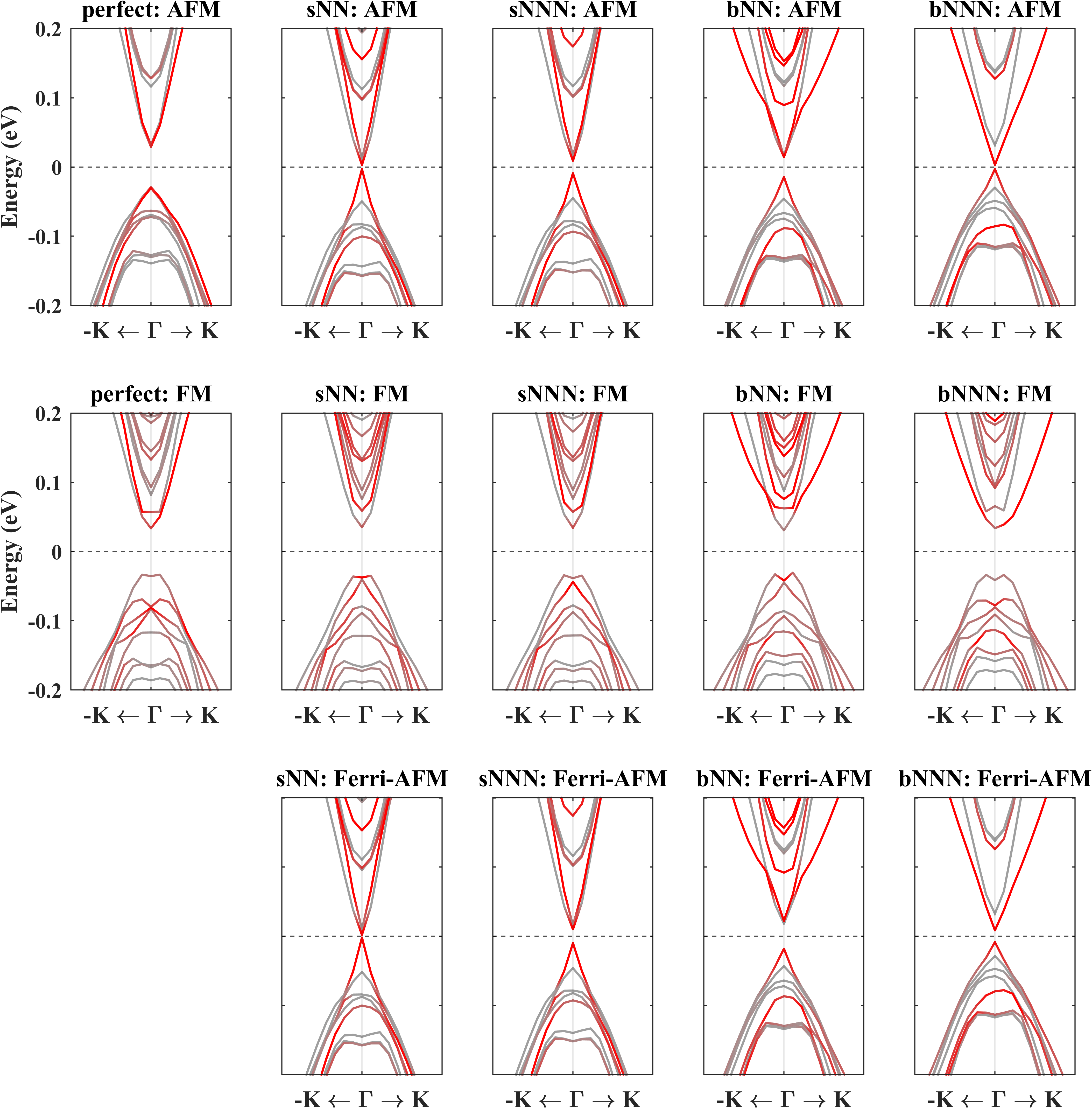}
\caption{\label{FigS4}
band structure of MnBi$_2$Te$_4$ surfaces with cation co-antisites.
The slabs have 5 septuple layers with an in-plane cell size  3$\times$3.
This figure is similar to figure 2 in the main text but for the $k$ path of $-K$-$\Gamma$-$K$ in the hexagonal Brillouin zone of the perfect surface.
}
\end{figure}

\begin{figure}
\centering
\includegraphics[width=0.95\linewidth]{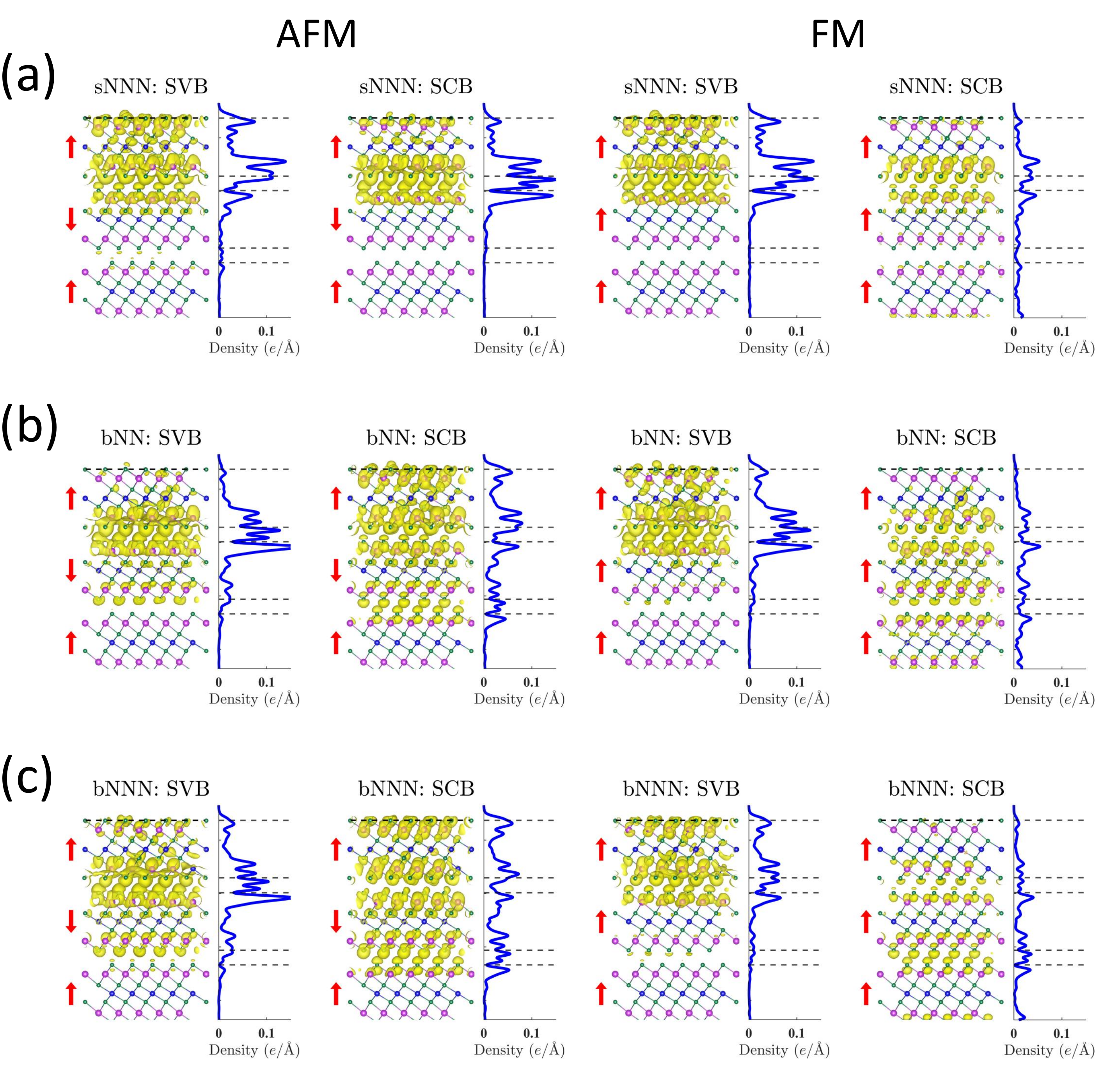}
\caption{\label{FigS5}
Distribution of top surface valence band (SVB) and bottom surface conduction band (SCB) charge densities of MnBi$_2$Te$_4$ surfaces with different co-antisites.
Similar to figure 3 in the main text but for the other three antisite configurations, (a) sNNN, (b) bNN, and (c) bNNN.
The defect configurations are shown in figure 1 in the main text.
}
\end{figure}

\begin{figure}
\centering
\includegraphics[width=0.75\linewidth]{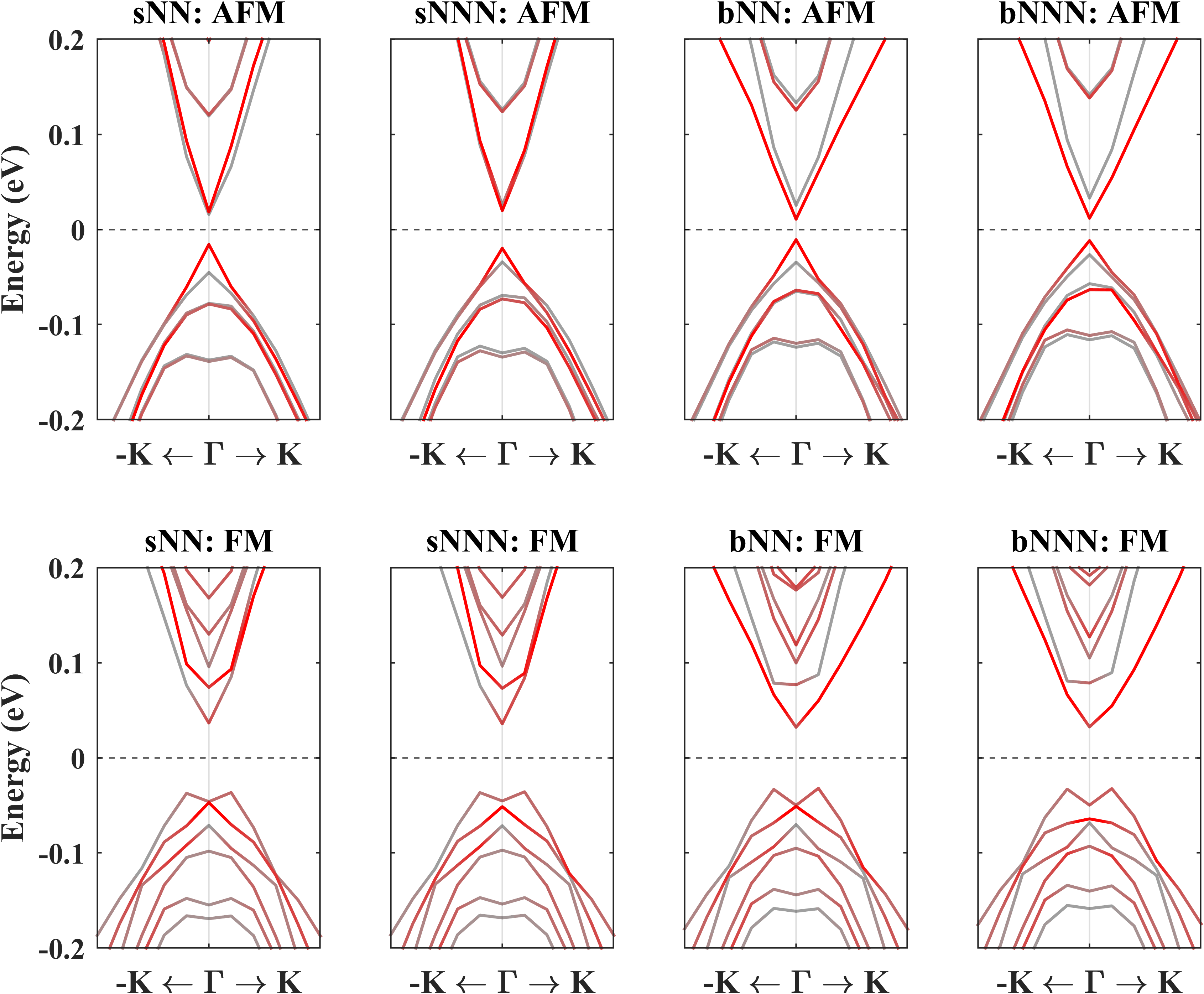}
\caption{\label{FigS6}
Similar to Fig. \ref{FigS4} but for a slab of 4 septuple layers with an in-plane cell size 4$\times$4 (lower antisite density).
}
\end{figure}

\begin{figure}
\centering
\includegraphics[width=0.95\linewidth]{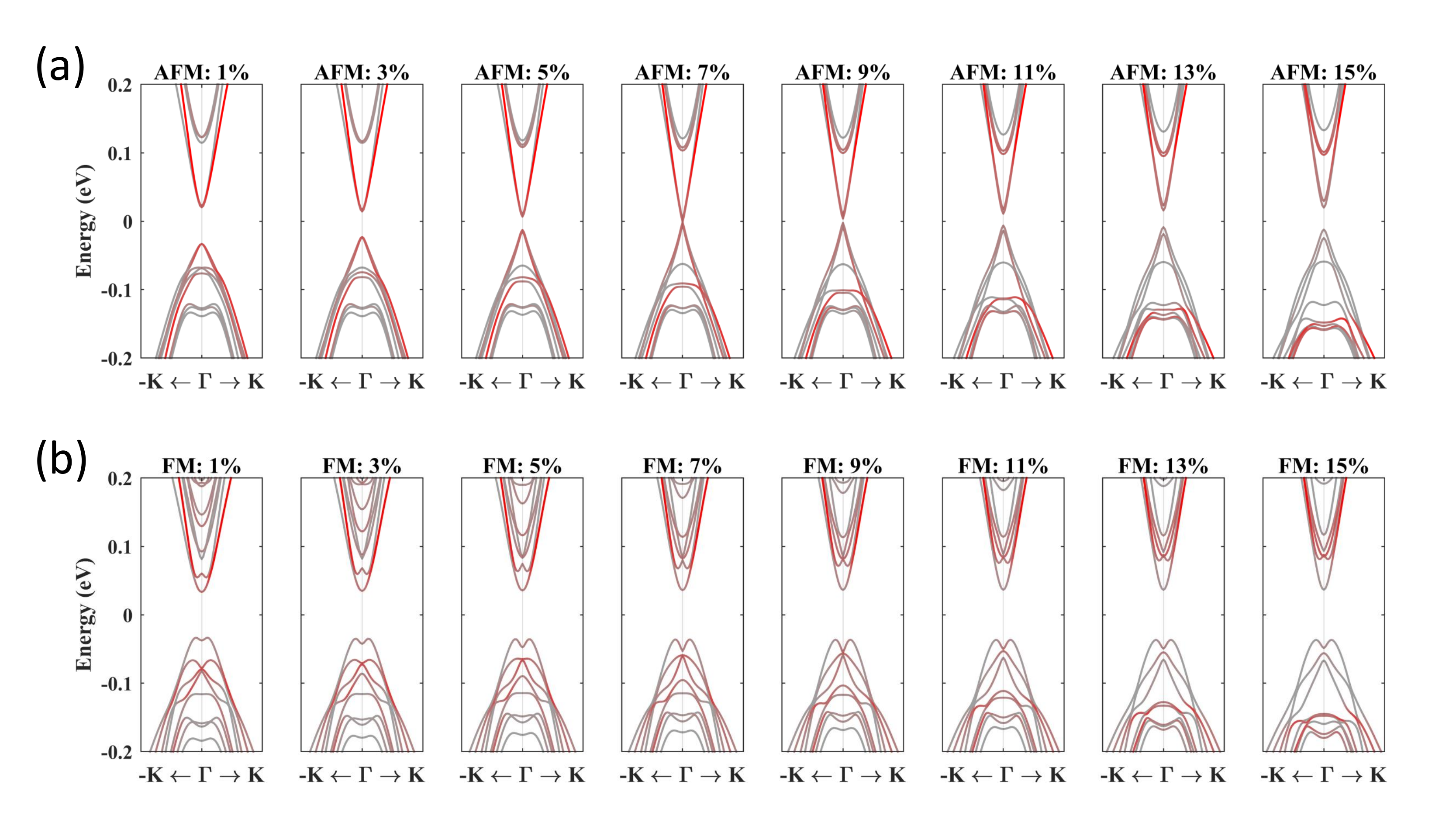}
\caption{\label{FigS7}
Band structure evolution of a 5-septuple-layer MnBi$_2$Te$_4$ slab (no surface defect) with the expansion of the surface van der Waals gap.
(a) and (b) show the band structure under AFM and FM, respectively, where the expansion percentage (compared to the bulk van der Waals gap of MnBi$_2$Te$_4$) is shown in the title of each panel.
The color red reflects the weight of the surface septuple layer in the band structure.
While the magnetic gap is closed at about 7\% expansion of the surface van der Waals under AFM, the magnetic gap is almost unchanged under FM.
}
\end{figure}

\begin{figure}
\centering
\includegraphics[width=0.75\linewidth]{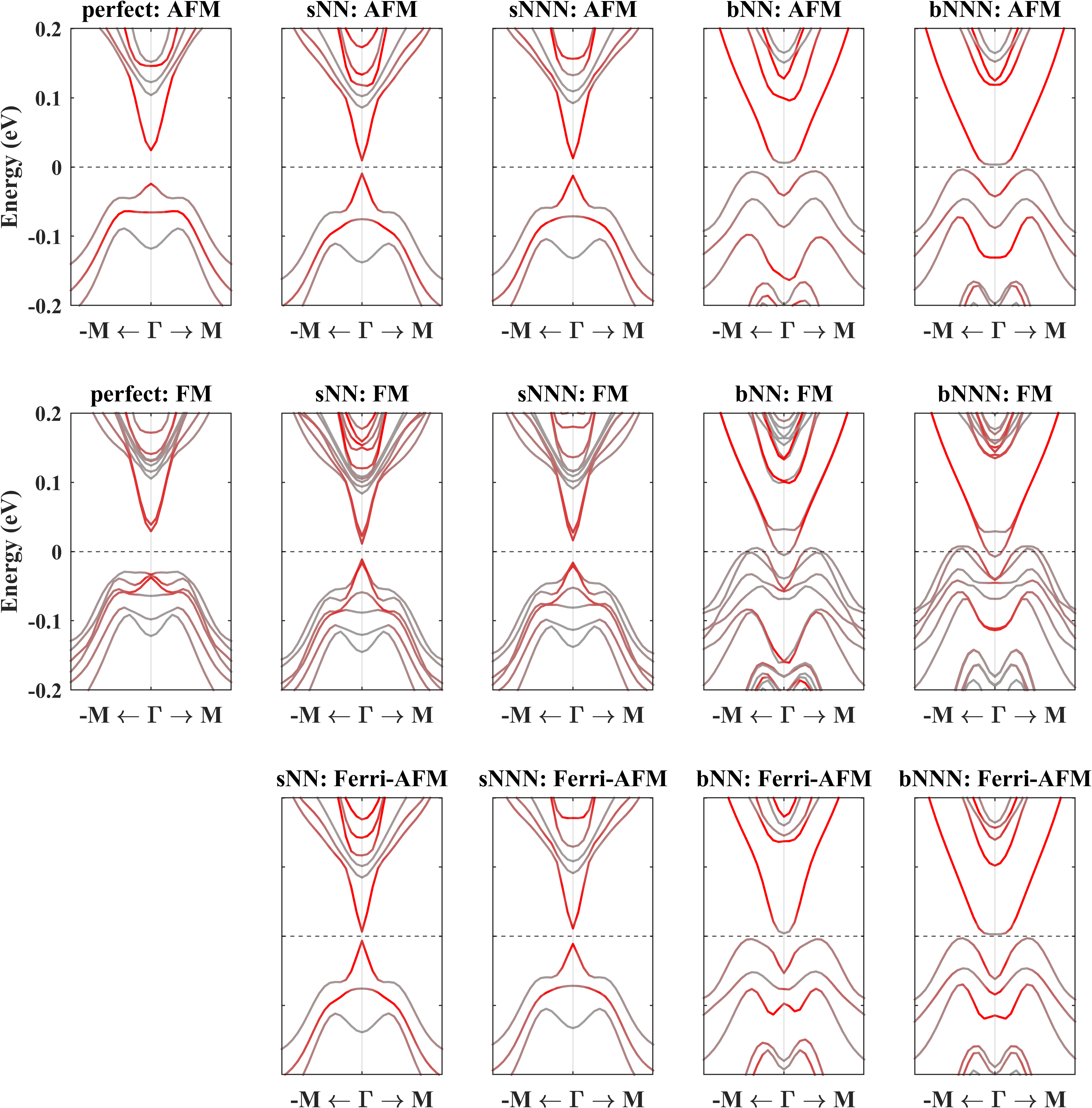}
\caption{\label{FigS8}
Band structure of MnBi$_2$Te$_4$-terminated surfaces of MnBi$_4$Te$_7$ with different surface cation co-antisites.
There are four septuple layers and three Bi$_2$Te$_3$ layers in the slabs.
The in-plane cell size is 2$\times$2.
The magnetic configurations are similar to cases of MnBi$_2$Te$_4$.
The magnetic gap of sNN and sNNN surfaces are decreased under both FM, AFM, and Ferri-AFM, which may indicate the insensitivity of the magnetic gap on the magnetic ordering under defects.
For bNN and bNNN surfaces, the band structure becomes more complex due to the strong hybridization between the surface MnBi$_2$Te$_4$ layer and the Bi$_2$Te$_3$ layer below, leading to the coverage of the magnetic gap by bulk bands (color grey).
The color red reflects the weight of the surface septuple layer.
}
\end{figure}

\begin{figure}
\centering
\includegraphics[width=0.95\linewidth]{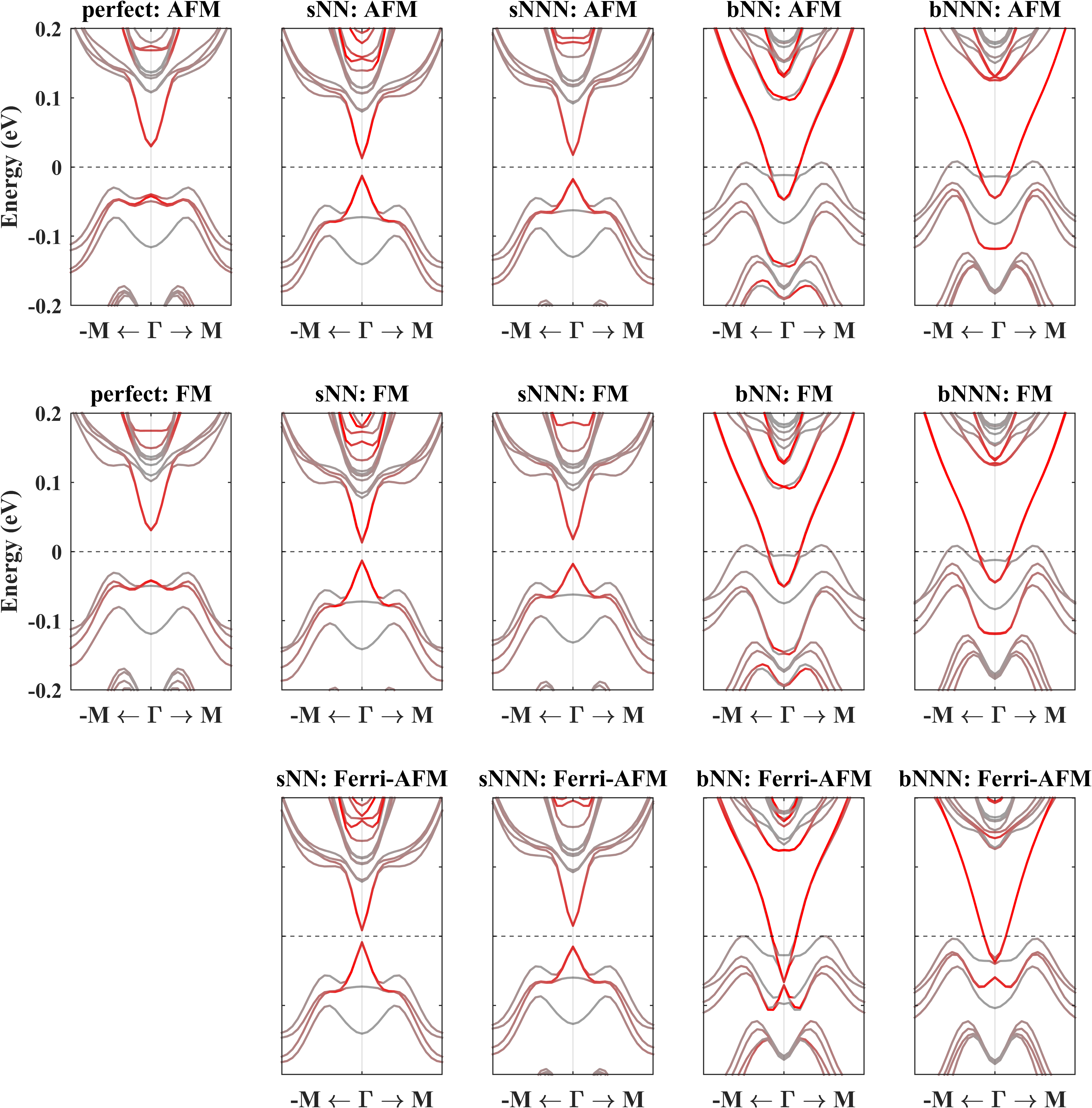}
\caption{\label{FigS9}
Similar to Fig. \ref{FigS8} but for MnBi$_2$Te$_4$-terminated surfaces of MnBi$_6$Te$_{10}$.
}
\end{figure}

\hspace*{\fill} \\
\hspace*{\fill} \\
\hspace*{\fill} \\
\textbf{Table of contents:}
\begin{itemize}
  \item Table \ref{Table_gap} shows the total energies of the different co-antisites on the MnBi$_2$Te$_4$ surface.
  \item Figure \ref{FigS1} displays the three magnetic configurations considered in this work.
  \item figure \ref{FigS2} shows the band structure of MnBi$_2$Te$_4$ with single surface antisite.
  \item Figure \ref{FigS3} shows the band structures of MnBi$_2$Te$_4$ slabs with high surface co-antisite density (in-plane cell size 2$\times$2).
  \item Figure \ref{FigS4} shows the band structures of MnBi$_2$Te$_4$ slabs with different surface co-antisites (in-plane cell size 3$\times$3) along the $-K$-$\Gamma$-$K$ line in the hexagonal Brillouin zone of the perfect surface.
  \item Figure \ref{FigS5} displays the effect of different surface co-antisites on the distribution of charge densities of the top surface valence and bottom surface conduction bands of MnBi$_2$Te$_4$.
  \item Figure \ref{FigS6} is similar to Fig. \ref{FigS4} but for lower densities of surface co-antisites (in-plane cell size 4$\times$4).
  \item Figure \ref{FigS7} contains the evolution of the band structure of a perfect MnBi$_2$Te$_4$ surface under surface van der Waals gap expansion.
  \item Figure \ref{FigS8} and Fig. \ref{FigS9} are similar to Fig. \ref{FigS3} but for MnBi$_2$Te$_4$-terminated surfaces of MnBi$_4$Te$_7$ and MnBi$_6$Te$_{10}$ respectively.
\end{itemize}
%%%%%%%%%%%%%%%%%%%%%%%%%%%%%%%%%%%%%%%%%%%%%%%%%%%%%%%%%%%%%%%%%%%%%%%%%%%%%%%%%%%%%%%%

\clearpage
%%%%%%%%%%%%%%%%%%%%%%%%%%%%%%%%%%%%%%%%%%%%%%%%%%%%%%%%%%%%%%%%%%%%%%%%%%%%%
\begin{table}
\centering
\caption{\label{Table_gap} Total energies of different surface cation co-antisites in MnBi$_2$Te$_4$ (relative to the total energy of the perfect surface under AFM). In the unit of meV.
}
\renewcommand\arraystretch{1.0}
\begin{ruledtabular}
\begin{tabular}{rccccc}
  surface & cell-size & FM & AFM & Ferri-AFM \\ %\multicolumn{3}{c}{Direction of \textbf{m}} \\
  \hline
  \multirow{3}{*}{sNN} &2$\times$2& 293 & 285 & 289 \\
                       &3$\times$3& 341 & 295 & 298 \\
                       &4$\times$4& 415 & 319 & \\
                       \specialrule{0em}{3pt}{3pt}
  \multirow{3}{*}{sNNN}&2$\times$2& 211 & 206 & 202 \\
                       &3$\times$3& 285 & 250 & 256 \\
                       &4$\times$4& 382 & 192 & \\
                       \specialrule{0em}{3pt}{3pt}
  \multirow{3}{*}{bNN} &2$\times$2& 296 & 295 & 301 \\
                       &3$\times$3& 372 & 327 & 330 \\
                       &4$\times$4& 422 & 338 & \\
                       \specialrule{0em}{3pt}{3pt}
  \multirow{3}{*}{bNNN}&2$\times$2& 242 & 236 & 231 \\
                       &3$\times$3& 312 & 278 & 285 \\
                       &4$\times$4& 386 & 305 & \\
\end{tabular}
\end{ruledtabular}
\end{table}
%%%%%%%%%%%%%%%%%%%%%%%%%%%%%%%%%%%%%%%%%%%%%%%%%%%%%%%%%%%%%%%%%%%%%%%%%%%%%

%%%%%%%%%%%%%%%%%%%%%%%%%%%%%%%%%%%%%%%%%%%%%%%%%%%%%%%%%%%%%%%%%%%

%%%%%%%%%%%%%%%%%%%%%%%%%%%%%%%%%%%%%%%%%%%%%%%%%%%%%%%%%%%%%%%%%%%

%%%%%%%%%%%%%%%%%%%%%%%%%%%%%%%%%%%%%%%%%%%%%%%%%%%%%%%%%%%%%%%%%%%

%%%%%%%%%%%%%%%%%%%%%%%%%%%%%%%%%%%%%%%%%%%%%%%%%%%%%%%%%%%%%%%%%%%

%%%%%%%%%%%%%%%%%%%%%%%%%%%%%%%%%%%%%%%%%%%%%%%%%%%%%%%%%%%%%%%%%%%

%%%%%%%%%%%%%%%%%%%%%%%%%%%%%%%%%%%%%%%%%%%%%%%%%%%%%%%%%%%%%%%%%%%

%%%%%%%%%%%%%%%%%%%%%%%%%%%%%%%%%%%%%%%%%%%%%%%%%%%%%%%%%%%%%%%%%%%

%%%%%%%%%%%%%%%%%%%%%%%%%%%%%%%%%%%%%%%%%%%%%%%%%%%%%%%%%%%%%%%%%%%

%%%%%%%%%%%%%%%%%%%%%%%%%%%%%%%%%%%%%%%%%%%%%%%%%%%%%%%%%%%%%%%%%%%

\clearpage
% \bibliography{Reference_MBT}